\documentclass[review]{elsarticle}

\usepackage[colorlinks,citecolor=blue,linktoc=all,linkcolor=cyan]{hyperref}
\usepackage{graphicx}

\usepackage[T1]{fontenc}
\usepackage{dsfont}               
\usepackage{mathrsfs}             
\usepackage{slashed}              
\usepackage{amsmath}
\usepackage{amssymb}
\usepackage{amsbsy}
\usepackage{amsfonts}

\numberwithin{equation}{section}
\numberwithin{table}{section}
\numberwithin{figure}{section}
\usepackage{xcolor}
\usepackage{threeparttable}
\usepackage{longtable}

\journal{Progress in Particle and Nuclear Physics}

\usepackage{titlesec}
\usepackage{sectsty}
\titleformat{\section}{\normalfont\Large\bfseries}{\thesection}{1em}{}
\titleformat{\subsection}{\normalfont\large\bfseries}{\thesubsection}{1em}{}
\titleformat{\subsubsection}{\normalfont\normalsize\bfseries}{\thesubsubsection}{1em}{}

\begin{document}
\sloppy
	
	\begin{frontmatter}
		
		\title{ \vspace{1cm} Decay spectroscopy of heavy and superheavy nuclei}

		\author[mymainaddress,mysecondaryaddress]{Dieter Ackermann}
		

		\ead{d.ackermann@ganil.fr}
		
		\address[mymainaddress]{Grand Acc\'{e}l\'{e}rateur National d'Ions Lourds -- GANIL, CEA/DRF-CNRS/IN2P3, 55027, F-14076 Caen, France}

		\begin{abstract} After more than half a century since the first predictions of the so-called {\it "island of stability of superheavy nuclei"}, exploring the limits of nuclear stability at highest atomic numbers is still one of the most prominent challenges in low-energy nuclear physics. 
        These exotic nuclear species reveal their character and details of some of their properties through their induced or spontaneous disintegration. 
        
        The achievements in the field of superheavy nuclei (SHN) research, which involves studying the production and decay of the heaviest nuclear species, have been reported in a number of review papers. In the introduction of this paper, references are provided to review papers, summarizing the many aspects of SHN research in other disciplines, like chemistry, atomic physics, and earlier work on nuclear structure, including in-beam spectroscopy, and superheavy element (SHE) synthesis.
        
        This review is an attempt to summarize the experimental progress that has been made in recent years by employing the versatile tool park of Decay Spectroscopy After Separation (DSAS) for the heaviest isotopes from $Z$=99 (einsteinium) to $Z$=118 (oganesson). DSAS, with its major instrumentation components heavy-ion accelerator, separator and decay detection, is the only way to access the heaviest nuclei up to oganesson. While in-beam $\gamma$-spectroscopy has reached $^{256}$Rf in terms of the highest atomic number $Z$ and mass number $A$, SHE chemistry succeeded to sort flerovium ($Z$\,=\,114) as the heaviest element into the periodic table. Laser spectroscopy and precise mass measurements are limited basically to the nobelium/fermium region, with high-precision Penning-trap mass-measurements being performed for $^{256}$Lr and $^{257}$Rf, and with the $^{257}$Db mass obtained, using a multi-reflection time-of-flight mass spectrometer (MRToF MS).
        
        Apart from a brief introduction of the method (DSAS) and some nuclear structure features of SHN, the experimental findings reported in literature are summarized in this review, including a table listing the major decay properties, providing a comprehensive collection of references to experimental publications for each known isotope and isomeric state.

        \end{abstract}		
		
	\end{frontmatter}
	
	\newpage
	
	\thispagestyle{empty}
	\tableofcontents
	


\twocolumn	

\newpage
\section{Dedication}\label{Dedication}

A substantial part of my understanding of the nature of these fascinating and unique objects at the upper right end of the Segr\`{e} chart, the superheavy nuclei was formed in the common research activities that I pursued together with {\bf Christophe Theisen}, working on our common publications and in our numerous discussions. Writing our review paper on {\it "Nuclear structure features of very heavy and superheavy nuclei—tracing quantum mechanics towards the ‘island of stability’"}~\cite{Ackermann2017} was more pleasure than effort thanks to a fruitful and rewarding exchange of ideas with him. With his infinite wisdom and profound knowledge of all aspects of nuclear structure and reaction mechanism in heavy ion collisions, not only for the heaviest nuclear species, attacking new frontiers having Christophe as a mate, was pleasure and joy. We had exciting visions and promising projects together, based and initiated often based on his ingenious ideas and foresight, some of which just commencing now when he went missing. Christophe left us in March 2024, way too early at the height of his impressive career, in the middle of his creative and productive life. Continuing on the path he has prescribed, following his visions, we will do our best to honor his legacy. We will miss him infinitely. 
\begin{figure}[hbt]
\begin{center}
\includegraphics[width=\columnwidth]{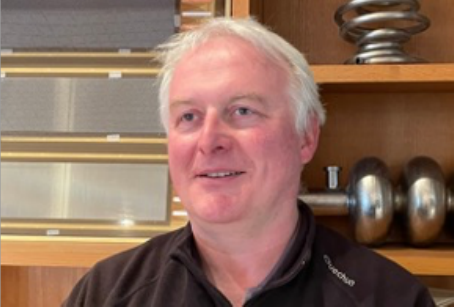}%
\end{center}
\caption{Christophe Theisen
\\ 
{\em (Photo with permit from V\'{e}ronique Theisen)}  
\label{fig:Christophe}}
\end{figure}

I dedicate this paper to my colleague and dearest friend {\bf Christophe Theisen}.

\newpage
\section{Introduction}\label{intro}
\begin{figure*}[t]
\hspace{0.5 cm}
\begin{center}
\includegraphics[width=\textwidth]{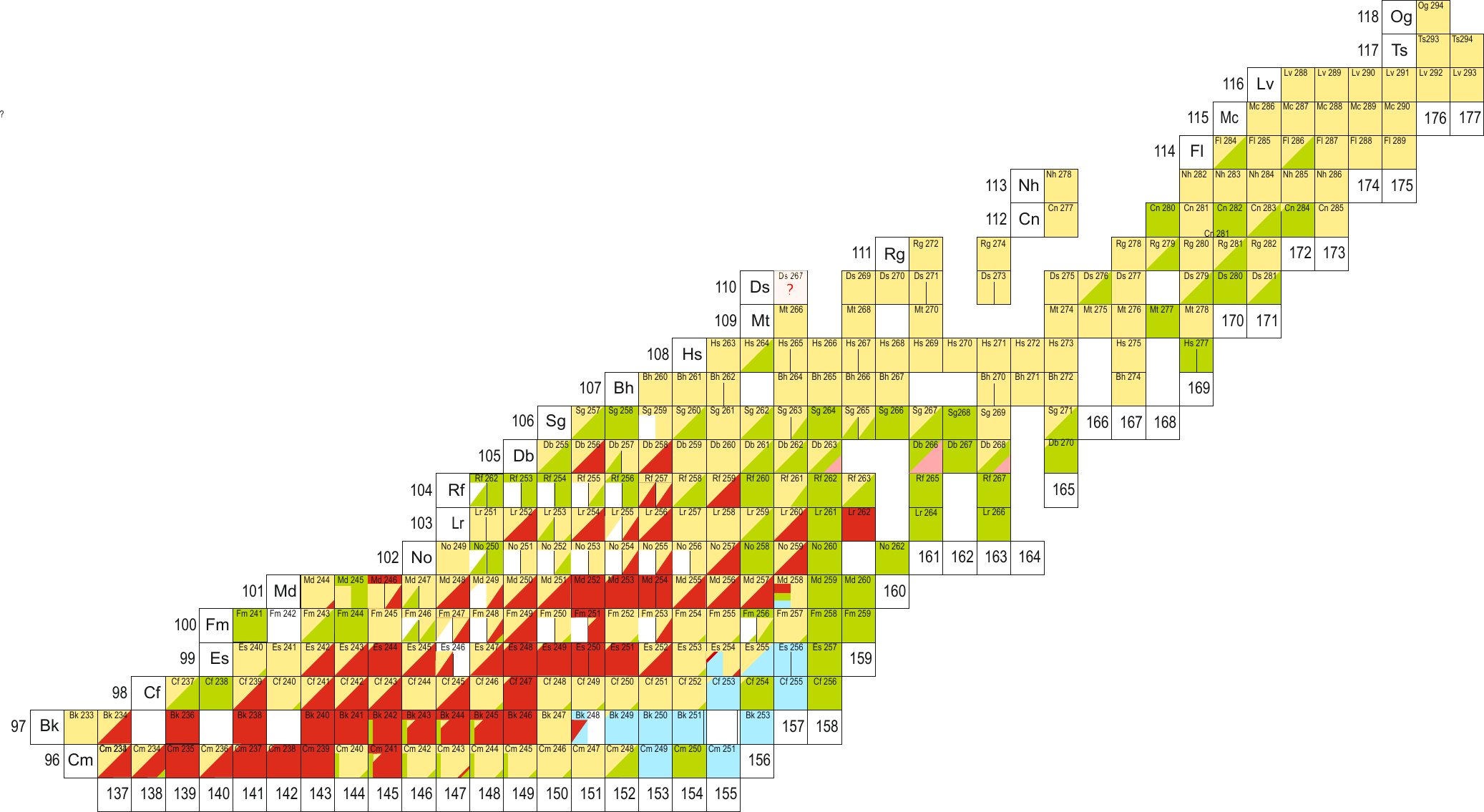}%
\end{center}
\caption{Excerpt of the chart of nuclides showing the heaviest nuclei observed ($Z$=96-118 and $N$=137-177). 
The colors indicate the decay mode and the subdivision in areas of different sizes the relative decay probabilities. 
Yellow denotes $\alpha$-decay, green spontaneous fission (SF), red $\beta^+$ or electron capture ($EC$) and blue $\beta^-$ decay. 
The scheme is adopted from~\cite{Soti2024}. 
For $^{267}$Ds a faint yellow has been chosen as the observation of its $\alpha$ decay is uncertain (see Ref.s~\cite{Ghiorso1995,Karol2001}). 
For the $\beta$ (EC) decay of $^{266,267,268}$Db a faint red has been chosen as the competition between $SF$ and $EC$ is being debated (see subsection~\ref{Db}).
\label{fig:Nchart96_118}}
\end{figure*}

The quest to explore the limits of stability of nuclear matter towards the highest atomic number $Z$ and the so-called {\it ``island of stability''} of spherical superheavy nuclei (SHN) is at present as popular as it was in the mid-sixties, when as a consequence of the nuclear shell model, introduced by Maria Goeppert-Mayer~\cite{Goeppert-Mayer1949,Goeppert-Mayer1950}, and Otto Haxel, Hans Jensen and Hans Suess~\cite{Haxel1949}, the first predictions for the next closed proton and neutron shells beyond $^{208}$Pb were proposed by Meldner~\cite{Meldner1967} and  Sobiczewski~\cite{Sobiczewski1966}. 
Already at that time, the occurrence of nuclear deformation and its major consequences for nuclear stability of these heavy nuclei had been recognized by Strutinsky~\cite{Strutinsky1966,Strutinsky1967}.
Beyond the desire and the reward for the discovery of a new chemical element or a new nuclear isotope, the nuclear structure features of these exotic species at the limit of nuclear stability and its quantum mechanical origin, as well as their chemical and ground-state properties, were the traditional driving forces for manifold activities throughout the last half-century.  

The advancement of the field was periodically reported in a number of reviews for the various disciplines involved, like for the synthesis of new superheavy elements (SHE) the recent review by Oganessian, Sobiczewski and Ter-Akopian~\cite{Oganessian2017}, or concerning the synthesis and search for even-$Z$ elements by Hofmann et al.~\cite{Hofmann2016}. 
The observation of three "promising" $\alpha$ decay signals in the latter has been critically discussed in Ref.~\cite{Hessberger2017a}.
 
SHE research also has a long tradition as one of the major subjects in modern nuclear chemistry. 
The determination of characteristic chemical properties has the potential to identify the atomic number of a hitherto unknown species. 
In a comprehensive and detailed overview, T\"{u}rler and Pershina reviewed the state of the research on the chemical properties of transactinide elements ($Z\geq$104) up to flerovium ($Z$=114)~\cite{Tuerler2013}.

Beyond the chemical applications of the properties of the atom, also atomic physics' methods like precise mass measurements and laser spectroscopy have become more and more fashionable in the last two decades. 
{\it "The quest for superheavy elements and the limit of the periodic table"} is the title of a recent review of the state and perspectives of SHE discussing in addition to the advancement in SHE chemistry, also aspects of atomic and nuclear structure features of superheavy atoms and nuclei~\cite{Smits2024}.
The recent progress in laser spectroscopy of actinides, referring to the potential of this type of investigation for the heaviest nuclei, has been summarized by Block,  Laatiaoui and Raeder
in 2021~\cite{Block2021}, while the capabilities of laser spectroscopic methods for the investigation of nuclear structure have been detailed and emphasized by Campbell, Moore and Parson~\cite{Campbell2016}.
Ion traps and Multi-Reflection Time of Flight Mass Spectrometers (MRTOF-MS) contribute substantially to the investigation of nuclear ground state properties of actinides and SHN, providing nuclear binding energies by high-precision mass measurements with mass resolution values of 30~keV to 100~keV~\cite{Block2019,Block2019a}.
In a paper titled {\it "Smooth trends in fermium charge radii and the impact of shell effects"}, Warbinek et al.~report on the most advanced results of systematic laser spectroscopy studies of the isotopic chains for nobelium and fermium, yielding fundamental findings concerning ground state properties like charge radii, and nuclear structure features like nucleon density distributions and the shell structure around the deformed proton and neutron shell gaps for $Z$~=~100 and $N$~=~152 at $^{252}$Fm~\cite{Warbinek2024}.

In a dedicated special issue of the journal Nuclear Physics A, reviews involving basically all disciplines pursuing SHE/SHN research had been collected in 2015~\cite{NPA944}. 
The list of papers comprises experimental as well as theory contributions, including for atomic physics mass measurements, laser spectroscopy and atomic/electronic structure theory, for chemistry liquid and gas-phase methods, and for the nuclear physics partition synthesis and reaction studies, as well as nuclear structure research. 

A similar collection of review papers on SHE/SHN research was published recently in the European Physical Journal A, covering topics like nuclear spectroscopy, experimental reaction studies, reaction and nuclear structure theory, SHE synthesis, stellar nucleosynthesis, and facilities and instrumentation for SHN/SHE research \cite{EPJA_SHE_2025}.

Focusing on Decay Spectroscopy After Separation (DSAS) of SHN, the present paper is dedicated to the nuclear structure of SHN. It is partly an update of a recent review, where Christophe Theisen and I were reporting on in-beam spectroscopy as well as on DSAS~\cite{Ackermann2017}. 
In this previous work we presented the history of the discovery and synthesis of chemical elements starting from the discovery of radium and polonium by Marie Curie~\cite{Curie1898,Curie1898a}, and extending to the synthesis of SHE during the first decade of this millennium with oganesson being the highest $Z$ and tennessine~\cite{Oganessian2010} the up to now last element to be synthesized~\cite{Oganessian2006}. We discussed the theoretical background driving the search for new elements and their properties, introduced the various nuclear decay processes and presented in detail spectroscopic methods and at that time accumulated knowledge on SHN nuclear structure by DSAS and in-beam spectroscopy. 
We discussed selected nuclear structure theory approaches and reported on the advances in experimental instrumentation. 
Here I will only briefly refer to the more detailed discussion of the historical and theoretical background in our previous work, while I will focus essentially on the recent experimental achievements and future perspectives in instrumentation.

The status of the field of SHN is illustrated by the representation of the presently known isotopes in Fig.~\ref{fig:Nchart96_118} as an update of Fig.~19 in Ref.~\cite{Ackermann2017}, which shows the upper right part of the chart of nuclides using the color scheme for the decay mode indication of Ref.~\cite{Soti2024} with 262 isotopes for the elements from $Z$=96 (curium) to $Z$=118, 18 more than in our earlier review, the lightest one being $^{236}$Bk decaying by electron capture ($EC$), which is not discussed here.

One of the nuclei missing in Fig.19 of Ref.~\cite{Ackermann2017} is $^{267}$Ds.
As it is still included in the "Karlsruher Nuklidkarte"~\cite{Soti2024} and listed in various other sources, I also included it here. It is, however, presented in a faint yellow as the observation and assignment of its $\alpha$ decay~\cite{Ghiorso1995} is uncertain (see Ref.~\cite{Karol2001}). 

Two more darmstadtium isotopes, $^{275,276}$Ds were observed at the new SHE-factory of Flerov Laboratory of Nuclear reactions (FLNR) of the Joint Institute for Nuclear Research (JINR) in Dubna, Russia~\cite{Gulbekian2019} as 4$n$ and 5$n$ fusion-evaporation channels in the reaction $^{48}$Ca+$^{232}$Th~\cite{Oganessian2023,Oganessian2024}.
Together with $^{276}$Ds, also its $\alpha$-decay daughter products, $^{272}$Hs and $^{268}$Sg terminating the decay sequence by $SF$, were observed for the first time in this experiment.
A third new darmstadtium isotope was reported in 2021 by S{\aa}mark-Roth with $^{280}$Ds as the termination of the $^{288}$Fl decay chain for the small $\alpha$-decay branch of $^{284}$Cn~\cite{Samark-Roth2021}. 
For more details see subsection~\ref{Ds} and related subsections.

Three additions to the chart of nuclides belong to the $\alpha$-decay chain of $^{244}$Md, with its daughter $^{240}$Es and grand daughter isotope $^{236}$Bk, the discovery of which was claimed in 2020 by two different research teams. 
The debate about its properties and different assignments will be discussed in subsection~\ref{Md}~\cite{Pore2020,Khuyagbaatar2020,Hessberger2021}. 
Its daughter and granddaughter isotopes, $^{240}$Es and $^{236}$Bk, were reported earlier in 2017 by Konki et al.~\cite{Konki2017}.

Pushing also towards the limits of stability at the neutron-deficient border of the Segr\`{e} chart, the discovery of the respective lightest isotopes were reported: in 2022 for lawrencium, $^{251}$Lr, at the Argonne Gas-Filled Analyzer AGFA (in 2024 confirmed at the Berkeley Gas-filled Separator BGS~\cite{Pore2024a}); and in 2025 for rutherfordium, $^{252}$Rf~\cite{Khuyagbaatar2025}, and for seaborgium, $^{257}$Sg~\cite{Mosat2025}, at the TransActinde Separator and
Chemistry Apparatus TASCA of GSI/FAIR, Darmstadt.

In the above-mentioned first series of experiments performed at the FLNR JINR SHE Factory, five more isotopes were observed for the first time: i) the most neutron-deficient moscovium isotope $^{286}$Mc~\cite{Oganessian2022d}, ii) $^{264}$Lr, populated by the newly revealed $\alpha$ branch of $^{268}$Db in the $^{288}$Mc decay chain~\cite{Oganessian2022c,Oganessian2022d}; iii) -- v) the two newly synthesized livermorium isotopes, $^{288,289}$Lv, with the decay chain of the lighter of the two now extending to the new copernicium isotope $^{280}$Cn, thanks to an $\alpha$ branch revealed in $^{284}$Fl~\cite{Oganessian2025}.


More details on these sixteen recently discovered isotopes will be given in the main section of this review in section~\ref{prop}. 
In this section, the major decay properties of the 204 known isotopes of the elements from einsteinium to oganesson are listed in Table~\ref{tab:isotope_list} together with references to all relevant publications.

For the $\beta$ decay($EC$), of $^{263,266,268}$Db a faint red has been chosen as the competition between spontaneous fission ($SF$) and $EC$ is being debated (see subsections~\ref{sf-b} and \ref{Db}). 

After the discovery of $^{250}$No by Belozerov et al.~in 2003~\cite{Belozerov2003} and the affirmation that both states, initially assigned to $^{249}$No and $^{250}$No, were to be ascribed to the ground state and a $K$-isomeric state of $^{250}$No by Peterson et al.~\cite{Peterson2006}, Kallunkathariyil et al.~reported in 2020 the detection of an additional $K$-isomeric state~\cite{Kallunkathariyil2020} at the gas-filled separator RITU of the University of Jyv\"{a}skyl\"{a} cyclotron laboratory (JYFL) (see subsection~\ref{No}). 

Referring to the various disciplines connected to the properties of the heaviest atoms and nuclei, I recall here the definitions we have formulated in our previous review for the categories superheavy elements: SHE, and superheavy nuclei: SHN~\cite{Ackermann2017}. 
Atomic physics and, in particular, nuclear chemistry are predominantly interested in atomic properties of these heavy objects and mainly talk about SHE, defining them as the transactinides with $Z\geq$104.
For the community interested in their nuclear structure, the quantum mechanical relevance is important.
That leads to the definition of SHN as those nuclides, around and beyond fermium, where the liquid drop fission barrier drops below the zero-point-motion energy of 0.5~MeV. 
These species owe their existence solely to quantum mechanical effects, which makes them an ideal laboratory to investigate and define the constraints of the underlying interaction, the strong nuclear force.  
The access to these interesting objects is strongly governed and limited by low and with increasing $Z$ ever-decreasing production cross-sections, demanding substantial efforts in method and instrumentation development, which we also addressed in detail in Ref.~\cite{Ackermann2017}. 
In this review, based on this earlier work, I will start with a brief reminder on the spectroscopic methods employed for DSAS in section~\ref{DSAS} and the specific nuclear structure features in section~\ref{specific} on the decay properties of the heaviest nuclei. 
In the main part in section~\ref{prop}, I will, after a brief overview, present the recent findings for the isotopes of all elements from $Z$=99 (einsteinium) to $Z$=118 (oganesson).
This review will be concluded in section~\ref{outlook}, presenting the technological progress pursued by new facilities, being presently constructed, coming online soon, or being operational since recently.




\newpage
\section {Decay spectroscopy after separation - DSAS}\label{DSAS}
\begin{figure*}[htb]
\begin{center}
\includegraphics[width=0.91\textwidth]{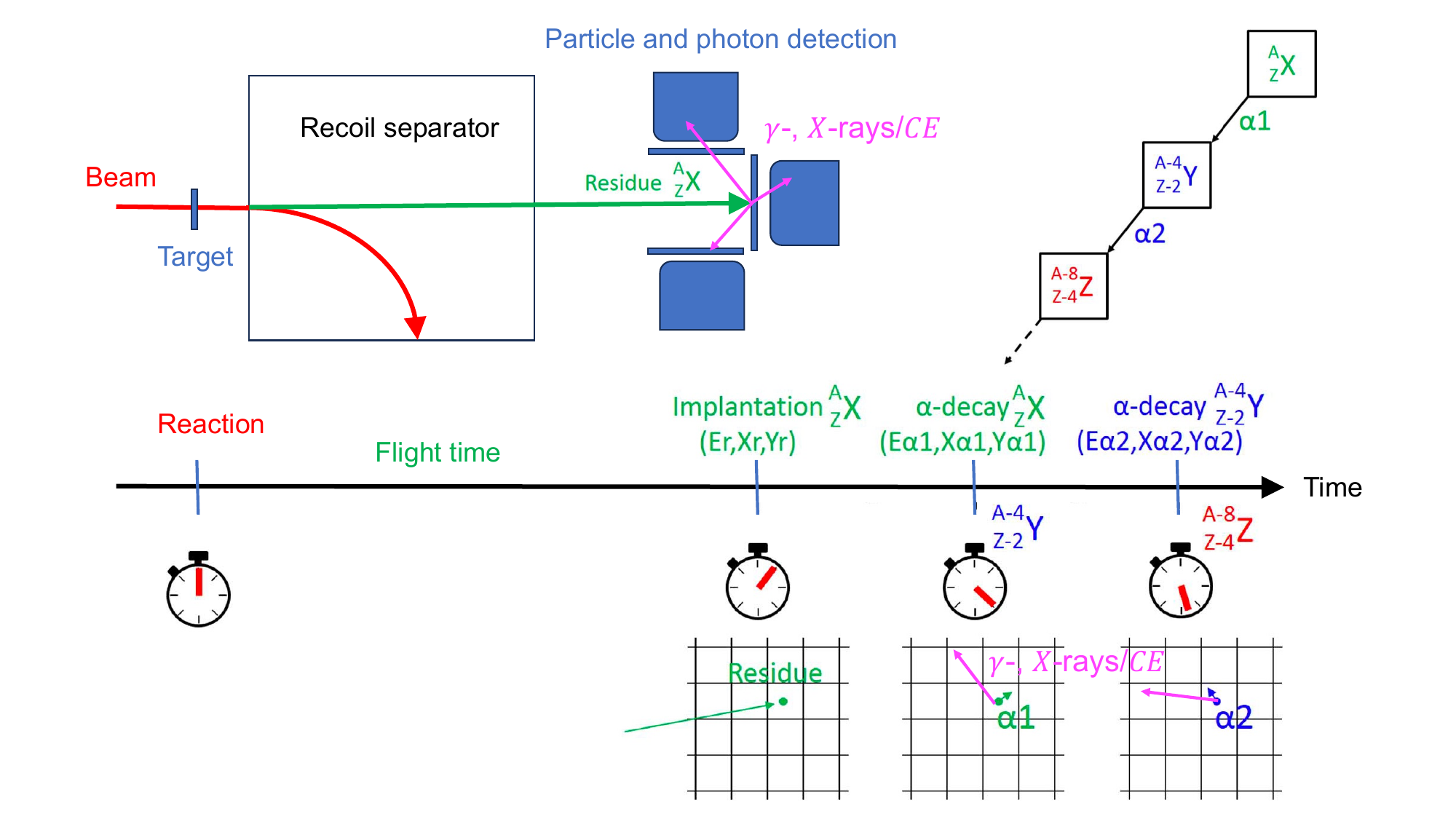}%
\end{center}
\caption{Decay spectroscopy after separation and genetic correlations: the recoiling nucleus $_Z^AX$ is implanted in a position-sensitive detector at position ($X_r$, $Y_r$). 
It subsequently decays via $\alpha$ emission in its neighborhood at position ($X\alpha1$, $Y\alpha1$) to the daughter nucleus ($_{Z-2}^{A-4}Y$) which itself decays to the granddaughter ($_{Z-4}^{A-8}Y$) at position ($X\alpha2$, $Y\alpha2$). 
In addition to the recoils and $\alpha$ particles $\gamma$-, $x$-rays and $CE$s are detected in coincidence. The technique allows correlations in position and/or time of the recoil implantation and its subsequent decay due to the inclusive detection of the particles and photons involved. (Figure and caption are taken from ref.~\cite{Ackermann2024})
}
\label{fig:Fig.GenCor}
\end{figure*}
The general production mechanism for most of the nuclei discussed here is heavy ion fusion-evaporation leading to so-called evaporation residues (ER) with partly extremely low cross-sections for which highly efficient separation schemes are needed to suppress the primary beam and products from unwanted reactions to provide the highest sensitivity for those rare events. 
A variety of different ion-optical separators is used at the various laboratories, being mainly of two types: i) vacuum separators like the velocity filters SHIP of GSI~\cite{Muenzenberg1979} and SHELS of FLNR JINR~\cite{Yeremin2015}, or mass spectrometers like the FMA  at ANL in Lemont, IL, U.S.A.; ii) gas-filled separators like RITU~\cite{Leino1995} at the cyclotron laboratory of the University of Jyv\"{a}skyl\"{a} in Finland, or the many gas-filled separators like DGFRS~\cite{Tsyganov1999} of FLNR JINR in Dubna, Russia, GARIS at RIKEN in Wako, Japan~\cite{Morita1992}, BGS, equipped with the FIONA mass separator, at LBNL, Berkeley, CA, U.S.A.~\cite{Gates2022}, TASCA at GSI/FAIR~\cite{Semchenkov2008}, and AGFA~\cite{Seweryniak2013} at ANL, or in the near future, also the combined separator and mass spectrometer installation S$^3$ of GANIL, presently being commissioned at SPIRAL2 (see also section~\ref{outlook}). 
For a detailed discussion of ion-optical separation principles and devices see our previous review~\cite{Ackermann2017}.

After being separated from the primary beam, the decay properties of the ER are studied employing comprehensive particle and photon detectors.
Alpha particles, electrons or fragments from $SF$ are typically registered by silicon detector arrays with varying granularity, while germanium detectors collect $\gamma$ and $x$-ray quanta emitted in the decay process of metastable states possibly populated in the ER, and from excited states of subsequent decay products. 
This method of decay spectroscopy after separation is schematically shown in Fig.~\ref{fig:Fig.GenCor}. 

The detection principle of $\alpha$-decay chains and genetic correlations, a powerful tool for particle identification for hitherto unknown species, is shown in this figure which is taken from ref.~\cite{Ackermann2024}, where more details on this powerful and well-established tool to investigate rare activities, including single-event detection, are given~\cite{Hofmann2000}.

While emitted $\alpha$ particles from g.s.-to-g.s. decays and fission fragments, correlated to the ER, implanted in the focal plane silicon detector, or the previous decay in a decay sequence, are most important for the identification of the decaying isotope, $\gamma$ rays and conversion electrons ($CE$), together with $alpha$ decay into excited decay-daughter states, yield detailed spectroscopic information on their low-lying nuclear structure. 
$CE$s have recently also proven to be a powerful tool for additional background suppression (see, e.g., ref.~\cite{Bronis2022} as well as the detection $x$-rays which in addition can also serve for efficient Z-identification.
The sensitivity of this correlation method down to lowest count rates, including single event detection, is the key for an efficient accumulation of nuclear structure data for the heaviest nuclear species over the last decades.
Highlights are, e.g., the discovery of the heaviest $K$ isomers $^{270m}$Ds~\cite{Hofmann2001} and $^{266m}$Hs~\cite{Ackermann2015b}, or the $\alpha$-$\gamma$ decay spectroscopy of odd-even nuclei like $^{247}$Md~\cite{Antalic2010, Hessberger2022}, $^{255}$Lr~\cite{Chatillon2006, Antalic2008} or even-odd nuclei $^{257}$Rf~\cite{Hessberger2016, Hauschild2022}.
The important role the unpaired nucleon is playing for exotic nuclear structure features has been underlined in ref.~\cite{Ackermann2023}, and, in particular, for the creation of high-spin/high-$K$ isomeric states in ref.~\cite{Ackermann2024} (see also subsections~\ref{spl} and ~\ref{K-iso}).

Similar to $\alpha$ decay, highly excited states in heavy nuclei can also be populated by $\beta$ decay, de-exciting by a series of internal transitions proceeding via $CE$ and $\gamma$ emission. 
However, as $\beta$ decay, proceeding by $EC$ or $e^+$ (and $e^-$) emission, is difficult to detect, $x$-rays and $CE$s, being emitted from the populated daughter states, play a major role for its identification, as in the case of the $^{253}$Md $\beta$ decay, revealing the 11/2$^-$[725] isomer in $^{253}$Fm~\cite{Antalic2011}, or the first attempt to identify $EC$-decay of an isomeric state in $^{257}$Rf~\cite{Hessberger2016}.

Short lifetimes of nuclear states in heavy nuclei, and, in particular, in nuclear isomers pose a major challenge to the separation and detection instrumentation. 
Despite optimized installation, like, e.g., in the case of the compact design of the gas-filled separator of ATLAS ANL, AGFA, typical flight paths of the order of $\mu$s lead to losses in the detection of $\mu$s isomers.
Lifetimes of this order, in addition, call for special measures to separate the various electronic signals from $ER$ implantation and the subsequent decays.
Thanks to the advancement in electronics development in recent years, the required time resolution is achieved by the application of flash-ADCs and pulse-shape analysis.
So-called "digital" data acquisition systems~\cite{Jordanov1994}, provide this functionality which is demonstrated by successful isomer identification in $^{254}$Rf~\cite{David2015,Khuyagbaatar2020b} and $^{255}$Rf~\cite{Khuyagbaatar2020b},
or $^{250}$No~\cite{Kallunkathariyil2020, Khuyagbaatar2022} (see also subsections~\ref{sf-b}, \ref{Rf}, and Fig.s~\ref{fig:PSA_el}, \ref{fig:PSA_fission}).
More details on the role of $CE$ and pulse shape analysis are discussed in subsection~\ref{sf-b} on the competition of $SF$ and $\beta$ decay. 
The advanced DSAS technologies have the potential to reveal the hitherto almost completely missing information on $\beta$ decay properties of nuclides for all elements beyond dubnium (see Fig.~\ref{fig:Nchart96_118}).






\newpage

\section{Specific nuclear structure features}\label{specific}
\begin{figure*}[h]
\begin{center}
\includegraphics[width=0.7\textwidth]{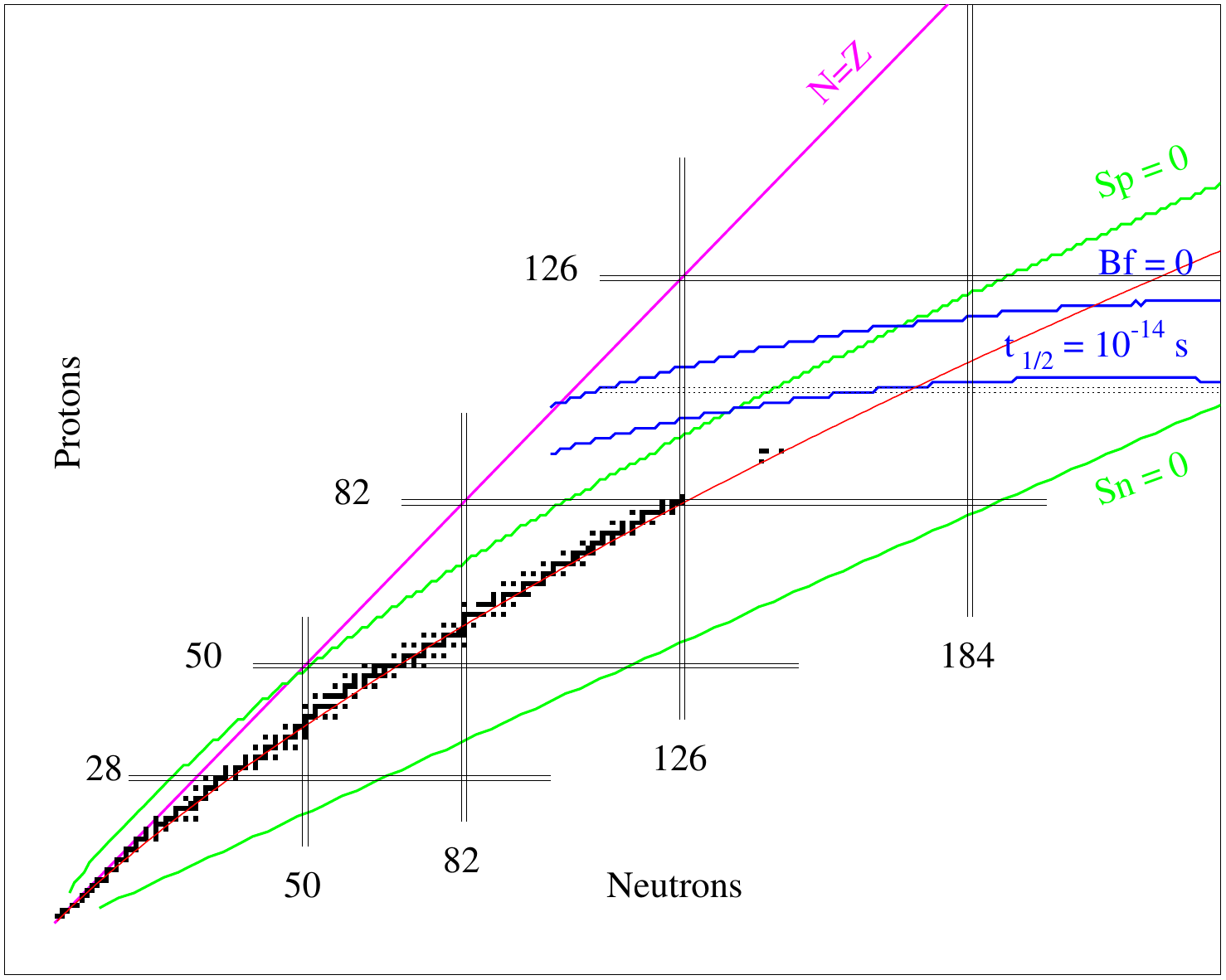}
\end{center}
\caption{Nuclear chart and its limits of stability deduced from the liquid-drop model.
The green lines correspond to the neutron $S_n=0$ and proton $S_p=0$ drip lines, with $S_n$ and $S_p$ denoting the neutron and proton separation energies, respectively.
The blue lines $B_f = 0$ and $t_{1/2} = 10^{-14}$ s correspond to a vanishing fission barrier
and a spontaneous fission half-life of $10^{-14}$ s, respectively.
The red line corresponds to beta-stable nuclei.
Besides magic numbers shown with lines, the horizontal dotted line corresponds to $Z=104$.
(Figure and caption are taken from Ref.~\cite{Ackermann2017})}
\label{fig.chart_drop}
\end{figure*}


As mentioned in the introduction, the heaviest nuclei owe their existence to quantum mechanics and shell effects as the macroscopic barrier against fission, which can be estimated using the liquid drop model (see e.g.~\cite{Bohr1975}, Appendix A2), vanishes with increasing atomic number $Z$.
Within the liquid drop model the nuclear binding energy is described by the Bethe-Weizs\"{a}cker formula:

\begin{eqnarray}
M(A,Z)  &= &a_V A \ - \ a_C Z^2/A^{1/3} \ - \ a_S A^{2/3} \nonumber  \\ 
                &+& \ a_A (N-Z)^2/A \ - \ \delta(A,Z)
\end{eqnarray}
with
\begin{equation}
\delta(A,Z) =
   \left \{
   \begin{array}{l l}
      +\delta_0  & N \ \rm{and} \ Z \ \rm{even} \\
      0 & A \ \rm{odd} \\
      -\delta_0 & N \ \rm{and} \ Z \ \rm{odd}
   \end{array}
   \right \}
\end{equation}
and $\delta_0 = a_P \ A^{-1/2}$.

In our previous review~\cite{Ackermann2017} (section 1.2.1) we discussed the limits of stability according to the liquid drop model and summarized it in Fig.~\ref{fig.chart_drop}.
As can be seen from the blue lines in Fig.~\ref{fig.chart_drop}, a definition of SHN based on the vanishing liquid drop fission barrier does not result in a defined atomic number like for the chemist's definition of SHE with Z=104, but refers rather to an area between fermium ($Z$=100) and rutherfordium ($Z$=104) depending also on the neutron number (blue lines in Fig.~\ref{fig.chart_drop}). For more details refer to Ref.~\cite{Ackermann2017}.

The nuclear structure features of nuclei in the region spanning from fermium to darmstadtium are presently in the focus of decay spectroscopy after separation (DSAS)~\cite {Ackermann2017,Ackermann2024,Asai2015}. DSAS offers efficient tools to study, in particular, single particle configurations at a given deformation, metastable states and the competition of decay modes ($\alpha$, $\beta$, $SF$ and internal transitions ($IT$)), with the potential to explore the limits of stability in terms of highest $Z$ and extreme isospin (mainly towards low neutron numbers).  

In the vicinity of low-level-density areas and large energy gaps in the plane spanned by single-particle energy levels (SPLs) and deformation in a Nilsson diagram 
representation (see e.g. Ref.~\cite{Chasman1977}), the properties of the single-particle states can be studied with reduced perturbation by mixing of other quantum states. 
The region of prolate-deformed nuclei around N=152 and Z=100 provides access to SPLs stemming from proton orbitals 
like, e.g., $f_{5/2}$ and $f_{7/2}$, and neutron orbitals like e.g. $d_{3/2}$ and $h_{11/2}$ which are predicted to form the spherical proton and neutron 
shell closures in the region of the so-called {\it ``island of stability''}~\cite{Chasman1977}. 
Investigating those SPLs can lead to findings that support conclusions on the development of shell gaps towards sphericity where those SPLs are degenerate at orbital energies.
Recent findings, reported in section~\ref{prop}, in this respect, will be discussed in the next subsection~\ref{spl}.

As mentioned above, the unpaired nucleon plays a major role here as a probe for the development of specific quantum states with deformation and single particle energy and the possible occurrence of metastable states. 
A specific class of those, i.e., $K$-isomers, and the advances in recent studies as reported in section~\ref{prop} will be discussed in subsection~\ref{K-iso}.

A comprehensive example for the intertwined decay properties leading to networks of communicating decay sequences, built on competing decay modes ($\alpha$-, $\beta$-decay and $SF$), is shown for the example of the $^{258}$Db decay network in Fig.~\ref{Fig:258Db_decay}.

\subsection{Single-particle states, deformation and shell gaps}\label{spl}

One of the major problems for the assignment and interpretation of single particle levels (SPLs) is the possible mixing with neighboring states in regions of high level density.  
This can be, at least partly, overcome in the vicinity of gaps in the level sequence like in the fermium/nobelium region with the Z\,=\,100 and N\,=\,152 shell gaps.
The most promising and possibly most conclusive studies are, therefore, carried out for isotopes along these proton and neutron numbers, as shown in the comprehensive nuclear structure compilation of Asai, Lopez-Martens and He{\ss}berger who, e.g., investigate isotonic trends for nuclei with neutron numbers 151 and 153.
One addition to this systematic investigation are the new results for $^{255}$No reported by Bronis et al.~\cite{Bronis2022} completing the isotonic trend of the 11/2$^-$[725] state for the N\,=\,153 isotones from curium to hassium (see subsection~\ref{No} and Fig.~\ref{fig:153_isotones}). 
Another striking feature of $^{255}$No is the presence of high-$K$ isomers in this nucleus as shown in Fig.~\ref{Fig:255No_Kessaci}, revealed consistently by two groups in independent experiments~\cite{Bronis2022,Kessaci2024}.
The above mentioned 11/2$^-$[725] state plays also a significant role in N\,=\,151 isotones and has been revealed in $^{253}$No, populated by the $\alpha$ decay from the analog state in $^{257}$Rf, where it is isomeric due to its low excitation energy as the second state above the ground state (see subsections~\ref{No} and \ref{Rf} and Ref.\cite{Hauschild2022}.)
A nice example, where the experimental assignment provides information on the location of the next proton and neutron shell closures beyond $^{208}$Pb, is the excitation energy $E^*$ assignment of two SPLs in $^{243}$Es which are stemming from orbitals defining a possible proton shell gap at $Z$\,=\,114 as proposed by theory.
The experimental findings for the low-lying structure in $^{243}$Es as shown in Fig.~\ref{fig:247Md-levels} contradict this prediction (see subsection~\ref{Es}). 
\begin{figure}[ht]
\begin{center}
\includegraphics[width=0.9\columnwidth]{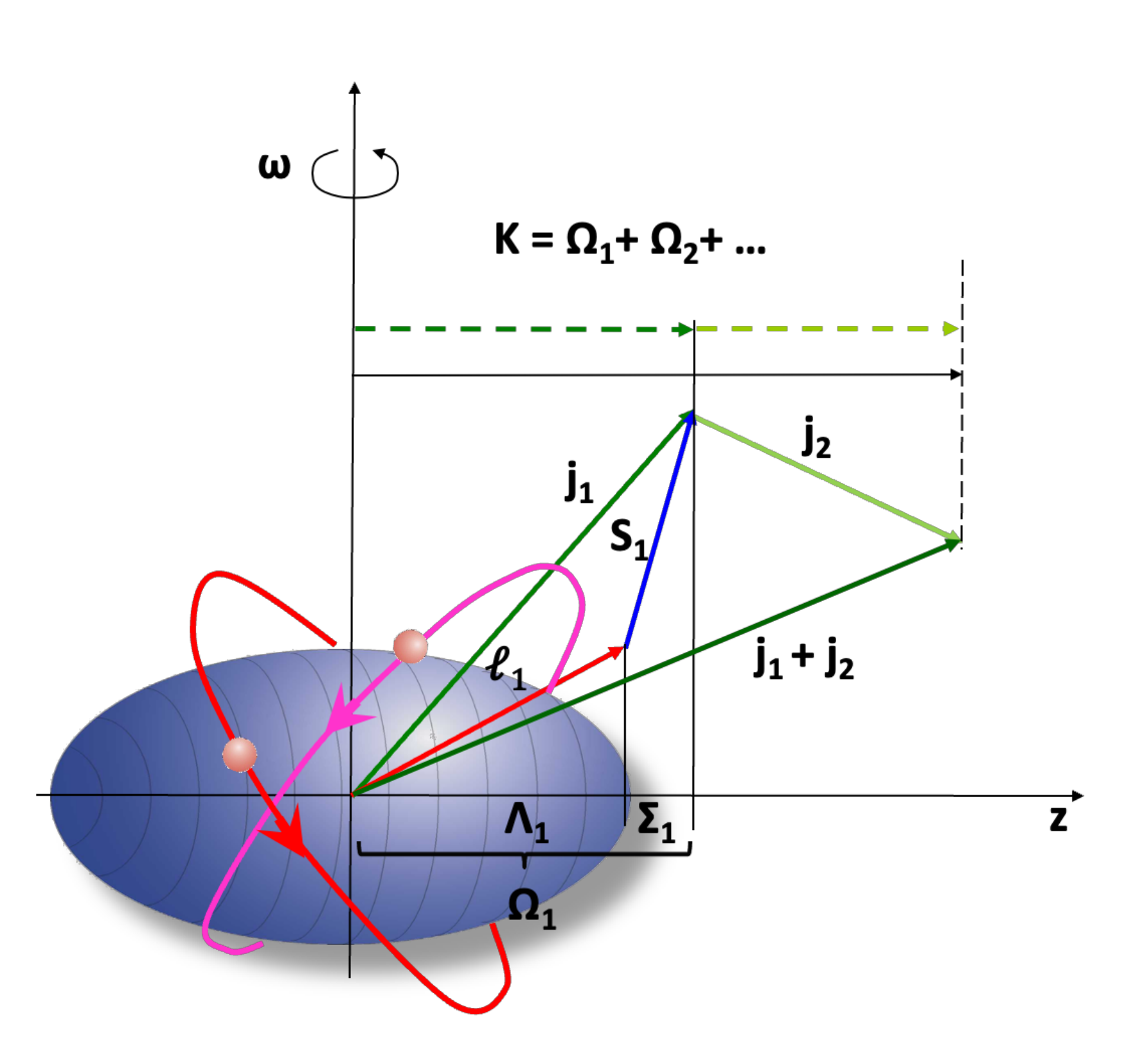}
\vspace{-1cm}
\end{center}
\caption{Definition of the $K$ quantum number as the total projection of the sum $j_i$ of the spin of the nucleon $S_i$ and the orbital angular momentum $l_i$ of all excited 2-quasiparticle states onto the symmetry axis of the nucleus.
(Figure and caption are taken from Ref.~\cite{Ackermann2017})}
\label{fig.K-def}
\end{figure}
\subsection{$K$-isomerism}\label{K-iso}
\begin{figure*}[ht]
\begin{center}
\includegraphics[width=0.7\textwidth]{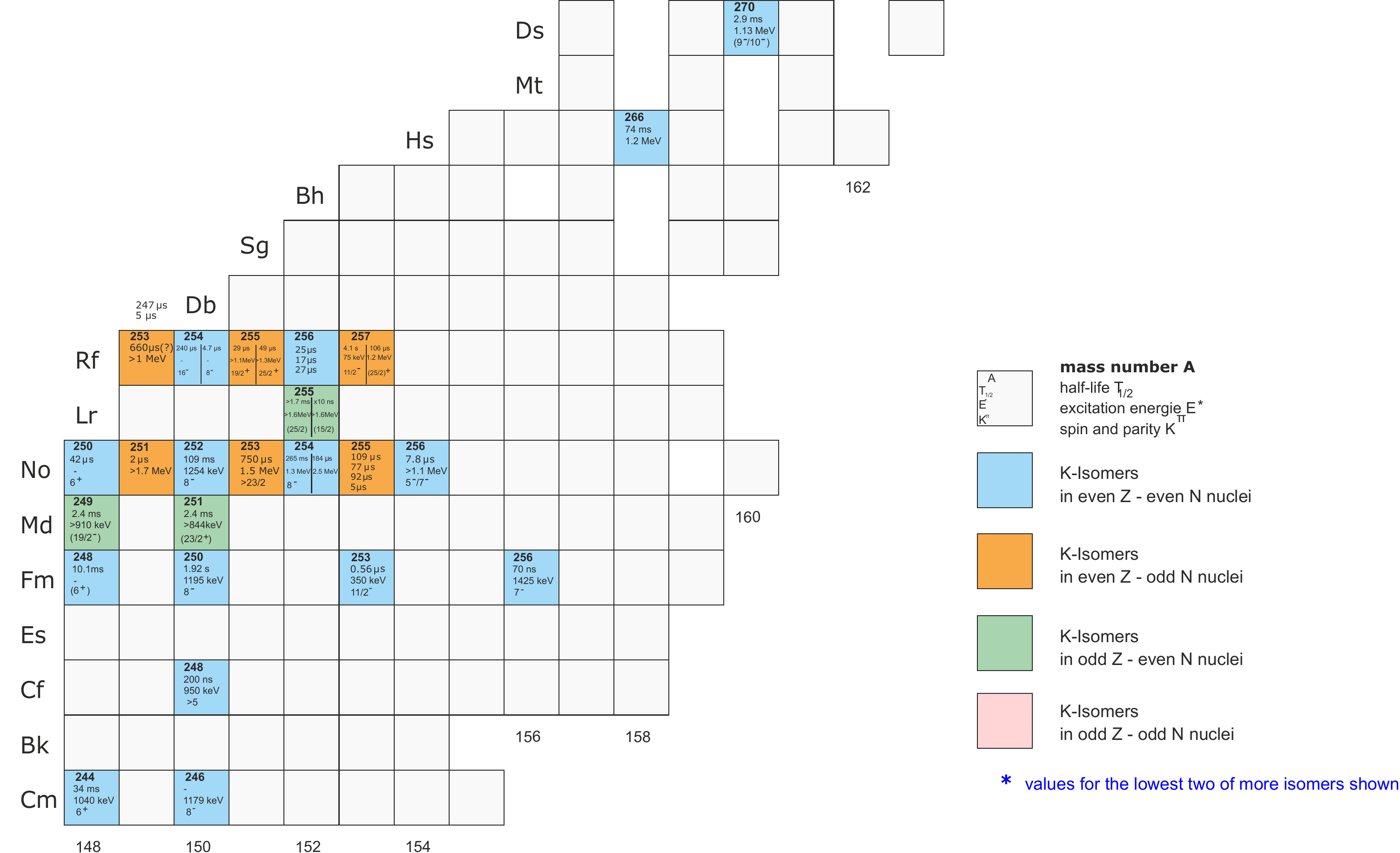}%
\end{center}
\caption{Update of Fig.~43 in Ref.~\cite{Ackermann2017}: Summary of $K$ isomers for the heaviest nuclei at and above $Z$=96. For detailed properties see table~\ref{tab:isotope_list} and references given therein.
\label{fig:K-Isomers_Zgt95}}
\end{figure*}
In a recent review, we discussed the role of metastable states in superheavy nuclei~\cite{Ackermann2024}, where more details can be found.
Here I restrict myself to one class of isomeric states which are particularly interesting for the region of deformed heavy and superheavy nuclei, focused on here.
The nuclei which are accessible for more detailed spectroscopic studies in the region around $Z$\,=\,100 and $N$\,=\,152 exhibit a substantial degree of deformation with a typical quadrupole deformation of $\beta_2 \approx 0.25$~\cite{Parkhomenko2004,Parkhomenko2005}.
Therefore, $K$-isomers play a dominant role in this region with the $K$ quantum number being the sum of the total spin projection on the symmetry axis of the deformed nucleus
\begin{eqnarray}
K  =  \sum_i \Omega_i & = & \sum_i (\Lambda_i + \Sigma_i)
\end{eqnarray}
with $\Lambda_i$ and $\Sigma_i$ being the projection on the nucleus' symmetry axis of the orbital angular momentum $l_i$ and the spin of the nucleon $s_i$.
This definition is schematically shown in Fig.~\ref{fig.K-def}.


The heaviest nuclei for which a $K$-isomer has been observed are $^{270}$Ds and its $\alpha$-decay daughter $^{266}$Hs, where the delay due to isomerism is observed in $\alpha$-decay times with the specific feature that the isomeric decay proceeds slower than the decay of the ground state~\cite{Hofmann2001,Ackermann2015b}.
An attempt to explain these findings by a phenomenological superfluid tunneling model succeeds in reproducing the experimental data when applying a reduction of the pairing gap by a factor of 0.6~\cite{Clark2018,Clark2024}.

Using the $CE$-$\gamma$ correlation reduces background efficiently, leading to the observation of numerous high-$K$ states in the nuclei beyond fermium. 
After initially being found in even-even heavy nuclei like, e.g., $^{254}$No, more and more $K$ isomers are reported in odd-even or even-odd nuclei.

For odd-$Z$, respectively odd-$N$ nuclei, the unpaired nucleon plays an important role. 
Indeed, 
together with the high-$j$ orbitals accessible around the proton and neutron shell gaps at $Z$=100 and $N$=152 as shown in Fig.~\ref{fig.249251Mdlevels} for $^{249,251}$Md, the unpaired nucleon helps to generate high-$K$ values which are a major ingredient for the formation of metastable states.
Nice examples are the recently discovered 3-quasiparticle isomers in $^{249,251}$Md~\cite{Goigoux2021} (see subsection~\ref{Md}) and $^{255}$No~\cite{Bronis2022,Kessaci2024}, including a suggested 5-quasiparticle excitation (see subsection~\ref{No}).

For most of the $K$-isomers shown in Fig.~\ref{fig:K-Isomers_Zgt95} and listed in Table~\ref{tab:K-isomers}, the prominent decay mode proceeds by internal transitions, i.e., by $\gamma$ emission or internal conversion. 
Apart from isomers decaying by $\alpha$ emission or internal transitions, also delayed $SF$ has been observed as for example for $^{258}$Rf, where the ground state and an excited state are populated by $\beta$-decay ($EC$) of $^{258}$Db, while the origin of the meta-stability has still to be clarified here~\cite{Hessberger2016a}. 
The scarce existing data for $K$ isomers decaying by $\alpha$-decay (apart from $^{270}$Ds and $^{266}$Hs, there is only  $^{178m2}$Hf) and SF ($^{246}$Fm only) is discussed in a recent review by R.M.~Clark~\cite{Clark2024}.

In Table~\ref{tab:K-isomers} and Fig.~\ref{fig:K-Isomers_Zgt95} the six even-odd nobelium and rutherfordium isotopes, and the three mendelevium and lawrencium isotopes, for which high-$K$ isomers have been reported, are presented with their major properties.
Up to date no high-$K$ isomer has been reported for an odd-odd nucleus with $Z$\,$\geq$\,96.

\onecolumn
\begin{center}
\begin{longtable}{lcccccl}


\caption
{
Extension of the table of known $K$ isomers in heavy and SHN from curium to darmstadtium, including odd-$Z$/$N$ and odd-odd isotopes and single-particle excitations with respect to earlier listings~\cite{Herzberg2008, Ackermann2015b}. $IT$ denotes internal transitions that can proceed via $\gamma$ emission or internal conversion. Note of caution: the configuration assignments in many cases are often based on systematics and model assumptions, rather than on safe experimental findings. For details of those assignments see the corresponding references. (Table and caption taken from~\cite{Ackermann2024}\label{tab:K-isomers})
}

\\
\hline
\\
state & K$^\pi$&T$_{1/2}$& E$_x$   & decay & configuration& Ref.\\
        &        &         &         & mode  & assignments  &          \\
\\
\hline
\\
$^{244m}$Cm& 6$^+$  & 34 ms   &1.040 MeV& $IT$   &
    				  5/2$^+$[622]$_\nu$ $\otimes$ 7/2$^+$[624]$_\nu$  & \cite{Hoff1984}\\
          &        &         &         &            &                & \cite{Hansen1963}\\
$^{246m}$Cm& 8$^-$  &  -      &1.179 MeV& $IT$   &
                      7/2$^+$[624]$_\nu$ $\otimes$ 9/2$^-$[734]$_\nu$& \cite{Multhauf1976}\\
$^{248m}$Cf&$\geq $5& $>$140 ns&  0.9 MeV  & (IT)          &    -            & \cite{Orlandi2022}\\
$^{248m}$Fm& (6$^+$)&10.1(6) ms&  -   & $IT$   &   two-quasiparticle       & \cite{Ketelhut2010}\\
          &        &         &         &            &                & \cite{Herzberg2011}\\
          &        &         &         &            &                & \cite{Liu2014}\\
$^{250m}$Fm& 8$^-$  & 1.92(5) s  &1.195 MeV& $IT$   &
                      7/2$^+$[624]$_\nu$ $\otimes$ 9/2$^-$[734]$_\nu$& \cite{Greenlees2008}\\
$^{253m}$Fm& 11/2$^-$ & 0.56(6) $\mu$s  &  $\approx$350 MeV& $IT$   &
                      11/2$^-$[725]$_\nu$& \cite{Antalic2011}\\
$^{256m}$Fm& 7$^-$  & 70(5) ns   &1.425 MeV& $IT$,SF&
                      7/2$^+$[633]$_\pi$ $\otimes$ 7/2$^-$[514]$_\pi$& \cite{Hall1989}\\
$^{249m}$Md& (19/2$^-$)     & 2.4(3) ms   &$\geq$0.910 MeV& $IT$&
                      7/2$^-$[514]$_\pi$ $\otimes$ 5/2$^+$[622]$_\pi$ &\cite{Goigoux2021}\\ 
          &                &             &&&$\otimes$ 7/2$^+$[624]$_\nu$& \\
$^{251m}$Md& (23/2$^+$)       & 1.37(6) ms  &$\geq$0.844 MeV& $IT$&
                      7/2$^-$[514]$_\pi$ $\otimes$ 7/2$^+$[624]$_\nu$ & \cite{Goigoux2021}\\ 
          &                &             &&&$\otimes$ 9/2$^-$[734]$_\nu$& \\          
$^{250m}$No$^a$&
               6$^{+}$ &  34.9$^{+3.9}_{-3.2}$   & $\approx$1.2 MeV    & $IT$,($SF)$     
              &   5/2$^+$[622]$\nu$ $\otimes$ 7/2$^+$[624])$\nu$  & \cite{Peterson2006}\\
          &        &         &         &            &                & \cite{Kallunkathariyil2020}\\
 $^{251m}$No& -  & $\approx$2 $\mu$s  &$>$1.7 MeV& $IT$  &
                      7/2$^+$[624]$_\nu$ $\otimes$ 9/2$^-$[734]$_\nu$ & \cite{Hessberger2006}\\
        &        &         &         &            &     & \cite{Lopez-Martens2022}\\

$^{252m}$No&      8$^-$  & 109(3) ms  &1.254 MeV& $IT$   &
                      7/2$^+$[624]$_\nu$ $\otimes$ 9/2$^-$[734]$_\nu$ & \cite{Sulignano2007}\\
          &        &         &         &            &                & \cite{Sulignano2012}\\
$^{253m}$No& $\geq23/2$  & 627(5) $\mu$s  & >1.44 MeV& $IT$   & 
            $9/2^-$[734]$_\nu$ $\otimes$ 7/2$^+$[624]$_\nu$& \cite{Streicher2006}\\ 
        &        &         &         &            & $\otimes$ 7/2$^+$[613]$_\nu$               & \cite{Lopez-Martens2007}\\
            &&&\multicolumn{2}{c}{and/or$^b$} &$9/2^+$[624]$_\pi$ $\otimes$ $7/2^-$[514]$_\pi$
                                                                    & \cite{Streicher2010}\\ 
          &        &         &         &            & $\otimes$ $9/2^-$[734]$_\nu$               & \cite{Antalic2011}\\
$^{254m1}$No$^c$&   8$^-$ &265(2)~ms &   1.296 MeV   &$IT$ & two-quasiparticle            & \cite{Herzberg2006}\\
          &        &         &         &            &                & \cite{Wahid2025}\\
$^{254m2}$No$^c$&   (10$^+$)&5(2) ns& 2.015 MeV&$IT$ & two-quasiparticle  & \cite{Wahid2025}\\
$^{254m3}$No$^c$&   (16$^+$)&184(3) $\mu$s& 2.933 MeV&$IT$ & four-quasiparticle  & \cite{Tandel2006}\\
          &        &         &         &            &                & \cite{Clark2010}\\
          &        &         &         &            &                & \cite{Hessberger2010}\\
          &        &         &         &            &                & \cite{Wahid2025}\\
$^{255m1}$No&  11/2$^-$ & 109(9) $\mu$s & 240-300 keV& $IT$   & 
             11/2$^-$[725]$_\nu$& \cite{Bronis2022}\\
    &  11/2$^-$ & 86(6) $\mu$s  & $\approx$200 keV& $IT$   & 
             11/2$^-$[725]$_\nu$& \cite{Kessaci2022}\\
$^{255m2}$No$^d$& 19/2-23/2 & 77(6) $\mu$s  & 1.4-1.6 MeV& $IT$   & 
            1/2$^-$[521]$_\pi$ $\otimes$ $9/2^+$[624]$_\pi$& \cite{Bronis2022}\\
          &        &         &         &            &      $\otimes$ $11/2^-$[725]$_\nu$&\\  
          &  21/2$^+$ & 2(1) $\mu$s  &  $\approx$1.3 MeV&  $IT$   & 
            $1/2^-$[521]$_\pi$ $\otimes$ $9/2^+$[624]$_\pi$& \cite{Kessaci2022}\\
        &        &         &         &            &      $\otimes$ $11/2^-$[725]$_\nu$&\\ 
$^{255m3}$No$^d$& $\geq$19/2 & $\geq$1.2$^{+0.6}_{-0.4} \mu$s  & $\geq$1.5 MeV& $IT$           & -  & \cite{Bronis2022}\\
            & 27/2$^+$ & 92(13) $\mu$s & $\geq$1.5 MeV& $IT$   &  
            $7/2^-$[514]$_\pi$ $\otimes$ $9/2^+$[624]$_\pi$& \cite{Kessaci2022}\\
        &        &         &         &            &      $\otimes$ $11/2^-$[725]$_\nu$& \\ 
$^{255m4}$No & - & 5(1) $\mu$s & $\geq$2.5 MeV& $IT$   &  
            five-quasiparticle& \cite{Kessaci2022}\\
$^{256m}$No& 5$^-$/7$^-$ & $7.8^{+8.3}_{-2.6} \mu$s  & $\geq$ 1.1 MeV&  $IT$   & $11/2^-$[725]$_\nu$ $\otimes$ $1/2^+$[620]$_\nu$& \cite{Kessaci2021}\\
            &&&\multicolumn{2}{c}{and/or$^e$} &$11/2^-$[725]$_\nu$ $\otimes$ $3/2^+$[622]$_\nu$\\ 
$^{255m2}$Lr& (15/2) & 
 10-100 ns  & $>$1.6 MeV& $IT$   & 
            $1/2^-$[521]$_\$pi$ $\otimes$ $7/2^-$[514]$_\pi$& \cite{Jeppesen2009a}\\
          &        &         &         &            &$\otimes$ $9/2^+$[624]$_\pi$            & \\
$^{255m3}$Lr$^f$& (25/2) & $\geq$1.70(3) ms  & $\geq$1.6 MeV& $IT$   & 
            $7/2^-$[514]$_\pi$ $\otimes$ $7/2^+$[624]$_\nu$
                                                                     & \cite{Hauschild2008}\\
          &        &         &         &            &$\otimes$ $11/2^-$[725]$_\nu$& \cite{Antalic2008}\\
          &        &         &         &            &                & \cite{Jeppesen2009a}\\
$^{253m}$Rf$^g$&  -  &0.66$^{+40}_{-18}$ ms& $\geq$ 1.02 MeV & $IT$	   &  - 
                                                                     & \cite{Lopez-Martens2022}\\
           &  -  &$\approx$0.6 $\mu$s& - & $IT$	   & - 
                                                                     & \cite{Khuyagbaatar2021}\\
                                                                     
$^{254m1}$Rf& 8$^-$    & 4.7(1.1) $\mu$s  & - & $IT,(SF)$   & $7/2^+$[624]$_\nu$ $\otimes$ $9/2^-$[734]$_\nu$
                                                                     & \cite{David2015}\\
$^{254m2}$Rf& 16$^+$  &247(73) $\mu$s & - & $IT,(SF)$   & $7/2^+$[624]$_\nu$ $\otimes$ $9/2^-$[734]$_\nu$
                                                                     & \cite{David2015}\\
&&&&                                             &$\otimes$ $7/2^-$[514]$_\nu$ $\otimes$ $9/2^+$[624]$_\pi$\\ 
$^{255m2}$Rf$^h$& 19/2$^+$ & $29^{+7}_{-5} \mu$s  & 1.103 MeV& $IT$  
                                                &9/2$^-$[734]$_\nu$ $\otimes$ $1/2^-$[521]$_\pi$ 
                                                                   & \cite{Mosat2020a}\\
          &        &         &         &            & $\otimes$ $9/2^+$[624]$_\pi$& \cite{Chakma2023}\\
$^{255m3}$Rf$^h$& 25/2$^+$ & $49^{+13}_{-10} \mu$s  & 1.303 MeV& $IT$  
                                                 &9/2$^-$[734]$_\nu$ $\otimes$ $7/2^-$[514]$_\pi$
                                                                    & \cite{Mosat2020a}\\
          &        &         &         &            & $\otimes$ $9/2^+$[624]$_\pi$ & \cite{Chakma2023}\\
$^{256m1}$Rf& 6,7    & 25(2) $\mu$s  &$\approx$1.12 MeV& $IT$   &
                      -                                               & \cite{Jeppesen2009}\\
$^{256m2}$Rf& 10$^+$  & 17(2) $\mu$s  &$\approx$1.4 MeV& $IT$   &
                     (9/2$^-$[734]$_\nu$ $\otimes$ 11/2$^-$[725]$_\nu$) & \cite{Jeppesen2009}\\
$^{256m3}$Rf& -  & 27(5) $\mu$s  &$\>$2.2 MeV& $IT$   &
                     -                                                & \cite{Jeppesen2009}\\
$^{257m1}$Rf&  11/2$^-$ & 4.1(4) s  & $\approx$75 keV& $IT$   & 
             11/2$^-$[725]$_\nu$& \cite{Berryman2010}\\
$^{257m2}$Rf& (21/2$^+$) & 106(6)\,$\mu$s  &1.151(11) MeV& $IT$   &
                     1/2$^-$[521]$_\pi$ $\otimes$ 9/2$^+$[624]$_\pi$ 
                                                                   & \cite{Qian2009}\\
          &        &         &         &            & $\otimes$ $11/2^-$[725]$_\nu$ & \cite{Berryman2010}\\
          &        &         &         &            &                & \cite{Rissanen2013}\\
$^{266m}$Hs& -      &$\approx$74 ms&$\approx$1.2 MeV&$\alpha$         &
				     -                                                & \cite{Ackermann2012}\\
          &        &         &         &            &                & \cite{Ackermann2015b}\\
$^{270m}$Ds& 9$^-$,10$^-$&$\approx$6 ms&$\approx$1.13 MeV&$\alpha$&
				     11/2$^-$[725]$_\nu$ $\otimes$ 7/2$^+$[613]$_\nu$ & \cite{Hofmann2001}\\
          &        &         &         \multicolumn{2}{c}{or}      &
                11/2$^-$[725]$_\nu$ $\otimes$ 9/2$^+$[615]$_\nu$ & \cite{Ackermann2015b}\\
\hline

\end{longtable}
\vspace{-0.3 cm}
\begin{tablenotes}\footnotesize
\item $^a$ values taken from Ref.~\cite{Kallunkathariyil2020}
\item $^b$ see Ref.~\cite{Antalic2011}
\item $^c$ the configurations for the two isomers are still under debate (see corresponding references and discussion subsection~\ref{No})
\item $^d$ obviously the assignments for $^{255m2}$No and $^{255m3}$No are inverted in Ref.s~\cite{Bronis2022} and \cite{Kessaci2022}
\item $^e$ see Ref.~\cite{Kessaci2021}
\item $^f$ values taken from Ref.~\cite{Jeppesen2009a}
\item $^g$ note: half-lives differ by three orders of magnitude for the two literature values from~\cite{Lopez-Martens2022} and \cite{Khuyagbaatar2021} 
\item[h]configurations and the values for $T_{1/2}$ and $E^*$ are taken from Ref.~\cite{Chakma2023}
\end{tablenotes}
\end{center}
\twocolumn

\subsection{Decay mode competition}\label{dec-comp}

The quantum character of nuclear states is often crucial for the decay-mode dependent transition probability due to different hindrance effects. 
The effect that high-spin values of nuclear states have on fission barriers, was predicted, e.g., already in 1994 by S.~Cwiok et al.~for the $^{271}$Rg~\cite{Cwiok1994}. 
They find for an 11/2$^+$ state a by 2.5~MeV higher fission barrier as compared to a 1/2$^-$ state, resulting typically in a by more than two orders of magnitude longer life-time. 

The decay probability of internal transitions mainly depends, apart from the transition ($\gamma$) energy, on multi-polarity, difference of initial and final state spin values and parities, and transition type (electric or magnetic). 
It can be estimated employing relations established for electric and magnetic transitions by V.F.~Weisskopf~\cite{Weisskopf1951}. 
The major driving parameter for $\alpha$ and $\beta$ decay is the energy difference between initial and final states while the decay into analog states of equal or similar spin and parity configurations is preferred. 
Parity change and larger spin differences lead to an increased reduction of the transition probability, a decay hindrance.
For $SF$ the fission barrier height is crucial.
In the region of heavy and superheavy nuclei, these various modes are in strong competition due to similar decay probabilities for the various decay modes.
As a consequence, experimentally observed branching ratios provide valuable information on the quantum numbers of the decaying states.
In the following sections, recent observations of decay mode competition between $SF$ and $\alpha$ decay (section~\ref{a-sf}), as well as between $SF$ and $\beta$ decay~\ref{sf-b} will be discussed.

\subsubsection{Competition between $SF$ and $\alpha$ decay}\label{a-sf}

Various theory predictions suggest a fission hindrance generated on the more or less complex quantum configuration of a nuclear state.
In a Nilsson-Strutinsky model approach, on the basis of an average Woods-Saxon potential and a monopole pairing residual interaction, \v{C}wiok et al.~predicted that fission is hindered for nuclei with odd nucleon numbers, in particular, for high-spin Nilsson configurations~\cite{Cwiok1994} (see introduction of this subsection). This fact is supported by experimental observations for heavy nuclear systems. 

\begin{figure}[h]
\begin{center}
\includegraphics[width=\columnwidth]{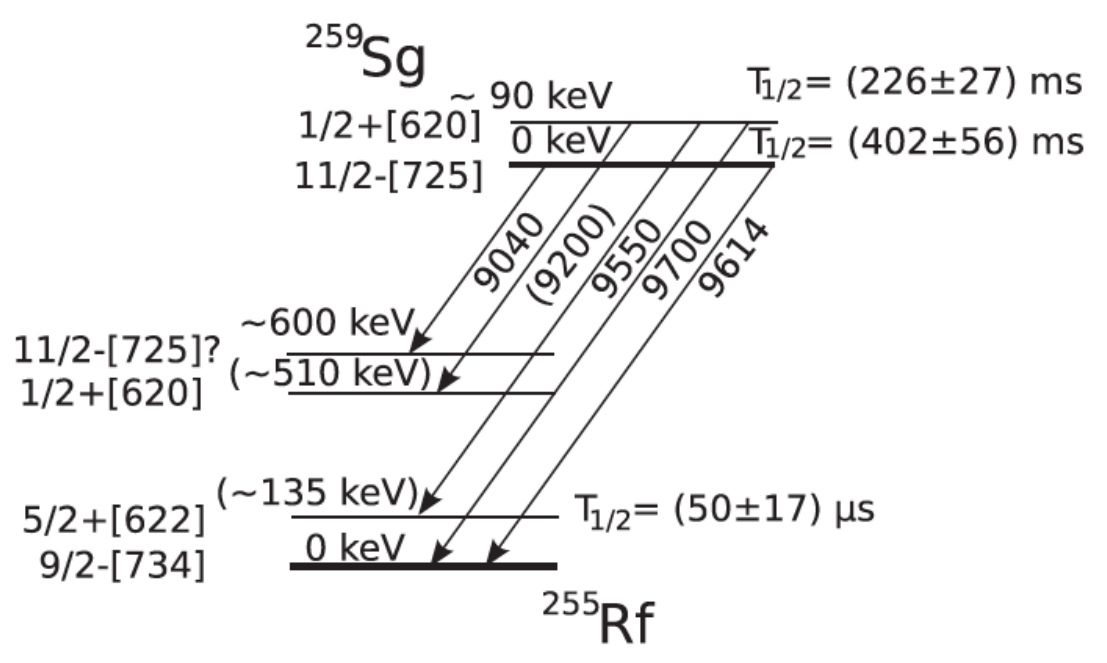}
\end{center}
\vspace{-0,5 cm}
\caption{Decay scheme for $^{259}$Sg$\longrightarrow^{255}$Rf as proposed by Antalic et al.~from Ref.~\cite{Antalic2015}. 
}
\label{fig:259Sg_levels}
\end{figure}

One example was reported earlier by Antalic et al.~regarding the decay properties of $^{259}$Sg~\cite{Antalic2015}.
For both its 11/2$^-$[725] ground state, $T_{1/2}$\,=\,411~ms, and the first excited 1/2$^+$[620] state, $T_{1/2}$\,=\,254~ms, $\alpha$ decay is observed while the $SF$ half-life is measured with 235~ms, suggesting that only the state with the low-spin configuration (1/2$^+$) is fissioning while fission from the high-spin negative-parity g.s.~is strongly hindered. 

The decay scheme for $^{259}$Sg$\longrightarrow^{255}$Rf is shown in Fig.~\ref{fig:259Sg_levels} (see also Fig.~\ref{fig:255Rf_levels} and section~\ref{Rf}). 
In the $N$\,=\,153 isotone series starting from $^{249}$Cm, at $^{259}$Sg the 1/2$^+$[620] Nilsson level is raised above the 11/2$^-$[725] state which is coming down from $\approx$\,350~keV and forming isomers in the lighter isotones with increasing half-lives (see Fig.~\ref{fig:153_isotones}). 
In turn the 1/2$^+$[620] level in $^{259}$Sg is becoming isomeric with its decay to the 11/2$^-$ g.s. being strongly hindered.

The recent results for the decay from g.s. and isomer in $^{247}$Md support a similar conclusion, as shown recently by He{\ss}berger et al.~by the comparison of the measured $SF$-$\alpha$ branching ratios which differ considerably for these two Nilsson states with similarly different configurations~\cite{Hessberger2022}.
While the 1/2$^-$[521] isomer decays with a probability of of 20(2)\% by $SF$, its 7/2$^-$[514] g.s. exhibits a fission probability of $8.6\times10^{-3}(10)$ (see subsection~\ref{Md}).

For the new data on the $^{273,275}$Ds and their decay products obtained at the FLNR JINR SHE-factory~\cite{Oganessian2024} Oganessian et al.~use similar arguments to produce scenarios, based on $SF$ hindrance arguments to construct in a first approximation decay sequences for both decay chains with decay mode, spin and level order assignments (see Fig.~\ref{fig:275Ds_decay} and section~\ref{Ds}).

\subsubsection{Competition between $SF$ and $\beta$ decay}\label{sf-b}
\begin{figure}[h]
\begin{center}
\includegraphics[width=\columnwidth]{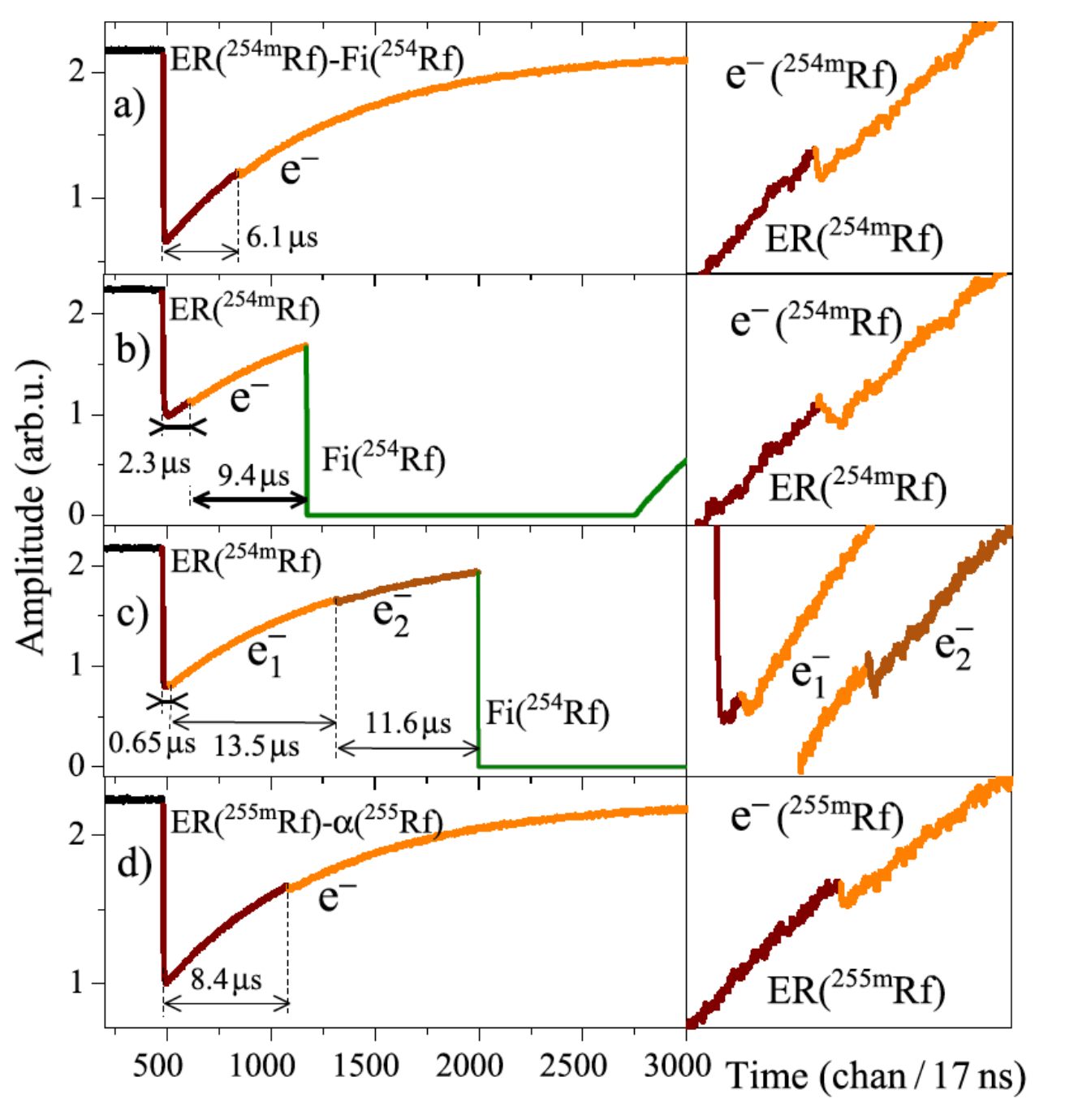}
\end{center}
\vspace{-0,5 cm}
\caption{Evaporation residue trace examples where a small-energy signal and/or fission were observed. Left panels show traces in a full time range (3000 channels) and right panels show the expanded view of traces, where the small-energy electron signals occurred. Corresponding time differences of signals are given relative to the preceding event. a) ER followed by fission of $^{254}$Rf in the analog data branch. A small-energy signal was detected after 6.1~$\mu$s. b) ER trace, in which both small-energy and fission signals associated with $^{254}$Rf were detected. c) ER trace similar to b) but with two small-energy signals. d) ER followed by $\alpha$ decay of $^{255}$Rf in the analog data branch. See text for details. (Figure and caption are taken from~\cite{Khuyagbaatar2020b})}
\label{fig:PSA_el}
\end{figure}

As mentioned in the introduction of this section, $\beta$ decay, in particular, $EC$ decay is difficult to detect for heavy and superheavy nuclei which is reflected in the almost lack lack of information on $\beta$-decay properties for isotopes with $Z$\,>\,105, beyond dubnium (see Fig.~\ref{fig:Nchart96_118}). 
The same figure illustrates, however, how competitive $\beta$ decay is for the heavy nuclei up to dubnium. 
This is also clearly demonstrated for the dubnium isotopes in a systematic analysis of $\beta$-decay half-lives as a function of $\beta$-decay energies for heavy isotopes from neptunium to dubnium in Fig.~\ref{fig:SHN_beta_sys} in section~\ref{Db}.
It shows that the measured half-lives deduced from the fission events that were attributed to the odd-odd dubnium isotopes at the end of the decay chains originating from the decay of moscovium and tennessine isotopes, $^{266,268}$Db, are competitive with $\beta$ decay half-lives deduced from their $Q_\beta$ values, raising some doubt about the $SF$ assignment (see subsection~\ref{Db}). Following this argumentation in Fig.~\ref{fig:Nchart96_118} the possible partial $SF$ decay is indicated by faint red colored triangles for these nuclides and, for similar reasons, also for $^{263}$Db (see subsection~\ref{Db}).

The difficulties in identifying $\beta$ decay in a decay chain can successfully overcome with the state-of-the-art DSAS detection systems with their advanced time and energy resolution capabilities, corroborated by reduced thresholds (see section~\ref{DSAS}). 
As shown for the $\beta$ decay of $^{258}$Db to $^{258}$Rf, $CE$s play also here a decisive role which led to the identification of two decay activities, one of them populating an excited state in the daughter nucleus~\cite{Hessberger2016a}.

An impressive example for the resolution of shortest times by advanced digital electronics is the trace analysis of a small electron signal on the tail of the large signal produced by the implantation of an ER. 
Fig.~\ref{fig:PSA_el} shows the detection of $CE$ signals on the tail of $ER$ signals which were successfully used by Khuyagbaatar et al.~to identify the isomers $^{254m}$Rf and $^{255m}$Rf (see subsection~\ref{Rf}).

Trace analysis can also be used to measure large fission fragment energies by the time-over-threshold method as demonstrated by Khuyagbaatar et al.~for the $^{254}$Rf decay in Fig.~\ref{fig:PSA_fission}~\cite{Khuyagbaatar2020b} together with the much lower $\alpha$ or even electron energies with a single preamplifier configuration (see subsection~\ref{Rf}).
The there quoted sample time of 17~ns allows for the time separation of processes in the sub-$\mu$s regime.

 \subsection{Some notes on advancement in theory}\label{theory}

 As mentioned in the introduction, the purpose of this article is to report on the progress of experimental findings in the regime of the heaviest nuclei, without the ambition to dive into the details of the various model approaches, beyond the discussion of the reported data. 
\begin{figure*}[ht]
\begin{center}
\includegraphics[width=\textwidth]{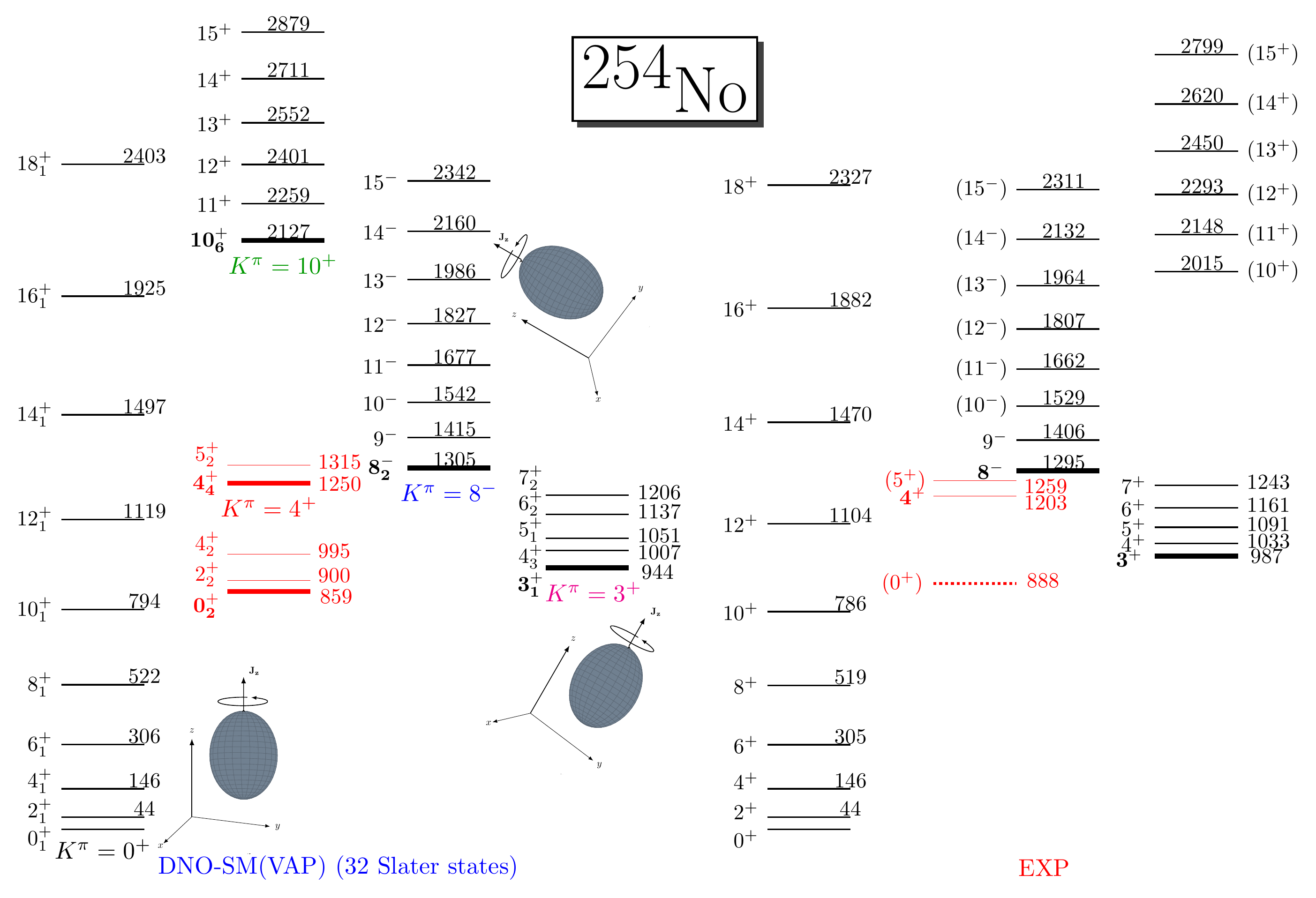}
\end{center}
\vspace{-0,5 cm}
\caption{Spectra of yrast and excited bands from DNO-SM(VAP) calculations in comparison with the experimental
data (EXP) taken from~\cite{Herzberg2006,Clark2010,Hessberger2010,Wahid2025,Forge2023}. 
The red levels are the newly proposed states from the recent study in~\cite{Forge2023,Forge2025}.
Tentative experimental spin-parity assignments are given between parentheses. 
Excitation energies are in keV. 
$K$ denotes the total angular momentum projection quantum number. 
The $K$-band heads are given in bold lines with the associated $K$ component which amounts up to $\approx$\,95\%, probing the axial symmetry preservation in the correlated wavefunctions. 
The ellipsoids, with different orientations (pointing to the appearance of different $K$ bands), thus illustrate the analogy to a classic rotating axial rigid rotor (cf.~pages 475-476~\cite{Ring1980}). 
(Figure and caption are taken from~\cite{Dao2025}; courtesy of F.~Nowacki)}
\label{fig:254No_SM}
\end{figure*}
 
 In section 4~{\it Theory lessons and exotic phenomena} of our previous review~\cite{Ackermann2017}, we had discussed some achievements of the various models, basically divided into the classical two approaches of macroscopic-microscopic and self-consistent models, the latter consisting of energy density functional as well as relativistic mean field approaches. 
 In this context, we discussed mainly the theoretical treatment of exotic shapes, which are certainly a central feature for nuclear matter, developing from the region of deformed nuclei around $Z$\,=\,100 and 108 and N\,=\,152 and 162 towards sphericity expected for the proton numbers 114, 120 or 126 and neutron numbers 172 and 182.

 Similarly, I chose to briefly discuss here in the following subsections two examples of recent theory developments, both published this year after the first submission of this review, which might be relevant for the progress in experimental SHN research envisaged for the coming years, in particular in view of the state-of-the-art and near-future facilities discussed in the last section of this review, section~\ref{outlook} Outlook.
 
 The first subject is a novel approach to investigating the superheavy nucleus $^{254}$No, which is a key isotope for DSAS and in-beam studies, as well as for experiments employing atomic physics methods like traps and laser spectroscopy, in the framework of a complete shell-model description by Dao and Nowacki~\cite{Dao2025}.  
 
 In the second subsection, I will summarize a theoretical study of the relevance of weak decays for SHN, with a first paper on {\it Weak decays in superheavy nuclei} published in May 2025 by Ravli\'{c} and Nazarewicz~\cite{Ravlic2025}, and a second one published at the end of June 2025, elaborating on the subject by discussing {\it Electron capture of superheavy nuclei with realistic lepton wave functions}, by  Ravli\'{c}, Schwerdtfeger and Nazarewicz~\cite{Ravlic2025a}. 
 This is well connected to the topic of $\beta$ decay and its competition with other decay modes as I discussed it in subsection~\ref{sf-b}. 

\subsubsection{$^{254}$No, SHN with even/odd nucleon numbers and the shell model}\label{254No_SM}

In our review from 2017~\cite{Ackermann2017}, we discussed in detail the prominent role of $^{254}$No that continues to be in the focus of many experimental and theoretical efforts up to date (see subsection~\ref{No} and the reference list in Table~\ref{tab:isotope_list}).
Beyond DSAS and nuclear in-beam studies, it attracted the interest of the atomic physics community, providing complementary information by the use of high precision mass measurements or advanced methods of laser spectroscopy~\cite{Block2010,Minaya-Ramirez2012,Raeder2018,Block2021}. 

The latest addition to the collection of theoretical approaches to understanding the nuclear structure of $^{254}$No is, at the same time, the first attempt of a complete shell model description by Dao and Nowacki~\cite{Dao2025}.
In this paper, the authors report on the first complete shell-model description of low-lying structures in superheavy nuclei like $^{254}$No, comparing in a first step their calculations for even-even, odd-A and odd-odd neighboring nuclei to experimental values with a rather decent reproduction quality. 

To this end, they use the novel Discrete Non-Orthogonal Shell Model with angular-momentum Variation After Projection (DNO-SM(VAP)), using an effective Kuo-Herling interaction in a generator coordinate method (GCM) framework on a basis of non-orthogonal wave functions, introduced by the authors in 2022~\cite{Dao2022}.

Describing the low-lying structure of odd-nucleon isotopes around $^{254}$No successfully, they obtain a remarkable reproduction of its experimental level scheme (see Fig.~\ref{fig:254No_SM}). 
In particular, they reproduce the experimental findings by Forge et al.~\cite{Forge2023} on the observation of a second 0$_2^+$ state and the 3$^+$-4$^+$ Gallegher-Moszkowski splitting~\cite{Gallagher1958}.

Regarding the discussion of the debated configuration of the long-lived 8$_2^-$ K-isomer, they propose that this state originates from the proton-neutron coupling ($[h 9/2_\pi$ $\otimes$ $j 15/2_\nu$]) with small mixing of other orbitals, as compared to the two-quasiproton configuration  $7/2^-$[514]$_\nu$ $\otimes$ $9/2^+624]_\nu$) proposed recently by Wahid et al.~\cite{Wahid2025} and earlier by He{\ss}berger et al.~\cite{Hessberger2010}, and the two-quasineutron configurations $9/2^-$[734]$_\nu$ $\otimes$ $7/2^+$[624]$_\nu$ or $9/2^-$[734]$_\nu$ $\otimes$ $7/2^+$[613]$_\nu$ as also earlier discussed by R.M.~Clark et al.~\cite{Clark2010} (see discussion in subsection~\ref{No}).

In this context, the 10$+$ band head and its decay into states of the rotational band above the 8$_2^+$ isomer play a major role.
Dao and Nowacki find this state at 2127~keV, as compared to 2015~keV from experiments, and claim a four-particle recoupling (($[h 9/2_\pi$ $\otimes$ $i 13/2_\nu]\times[j 15/2_\nu$ $\otimes$ $g 7/2_\nu]$) as its main origin.
Further details, especially on recent experimental findings are discussed in subsection~\ref{No}

\begin{figure*}[ht]
\begin{center}
\includegraphics[width=0.71\textwidth]{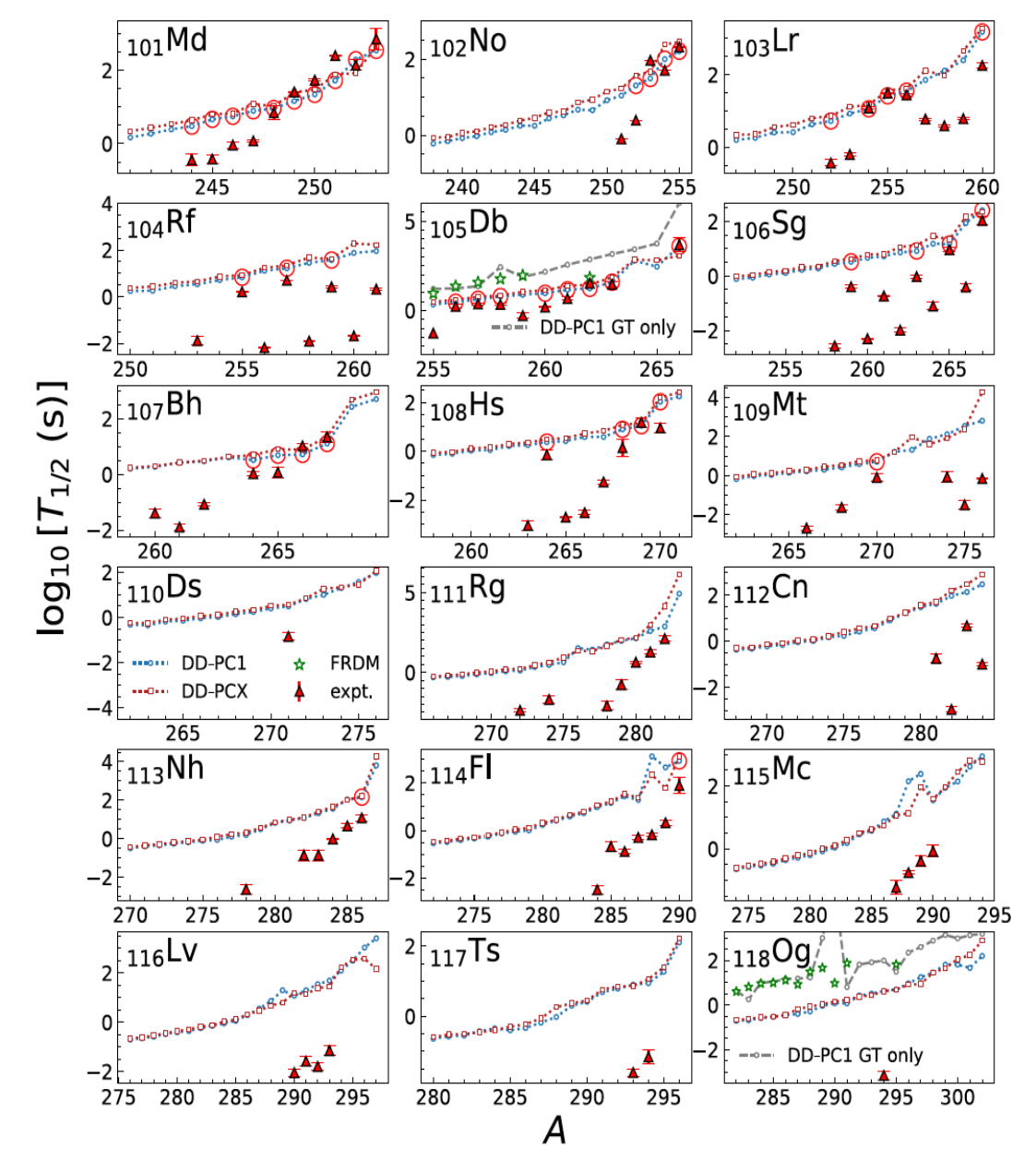}
\end{center}
\vspace{-0,5 cm}
\caption{The $\beta^+$/$EC$-decay rates for nuclei between $_{101}$Md up to $_{118}$Og. 
The RQRPA calculations with the DD-PC1 (open circles) and DD-PCX (open squares) functionals are compared with the available experimental data (triangles)~\cite{Kondev2021}, and microscopic-macroscopic (FRDM) calculations (stars)~\cite{Moeller1997}. 
Nuclei with $EC$ branching ratio $\mathcal{R}$ predicted to be larger than 5\% with respect to measured half-lives, consistently for both functionals, are marked with red circles. For $_{105}$Db and $_{118}$Og chains, we display the DD-PC1 EC rates with contribution from only allowed GT transitions (dashed line). 
(Figure and caption are taken from~\cite{Ravlic2025})}
\label{fig:EC_times_md-og}
\end{figure*}

\subsubsection{SHN and weak decays}\label{weak}

As pointed out in section~\ref{sf-b}, no experimental evidence has been registered on weak decays for nuclei beyond dubnium ($Z$\,=\,105). 
One of the reasons for this is that the prevailing weak decay mode for those relatively neutron-deficient isotopes is electron capture ($EC$), which can be detected only indirectly via the identification of the decay-daughter isotope.
How this can be achieved, e.g., by employing $CE$- and $x$-ray correlations, is discussed in more detail for $^{258}$Db in Ref.~\cite{Hessberger2016a} (see section~\ref{Db}).

On the theory side, as mentioned in the introduction of this subsection, Ravli\`{c} et al.~report on a novel approach to understand the weak decay of SHN. 
In the first paper cited, they present the results of calculations, based on relativistic nuclear density functional theory and the quasiparticle random-phase approximation. 
In comparison to earlier approaches, by P.~Sarriguren~\cite{Sarriguren2019,Sarriguren2020,Sarriguren2021,Sarriguren2022}, and P.~M\"{o}ller, J.~Nix and K.-L.~Kratz~\cite{Moeller1997}, where only allowed transitions (Gamov-Teller(GT)) were considered, first forbidden (FF) transitions, in particular, 1$^-$ transitions play an important role and can lead to competitive transition probabilities.
For $EC$/$\beta^+$ decay, with $EC$ being the dominating decay mode for SHN, on the neutron-deficient side, as well as for $\beta^-$ decay for the neutron-rich part of isotopic chains, the authors find the FF 1$^-$ transitions dominate for the investigated elements from mendelevium to oganesson.
In this range of the Segr\'{e} chart, employing the DD-PC1 and DD-PCX functionals for their EDF calculations, they obtain 45 isotopes for which the $EC$ decay has a branching greater than 5\%, of which only six are even-even, as shown in Fig.~\ref{fig:EC_times_md-og} where those isotopes are marked by red circles.
While most of those nuclides are found in the lighter elements, they extend beyond dubnium up to one in flerovium ($^{290}$Fl). 
Particularly interesting, regarding the dubnium isotopes, is the comparison to Fig.~\ref{fig:SHN_beta_sys}, with $EC$ found to be competitive for $^{263}$Db and even dominating (shorter EC half-life) for $^{268}$Db, supporting the suggestions in Fig.~\ref{fig:Nchart96_118}, where I marked possible $EC$ decay by pink triangles.
These findings support the experimental search for weak decay modes in SHN, as discussed briefly also above in subsection~\ref{sf-b}.

In the second article, mentioned in the introduction of this subsection, Ravli\`{c}, Schwerdtfeger and Nazarewicz~\cite{Ravlic2025a}, introduce an improved (realistic) lepton (electron) wave functions to better model the $EC$ decay process.
Instead of the standard lowest-order (LO) approximation normally employed, they use a single-particle approximation Dirac Wave Functions (DWF), as well as a many body ansatz with Dirac-Hartree-Fock Wave Functions (DHFWF) to calculate the electron radial wave functions (ERWFs). 
The nuclear response is calculated using relativistic density functional theory (DFT) and Quasiparticle Random Phase Approximation (QRPA).

After benchmarking their method for lighter to medium-mass nuclei from the iron, palladium and dysprosium isotopic chains, they apply their calculations based on more sophisticated electron wave functions to superheavy nuclei.
For the example of oganesson isotopes ($^{290,294}$Og), they obtain an $EC$-rate reduction of up to 40\%.

An experimental verification of these predictions is clearly desirable. 
An investigation employing $CE$ and x-ray spectroscopy as, e.g., applied by He{\ss}berger et al.~for the case of the $^{258}$Db $EC$-decay into the ground and an excited state of $^{258}$Rf~\cite{Hessberger2016a}, seems most promising.

\newpage
\section{Nuclear properties of the isotopes of the heaviest twenty elements}\label{prop}

\begin{figure*}[ht]
\begin{center}
\includegraphics[width=1\textwidth]{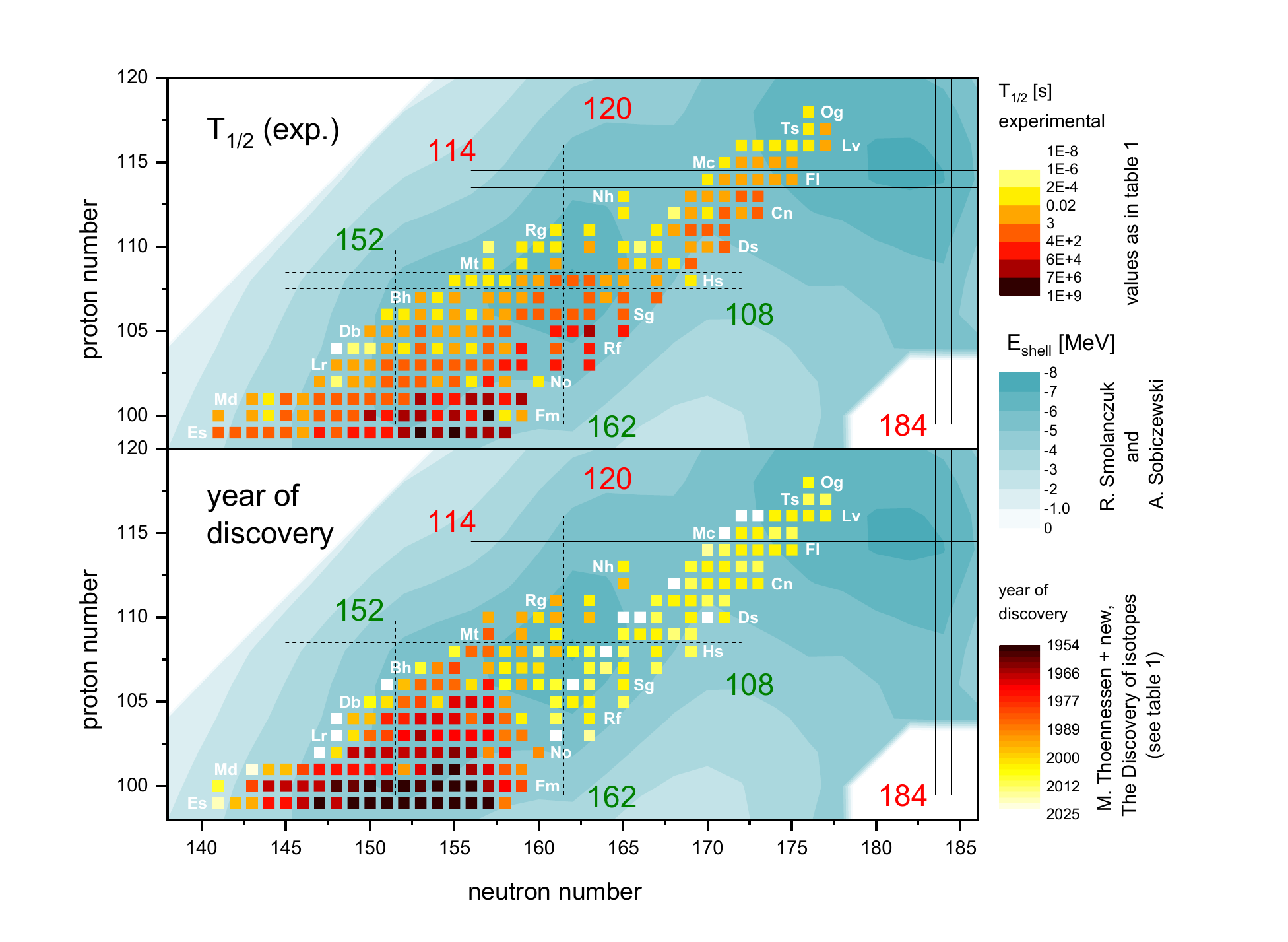}%
\vspace{-1cm}
\end{center}
\caption{Shell correction energies from Ref.s~\cite{Smolanczuk1995a,Smolanczuk1995} as a function of proton and neutron numbers with (upper panel) experimental half-lives $T_{1/2}$ as given in table~\ref{tab:isotope_list} and (lower panel) years of discovery from Ref.~\cite{Thoennessen2016}, including updates from the {\em Discovery of Nuclides Project} at \href{https://people.nscl.msu.edu/~thoennes/isotopes/}{https://people.nscl.msu.edu/$\sim$thoennes/isotopes/}, for the heaviest nuclei from einsteinium to oganesson. 
The predicted closed shells on the {\it ``island of stability''} for spherical superheavy nuclei SHN for protons at $Z$=114 and 120, and neutrons at $N$=184 are indicated by solid red lines.
The subshell closures or shell gaps for deformed shell-stabilized nuclei for protons at $Z$=108, and for neutrons at $N$=152 and 162, are indicated by dashed green lines.
\label{fig:E_shell_dates}}
\end{figure*}

The continuous history of synthesis and discovery of isotopes from einsteinium to oganesson starts in the mid-1950ies, and the most recent isotopes were discovered this year.
The lower panel of Fig.~\ref{fig:E_shell_dates} illustrates the development in terms of year of discovery for all 204 nuclides, of which five have been discovered in 2025. 
The background for both panels of Fig.\ref{fig:E_shell_dates} shows the shell-correction energies calculated in the framework of the macroscopic-microscopic model introduced by Adam Sobiczewski~\cite{Smolanczuk1995a,Smolanczuk1995}. 
The year of discovery, identified as the year of first publication, has been verified with the data reported in~\cite{Thoennessen2016}, including updates from the "Discovery of Nuclides Project" at:\\
\href{https://people.nscl.msu.edu/~thoennes/isotopes/}{https://people.nscl.msu.edu/$\sim$thoennes/isotopes/}.\\
The upper part of the same figure shows their half-lives which have been taken from the National Nuclear Data Center (NNDC) of Brookhaven National Laboratory (BNL) via its website:  \href{https://www.nndc.bnl.gov/}{https://www.nndc.bnl.gov/}, verified with NUBASE2020 data evaluation~\cite{Kondev2021} and updated with recent publications as listed in Table~\ref{tab:isotope_list}.

\begin{figure*}[ht]
\begin{center}
\includegraphics[width=0.9\textwidth]{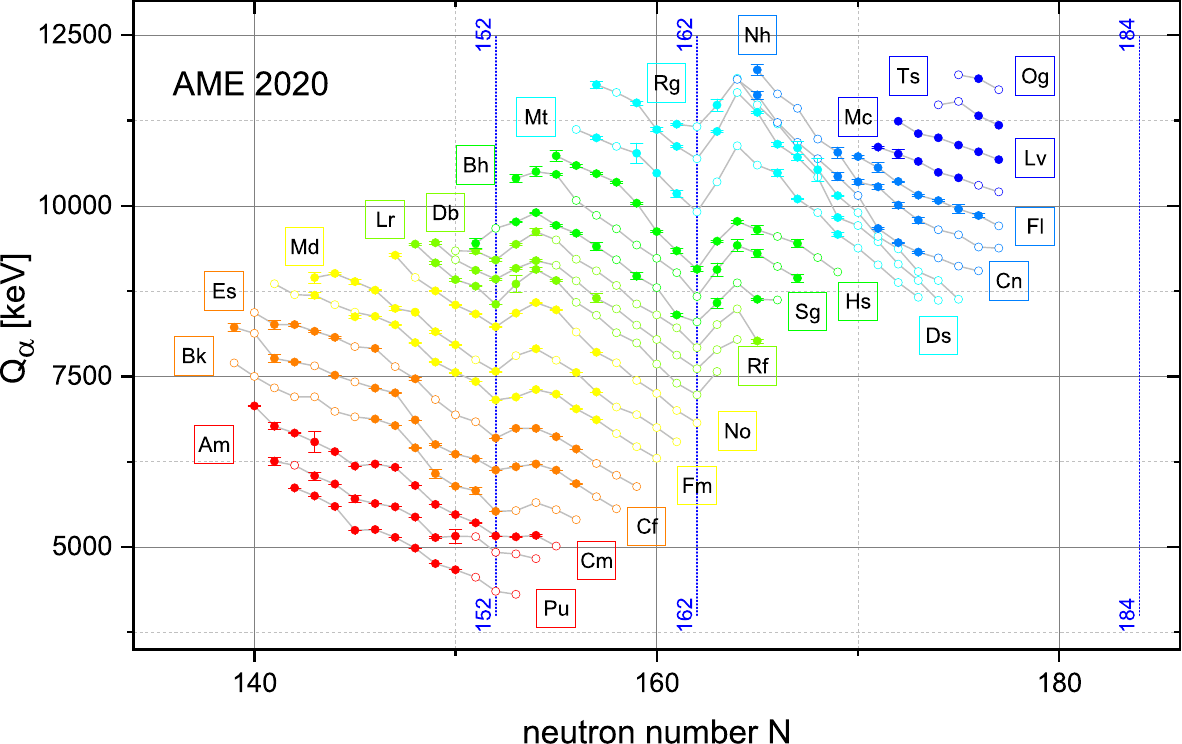}%
\end{center}
\caption{$Q_\alpha$ values from Ref.~\cite{Wang2021} and from recent experimental data (see Table~\ref{tab:isotope_list}) as a function of neutron number for isotopes from $Z$=94 to 118, indicated by their chemical Symbols.
Full circles denote experimental and open circles denote theoretical values. Blue dotted lines mark the predicted neutron shell closure at $N$=184 and the deformed neutron shell gaps at $N$=152 and 162, respectively.
\label{fig:Qa_94_118}}
\end{figure*}

\begin{figure*}[ht]
\begin{center}
\includegraphics[width=0.9\textwidth]{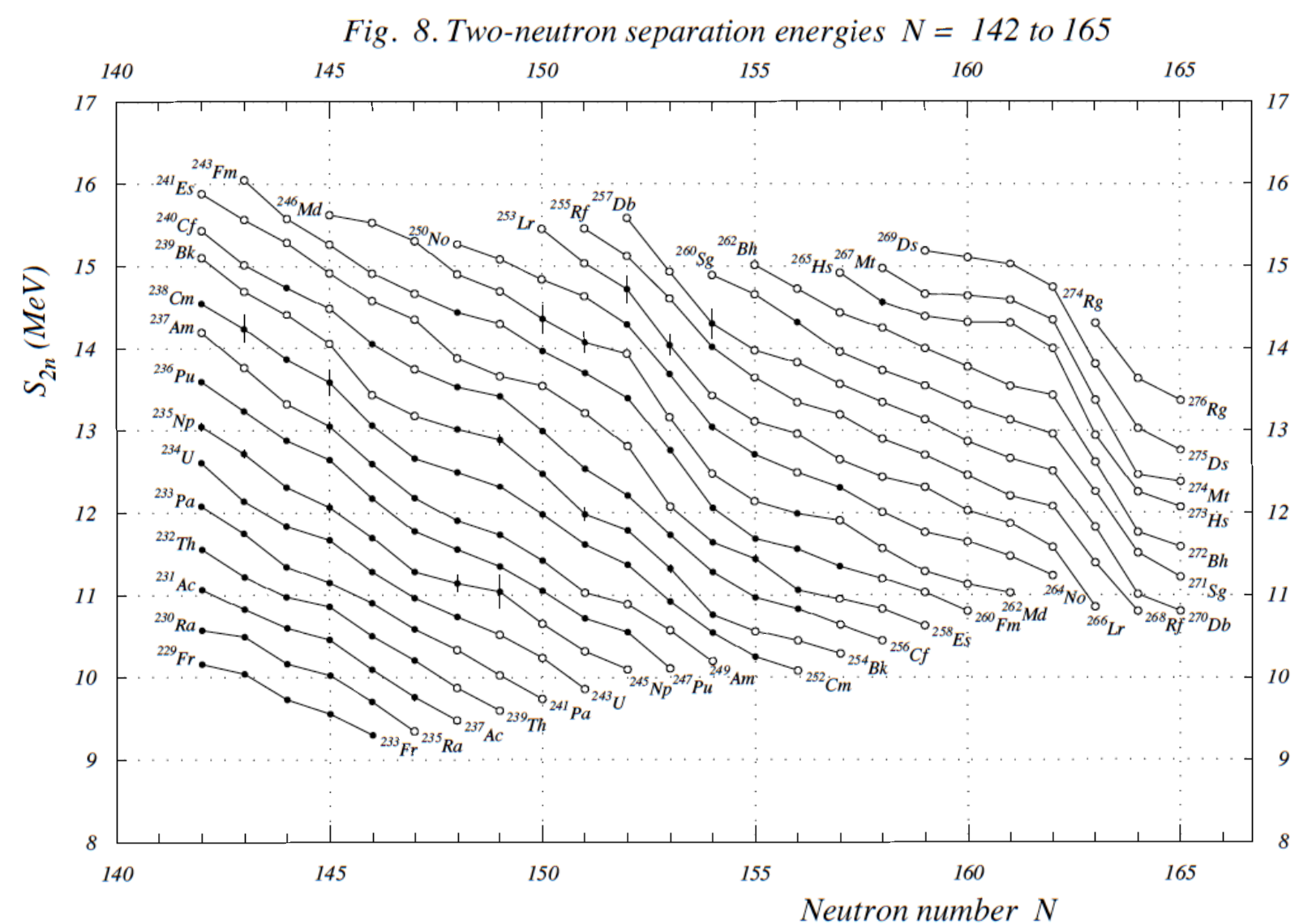}%
\vspace{-0.5 cm}
\end{center}
\caption{Two-neutron separation energy values $S_{2n}$ of even-$Z$ elements from francium ($Z$\,=\,87) to roentgenium ($Z$\,=\,111), and from $N$\,=142 to 165.
(Figure taken from Ref.~\cite{Wang2021})
\label{fig:S_2n}}
\end{figure*}

In both panels, shell closures, predicted by the various models, for protons ($Z$=114 and 120) and neutrons ($N$=184) are indicated, together with the shell gaps of deformed SHN at $Z$=108 and $N$=152 and 162.
From the shown nuclear half-lives only, partly due to the large range spanning from 10$^{-7}$~s to 472 days, the location of these assignments is not obvious at first glance. 
They become more obvious from the combined theoretical (open circles) and experimental $Q_\alpha$ values as a function of neutron number $N$ presented in Fig.~\ref{fig:Qa_94_118} with pronounced minima for a broad elemental range at the deformed neutron shell gaps.
In addition, large gaps between the isotopic lines in the vicinity of $N$=162 are visible at $Z$=108. 
A similar, less pronounced effect is present around fermium at $N$=152, which confirms the shell gap at $Z$=100 as shown in the Nilsson representation of single-particle levels as a function of deformation in Ref.s~\cite{Bohr1975} (Fig.~5.4) and~\cite{Chasman1977} (Fig.~4) at quadrupole deformations around 0.25.
This is supported by the trend of two-neutron separation energies in Fig.~\ref{fig:S_2n}, which shows a more pronounced slope change for fermium isotopes around $N$=152 as compared to the neighboring even-$Z$ isotopic chains.
Another indication for a possible shell gap at $Z$=100 is provided by the trend of excitation energies $E^*$ of the first $2^+$ states of isotopes in the vicinity of $N$=152 (see subsection~\ref{Fm} and Fig.~\ref{fig:2+_92-104}). 

In Table~\ref{tab:isotope_list}, basic properties, like half-lives, ground-state spin and parities (if known), decay mode branching ratios, $Q_\alpha$ values and year of discovery for the 198 known isotopes from einsteinium to oganesson are listed together with the relevant publications.
In addition to the ground states, experimentally found isomeric states indicated as $^{Am}Z$ are also listed with their excitation energies $E^*$, if known. 
The following subsections will report on recent results for the isotopes of the elements from einsteinium to oganesson.
A comprehensive collection of papers since 2017 is organized in subsections for each isotope for which new data has been accumulated, with a more detailed discussion for specifically interesting cases.
For the highest Z elements, seaborgium and beyond, many isotopes are produced as members of decay chains of heavier ERs, with recent re-measurements of some of those decay chains, leading mainly to a reduction of uncertainties of their decay properties listed in Table~\ref{tab:isotope_list} with updated values with respect to data provided by the NNDC website and the NUBASE2020 data collection~\cite{Kondev2021}.
An updated overview of the decay modes is presented in Fig.~\ref{fig:Nchart96_118} in the introduction.

\begin{figure*}[ht]
\includegraphics[width=\textwidth]{Es-Db_decay_scheme_2024.pdf}
\caption{Decay scheme for odd-$Z$/even-$N$ isotopes from einsteinium to dubnium in the vicinity of the $N$=152 closed shell, updated from Ref.~\cite{Ackermann2017}. 
Data are taken from~\cite{Hatsukawa1989,Moody1993,Ninov1996,Gan2001,Hessberger2001,Gan2004,Chatillon2006,Hessberger2005,Hauschild2008,Antalic2008,Qian2009,Antalic2010,Hessberger2012,Hessberger2016,Pore2020,Khuyagbaatar2020,Hessberger2021,Hessberger2022}.
The measured $\alpha$-decay energies (not the $Q_\alpha$ values) are given for easier comparison with measured data.
For$^{243}$Es, the decay scheme has been updated with the new findings of He{\ss}berger et al.~\cite{Hessberger2022}, where the $\pi$3/2$^-[521]$ Nilsson level has found to be the ground state, and the $\pi$7/2$^+[633]$ lying above at an excitation energy of 10~keV, while the Antalic et al.~\cite{Antalic2010} considered still two possible scenarios (see previous version in Ref.~\cite{Ackermann2017}).
For the other einsteinium ground state assignments, the representation of~\cite{Hessberger2005} has been adopted here. 
For $^{245}$Md and $^{241}$Es $\alpha$ transitions are shown as reported in the original paper (\cite{Ninov1996}). 
For the complete transition data see Ref.~\cite{Hessberger2021}, and Fig.~\ref{fig.245Md_Ea} in subsection~\ref{Md}.}
\label{fig.Es-Db_decay_scheme}
\end{figure*}

\subsection{Einsteinium - $Z$=99}\label{Es}

For the element with atomic mass number 99, einsteinium, a total of 18 isotopes from $^{240}$Es to $^{257}$Es has been observed. 
For many odd-$Z$/odd-$A$ einsteinium isotopes, the assignment of the ground state spin and parity is still open, with an unambiguous assignment being hampered by the predicted small energy difference of the g.s. candidate configurations 7/2$^+$[633] and the 3/2$^-$[521]. This can be seen from Fig.~\ref{fig.Es-Db_decay_scheme} which shows the state of knowledge in a global decay scheme for odd-$Z$/odd-$A$ isotopes from einsteinium to dubnium as an update of the scheme shown in~\cite{Ackermann2017}. 
While the experimental activities concerning the heavier einsteinium isotopes date further back in time, the three lightest ones received quite some attention lately.
The lightest two isotopes are subject of a recent debate in the community, studying the nuclear properties of the heaviest nuclear species \cite{Pore2020,Khuyagbaatar2020,Hessberger2021}, which will be discussed in subsection~\ref{Md} together with the mendelevium isotopes from which they were populated via $\alpha$ decay.\\
\begin{figure}[ht]
\begin{center}
\includegraphics[width=0.5 \textwidth]{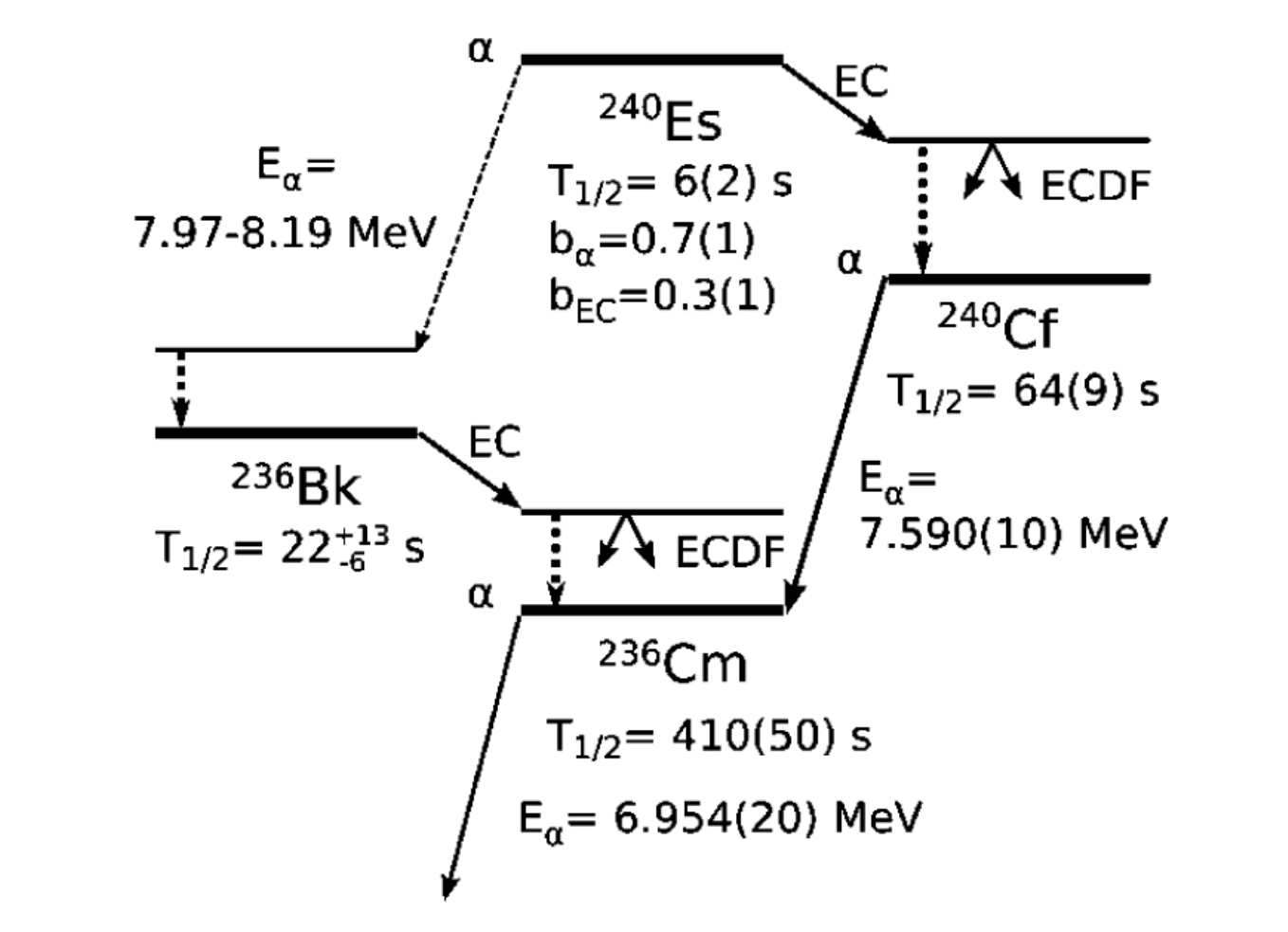}%
\end{center}
\caption{Decay scheme proposed for $^{240}$Es as proposed in Fig.~5 in Ref.~\cite{Konki2017}.
\label{fig:240Es_decay}}
\end{figure}

{\it $^{\it240}$Es}\\ 
$^{240}$Es is the most recent addition to the einsteinium isotopic chain and was first reported in 2017 by Konki et al.~\cite{Konki2017}, together with its $\alpha$-decay daughter nucleus $^{236}$Bk.   
For $^{240}$Es $\alpha$ decay and $EC$ with branchings of $b_\alpha = 0.7(1)$ and $b_{EC} = 0.3(1)$  and a half-life of $T_{1/2} = 6(2)$~s were established, while for $^{236}$Bk only $EC$ decay was observed, with a half-life of $T_{1/2} = 22^{+13}_{-6}$~s. 
In both cases the $EC$ decay leads to ($\beta$/$EC$-delayed) fission (ECDF) from excited states populated in their respective decay-daughter nuclei (see Fig.~\ref{fig:240Es_decay}). 
These findings complete a systematic picture of einsteinium and berkelium ECDF probabilities, showing an increasing linear trend as a function of the difference between the $EC$-decay $Q$-value and the fission barrier $Q_{EC}-B_{SF}$ towards decreasing neutron number for the light members of their respective isotopic chains, with $^{240}$Es being the lightest known einsteinium isotope. For a more detailed discussion of $\beta$-delayed fission see Ref.~\cite{Andreyev2013} and updates in Ref.~\cite{Andreyev2018}. \\

{\it $^{\it241}$Es}\\ 
The identification of $^{241}$Es was first reported in 1996, produced in the irradiation of $^{209}$Bi by $^{40}$Ar at the velocity filter SHIP at GSI ~\cite{Ninov1996} by $\alpha$ decay of $^{245}$Md. 
In addition to $\alpha$ decay, $EC$ was observed. 
As $^{241}$Es is part of the discussion in the debate mentioned in the introduction of this subsection which will be further discussed in subsection~\ref{Md}, it is attracting some interest presently, after it was first observed in 1996 together with $^{242}$Es as $\alpha$-decay products of $^{245,246}$Md~\cite{Ninov1996}. \\

\begin{figure*}[ht]
\begin{center}
\includegraphics[width=0.9\textwidth]{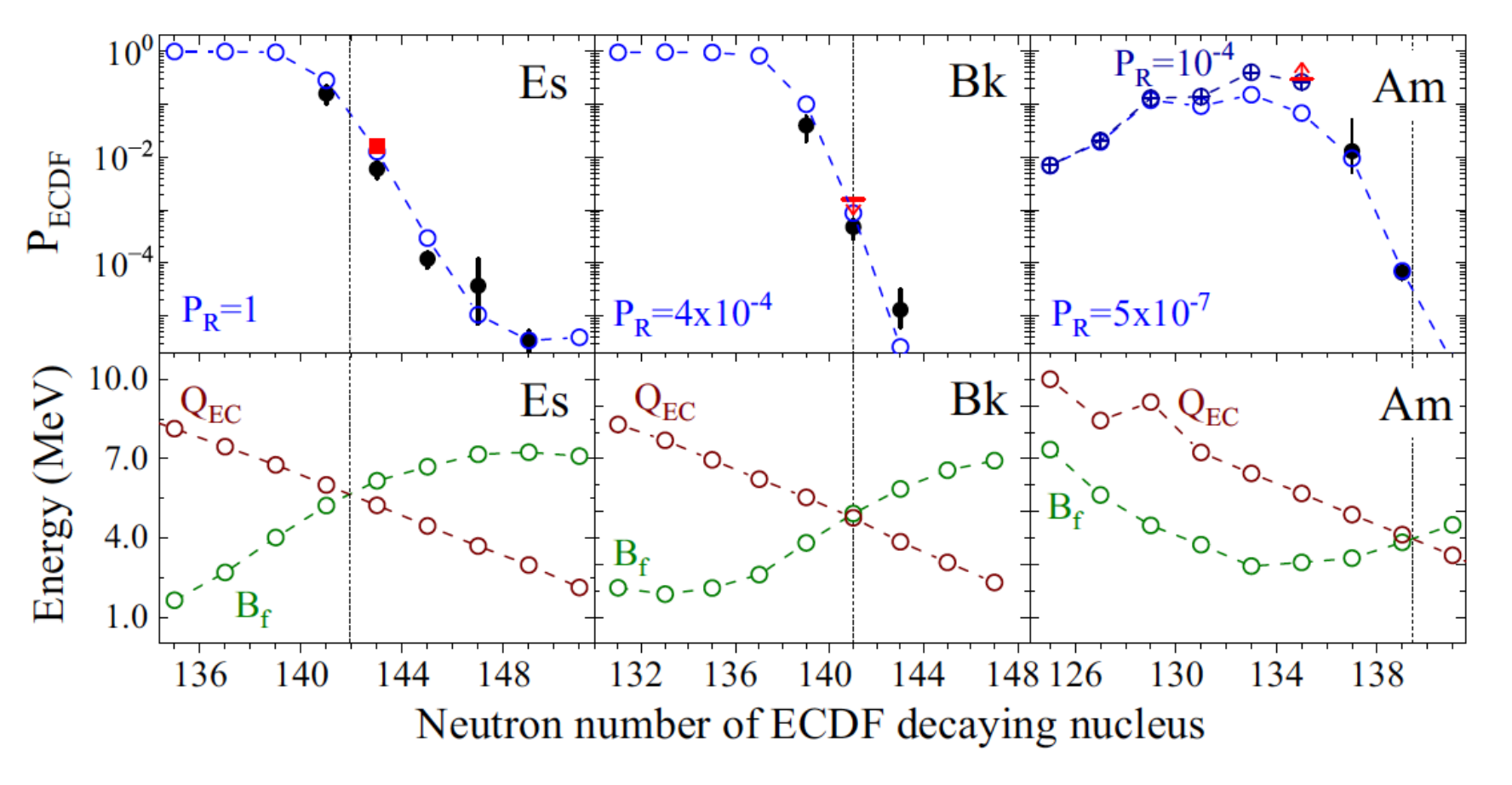}%
\vspace{-0.5 cm}
\end{center}
\caption{Upper panels: Probabilities of ECDF in Es, Bk, and Am isotopes. 
Solid and open circles mark the experimental~\cite{Andreyev2013a,Konki2017} and calculated~\cite{Khuyagbaatar2019a} $P_{ECDF}$, respectively. 
A retardation factor PR used for the description of the delayed fission probabilities according to Ref.~\cite{Khuyagbaatar2019a} and corresponding to calculated values (open circles) is given for each element. 
The presently measured value for $^{242}$Es is shown by a solid square. 
Arrows indicate the present upper limit for $^{238}$Bk and the recently reported lower limit for $^{230}$Am~\cite{WilsonGL2017}. 
Lower panels: Theoretical $B_f$ and $Q_{EC}$ values taken from Refs.~\cite{Moeller2009,Moeller1995} are shown as a function of the neutron number of the EC-decaying odd-odd nucleus. 
Dashed vertical lines mark the crossing of the $Q_{EC}$ and $B_{f}$ curves. 
(Figure and caption are taken from Ref.~\cite{Khuyagbaatar2024}.
\label{fig:ECDF_Es-Bk-Am}}
\end{figure*}

{\it $^{\it242}$Es}\\ 
In 2024, Khuyagbaatar et al.~reported on a new measurement of the decay properties of $^{242}$Es~\cite{Khuyagbaatar2024}.
In addition to 662 $\alpha$ decays, they observed 26 fission events, which they assigned to $\beta$-($EC$-)delayed fission (ECDF), expected to occur for heavy odd-odd nuclei with an enhanced probability, due to the fission hindrance for odd and odd-odd heavy nuclei.
The isotopic ECDF systematics regarding ECDF probabilities in relation to $Q_{EC}$ values and fission barrier heights for einsteinium in comparison to berkelium and americium isotopic chains are shown in Fig.~\ref{fig:ECDF_Es-Bk-Am}.\\

{\it $^{\it243}$Es}\\ 
In a new measurement, employing for the first time the direct production by the fusion-evaporation reaction $^{197}$Au($^{48}$Ca,$2n$)$^{243}$Es, the earlier observed decay properties of $^{243}$Es have been confirmed by Briselet et al.~\cite{Briselet2019}. 
This was part of a series of measurements performed at the cyclotron accelerator laboratory of the University of Jyv\"{a}skyl\"{a} (JYFL), using also $^{203,205}$Tl targets to populate the mendelevium isotopes $^{249,250}$Md, which will be discussed in subsection~\ref{Md}. 

\begin{figure}[ht]
\begin{center}
\includegraphics[width=\columnwidth]{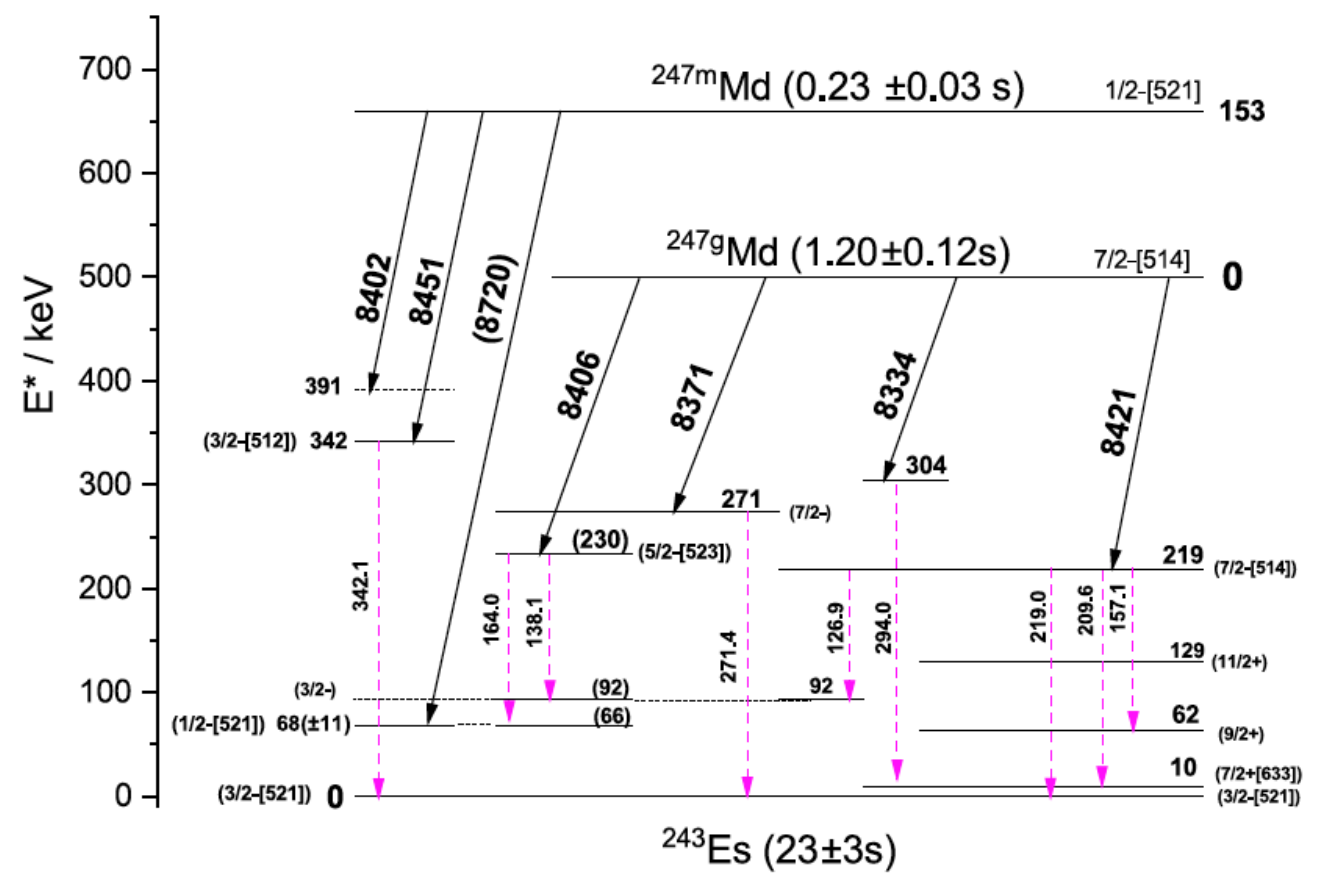}%
\vspace{-0.5 cm}
\end{center}
\caption{Proposed tentative decay scheme of $^{247}$Md.
Half-lives for $^{247gs,247m}$Md are from this work~\cite{Hessberger2022}, half-life of $^{243}$Es is taken from~\cite{Antalic2010}. Alpha transitions
are represented by full lines, $\gamma$ transitions by dashed lines.
(Figure and caption are taken from Ref.~\cite{Hessberger2022}).
\label{fig:247Md-levels}}
\end{figure}

A first tentative decay scheme for $^{243}$Es was proposed in Ref.~\cite{Antalic2010} from $^{247}$Md to $^{239}$Bk which is expected to decay almost exclusively by $EC$.  
New experimental information is now also available from a recent $^{247}$Md $\alpha$-$\gamma$ decay study (see also subsection~\ref {Md}), which provides important information for the above-mentioned ground state spin and parity assignment in a recent paper by He{\ss}berger et al.~\cite{Hessberger2022}.
Following the earlier work~\cite{Antalic2010}, strong arguments for the ordering of the 7/2$^+$[633] and the 3/2$^-$[521] states have now been presented, and the $^{243}$Es ground-state could be assigned as 3/2$^-[$521$]$ establishing the hitherto uncertain order in competition with the close-lying 7/2$^+[$633$]$. 
The excitation energy $E^*$ of the 7/2$^+[$633$]$ above g.s.~is given as 10~keV. 
With this, a total of six low-lying Nilsson states have been assigned tentatively (see Fig.~\ref{fig:247Md-levels}). 
The 7/2$^-[$514$]$ level at $E^*$=219~keV, as well as the 5/2$^-$[523], are populated by 8421~keV and 8406~keV $\alpha$ transitions from the $^{247}$Md 7/2$^-[$514$]$ g.s.
An $\alpha$ transition of 8402 keV populates the 3/2$^-$[512] state at $E^*$\,$\approx$\,350 keV.
Finally, the 1/2$^-[$521$]$ state which is populated by a 8720~keV $\alpha$ decay from the $^{247m}$Md 1/2$^-[$521$]$ state, has been placed at $E^*$=68(11)~keV above the 3/2$^-$[521] g.s. 

A particular conclusion is drawn by the authors, considering this low energy difference of $E^*$ of 68(11)~keV observed for the  3/2$^-[$521$]$ and 1/2$^-[$521$]$ Nilsson levels in  $^{243}$Es.
These two states have their origin in the 2f$_{7/2}$ and 2f$_{5/2}$ spin-orbit partners which, according to model predictions define the proton shell gap at $Z$=114 (see, e.g., Ref.~\cite{Chasman1977}).
This low energy difference of 68~keV was found to be in contradiction to an energy gap of >1~MeV of these two levels at sphericity, as predicted by some microscopic-macroscopic models (see, e.g., Ref.~\cite{Chasman1977}). 
The authors regarded this as a hint for the next spherical proton shell location being at Z=120 rather than at Z=114.\\

{\it $^{\it244}$Es}\\ 
In 2024 Pore et al.~reported on new data for its $\alpha$ (three events) and $\beta$ decay (four events)~\cite{Pore2024} which are in line with its previous observation in Ref.s~\cite{Shaughnessy2002} and~\cite{Hessberger2001}, respectively.
\\

{\it $^{\it245}$Es}\\ 
In the irradiation campaign of $^{203,205}$Tl at JYFL mentioned above, Briselet et al. confirmed earlier measurements of the $\alpha$-decay half-life of $^{245}$Es populated by $\alpha$ decay of $^{249}$Md~\cite{Briselet2019}.\\

{\it $^{\it246}$Es}\\ 
$^{246}$Es is the last member in the $^{258}$Db decay shown in Fig.~\ref{Fig:258Db_decay}. 
In a recent study, a tentative level scheme was presented on the basis of $\gamma$ rays observed in coincidence with $^{250,250m}$Md $\alpha$ decays~\cite{Vostinar2019} (see also subsections~\ref{Md}). 
Despite the scarce statistics, four distinct transitions could be tentatively assigned, two of them originating from the g.s. and two from the isomer. 
The $^{258}$Db decay is an instructive example for the competing decays ($\alpha$, $\beta$($EC$) and $SF$) building a decay network passing across the $N$=152 neutron shell gap and in the vicinity of $^{252}$Fm with passages from odd-odd to even-even nuclei granted by $EC$. Some of these features will be discussed in more detail for nuclei which are members of this decay network in the following subsections.\\

{\it $^{\it251}$Es}\\ 
Total kinetic energy (TKE) and mass distributions have recently been reported by A.~Pal et al.~for the fission of $^{251}$Es at two excitation energies, 47.6~MeV and 52.4~MeV in a comparative investigation of three nuclei, including also $^{256}$Fm and $^{257}$Md~\cite{Pal2021} (see also subsections \ref{Fm} and \ref{Md}).\\

\subsection{Fermium - $Z$=100}\label{Fm}

As discussed in the introduction of this section, there are indications for a shell gap for SPLs at $Z$=100 at quadrupole deformations around $\beta_2$ of 0.25. 
With the deformed neutron shell gap at $N$=152, $^{252}$Fm occupies a prominent position among the 19 experimentally known fermium isotopes. 
Decay spectroscopy studies after separation of this nucleus are, however, hampered by its half-life of 25.39(4)~h. 
Hence, our knowledge of its low-lying nuclear structure is limited to the first $2^+$ state.
With respect to literature data (Ref.~\cite{Hoffman1990}), Asai et al. established an improved value for the excitation energy $E(2^+)$, employing $\alpha$ decay of $^{256}$No~\cite{Asai2016}.
They observed two $\alpha$ decay lines, a main transition at $E_\alpha$\,=\,8446~keV and a second transition at $E_\alpha$\,=\,8402~keV, with improved resolution as compared to Ref.~\cite{Hoffman1990}, from which they deduced $E(2^+)$=42.1(13)~keV. 
This is the lowest $E(2^+)$ for the fermium isotopes in the vicinity of $N$=152 as well as for the $N$=152 isotones around fermium, supporting the assumption of a pronounced double shell gap in $N$ and $Z$ for $^{252}$Fm as shown in Fig.~\ref{fig:2+_92-104}.

\begin{figure}[ht]
\begin{center}
\includegraphics[width=\columnwidth]{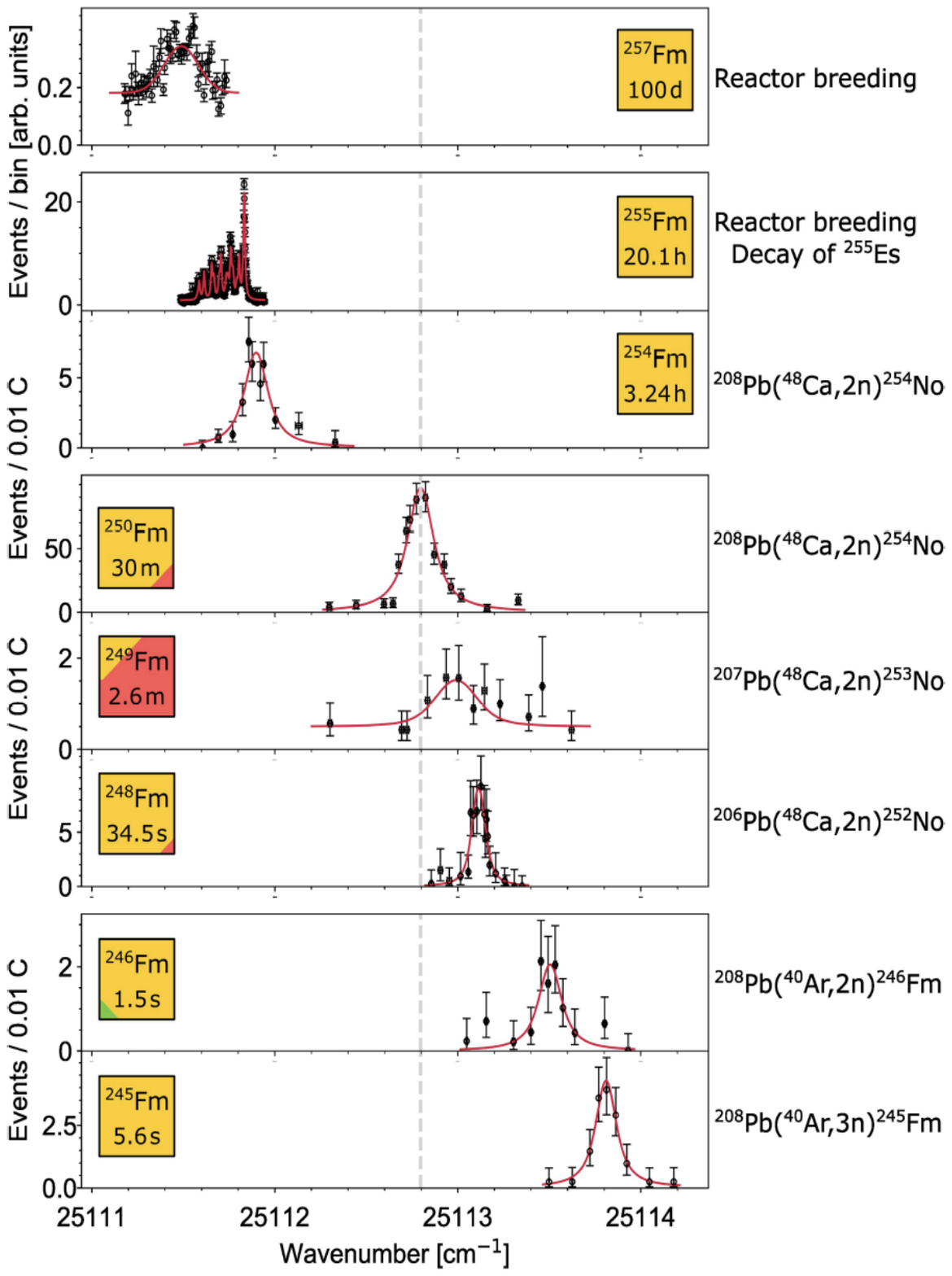}%
\vspace{-1 cm}
\end{center}
\caption{Laser resonances observed in studies of different fermium isotopes in online and offline measurements. Error bars show
statistical uncertainties (one standard deviation). 
The solid lines are fits to the data to extract the respective centroid wave number. 
The detected number of $\alpha$ events for the online investigated isotopes is normalized to the respective primary beam charge integral in units of 0.01~C. 
A more detailed zoom into the spectrum of $^{255}$Fm is shown as an inset in Fig.~1 in the main text (of Ref.~\cite{Warbinek2024}).
(Figure and caption are taken from the extended data of Ref.~\cite{Warbinek2024})
\label{fig:Fm_laser}}
\end{figure}

\begin{figure*}[ht]
\begin{center}
\includegraphics[width=0.8\textwidth]{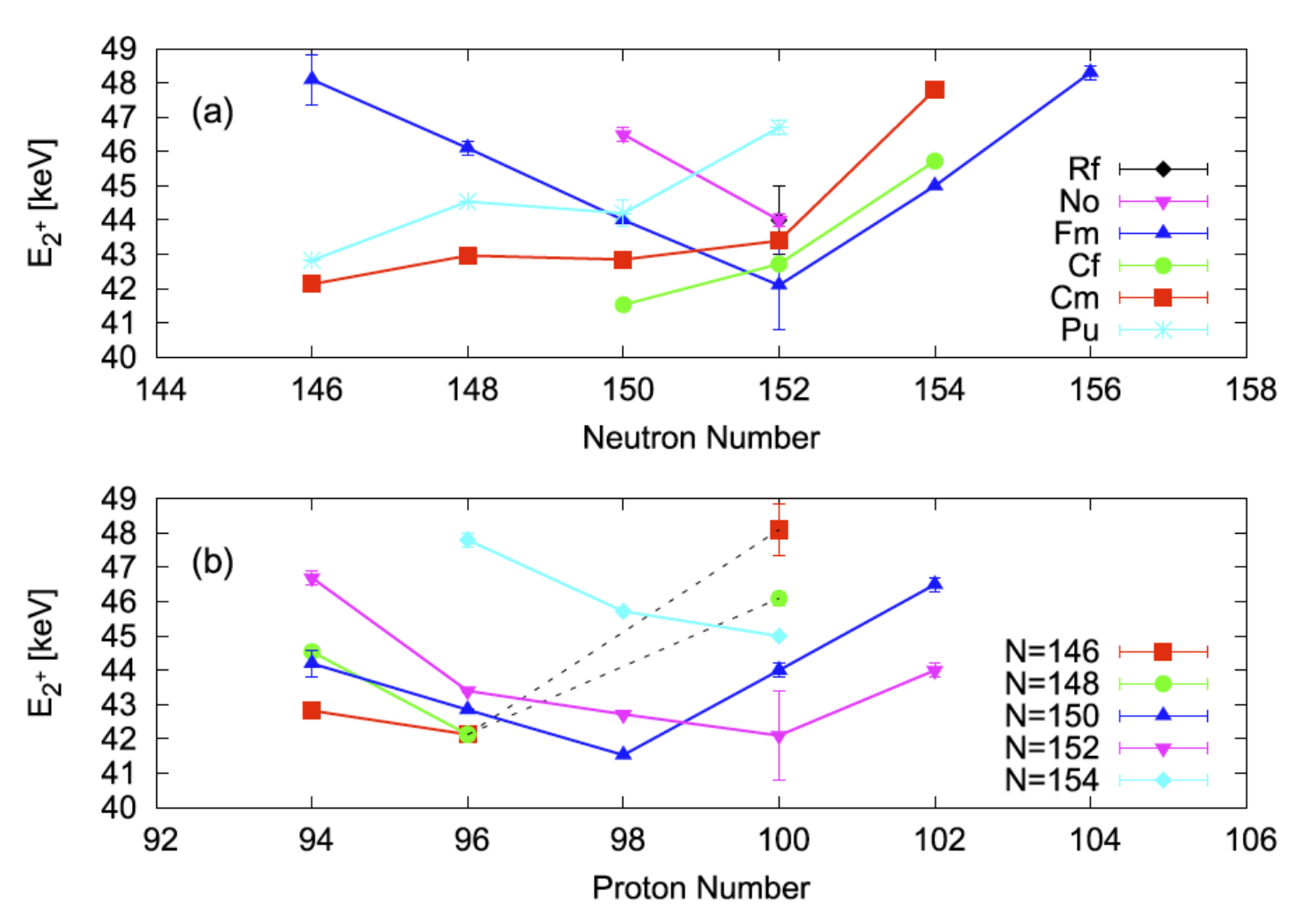}
\end{center}
\caption{Systematic behavior of the energies of the first 2$^+$ states in even-even nuclei. Upper panel: For isotopes of plutonium to rutherfordium with neutron number from 146 to 156. Lower panel: For isotopes with neutron numbers from 146 to 154 from plutonium to nobelium. The figure including the data points for $^{252}$Fm from Ref.~\cite{Asai2016} was taken from Ref.~\cite{Theisen2015}.
\label{fig:2+_92-104}}
\end{figure*}

Beyond the scope of this review, yet related via the accessible nuclear g.s. properties, are the recent advances in laser spectroscopy of the heaviest nuclei for which such studies were performed. 
The respective review and summary papers have been mentioned in the introduction.
Nevertheless, it is worthwhile to mention some of those achievements for particular cases.
As the nobelium isotopes, subject of pioneering efforts of precise mass measurements and laser spectroscopy, the fermium isotopes at the proton shell gap $Z$\,=\,100 and passing the $N$\,=\,152 neutron shell are part of those.
Therefore, the recent results on laser spectroscopy for fermium isotopes as reported by Warbinek et al.~\cite{Warbinek2024}, are shown in Fig.~\ref{fig:Fm_laser}, taken from this publication. 
It shows the laser resonance spectra, illustrating nicely the isotopic shift from which nuclear charge radii were deduced, for eight fermium isotopes between $^{245}$Fm and $^{257}$Fm.
Among the missing ones is the, for nuclear physics most interesting, $^{252}$Fm sitting on the intersection of the two subshell closures.
For further details I refer to Ref.~\cite{Warbinek2024}.\\

{\it $^{\it243,245}$Fm}\\ 
In a recent investigation of the $\alpha$-decaying isotopes $^{243}$Fm and $^{245}$Fm the literature values for decay time and energy could be improved~\cite{Khuyagbaatar2020a}, but, apart from a tentatively suggested first excited 1/2$^+$[631] state in $^{243}$Fm, which is based on systematic considerations, nothing is known regarding the excitation structure of these two nuclei.\\

{\it $^{\it244}$Fm}\\ 
For the decay of the new mendelevium isotope $^{244}$Md (see also section~\ref{Md}), Khuyagbaatar et al.~discuss in Ref.~\cite{Khuyagbaatar2020} the possible assignment of eight fission events which might partly be attributed to $^{244}$Fm fission populated by $^{244}$Md $EC$ decay (see also subsection~\ref{Md}). 
This will be further detailed in the next section on mendelevium isotopes in the context of the earlier-mentioned debate around the discovery of $^{244}$Md.\\

{\it $^{\it248}$Fm}\\
In a recent experiment at the BGS of LBNL, 22 $\alpha$ decays of $^{248}$Fm were detected, following the decay sequence of $^{256}$Db~\cite{Pore2024}. 
Two $\alpha$ decay energies were assigned with 7.830(47)~MeV and 7.950(47)~MeV, which differ from the literature values.
Nurmia et al.~had given two $^{248}$Fm $\alpha$ decay lines with $E_\alpha$\,=\,7.83(2)~MeV and 7.87(2), which were measured by employing a rotating-drum recoil detection setup, not being sensitive to electron-summing~\cite{Nurmia1967}, in contrast to the implantation detection used by Pore et al.~\cite{Pore2024} and in other studies (see~\ref{tab:isotope_list}). It is, however, unclear why the higher decay energy, differing by 120~keV from the lower one, was not reported in earlier investigations.
Half-lives could not be determined due to the population of $^{248}$Fm by the undetected $EC$ decay of $^{248}$Md (see subsections~\ref{Md} and \ref{Db}).\\

{\it $^{\it251}$Fm}\\ 
For the $N$=151 isotope $^{251}$Fm, low-lying 
level structure was proposed on the basis of $\alpha$-decay energy differences by Asai et al.~\cite{Asai2011}. 
Employing combined $CE$ and $\gamma$ spectroscopy, Rezynkina et al.~\cite{Rezynkina2018} found 
the new transitions of 192~keV from the 1/2$^+$ state to the 5/2$^-$ 23.7(11)~$\mu$s-isomer and of 200~keV from the isomer to the 9/2$^-$ ground state, confirming the decay scheme established in the earlier investigations. 
The isomer decay time was deduced from $CE$ time distribution.
In a systematic analysis of the neighboring in  $N$\,=\,151 and 150 isotones, and in comparison with theory, the authors discuss particle-phonon mixing and its important role in the $Z$\,=\,100 and $N$\,=\,152 region.

The 11/2$^-$[725] Nilsson state plays a major role for N\,=\,151 and 153 isotones.
For the $N$=151 isotones, it was observed earlier in $^{255}$Rf at $\approx$\,600~keV~\cite{Antalic2015} and very recently in $^{253}$No at 740 keV~\cite{Hauschild2022}. 
$^{251}$Fm would be the next lighter $N$\,=\,152 isotone, where the low-lying structure has been investigated up to $\approx$\,600~keV.
With $^{251}$Fm this would extend the observed trend of 11/2$^-$[725] state location to three nuclei, while for the $N$=153 isotones it has been observed for all members of the isotonic chain from curium to seaborgium~\cite{Asai2015} with the recent addition for $^{255}$No~\cite{Bronis2022} (see section~\ref{No} and Fig.~\ref{fig:153_isotones}), decreasing monotonously from $\approx$~400~keV to the ground state, replacing the 1/2$^+$[620] state in $^{259}$Sg. 
Due to the large spin difference and the decreasing excitation energy the lifetime of the 11/2$^-$[725] state increases towards higher $Z$ forming a 4.9~s isomer in $^{257}$Rf.
For more details and a comparison with theory see~\cite{Asai2015}.\\

{\it $^{\it256}$Fm}\\ 
Total Kinetic Energy (TKE) and mass distributions have recently been reported by A.~Pal et al.~for fission of $^{256}$Fm at an excitation energy of 47.6~MeV in a comparative investigation of three nuclei, including also $^{251}$Es and $^{257}$Md~\cite{Pal2021} (see also subsections \ref{Es} and \ref{Md}).\\

\begin{figure}[ht]
\begin{center}
\includegraphics[width=\columnwidth]{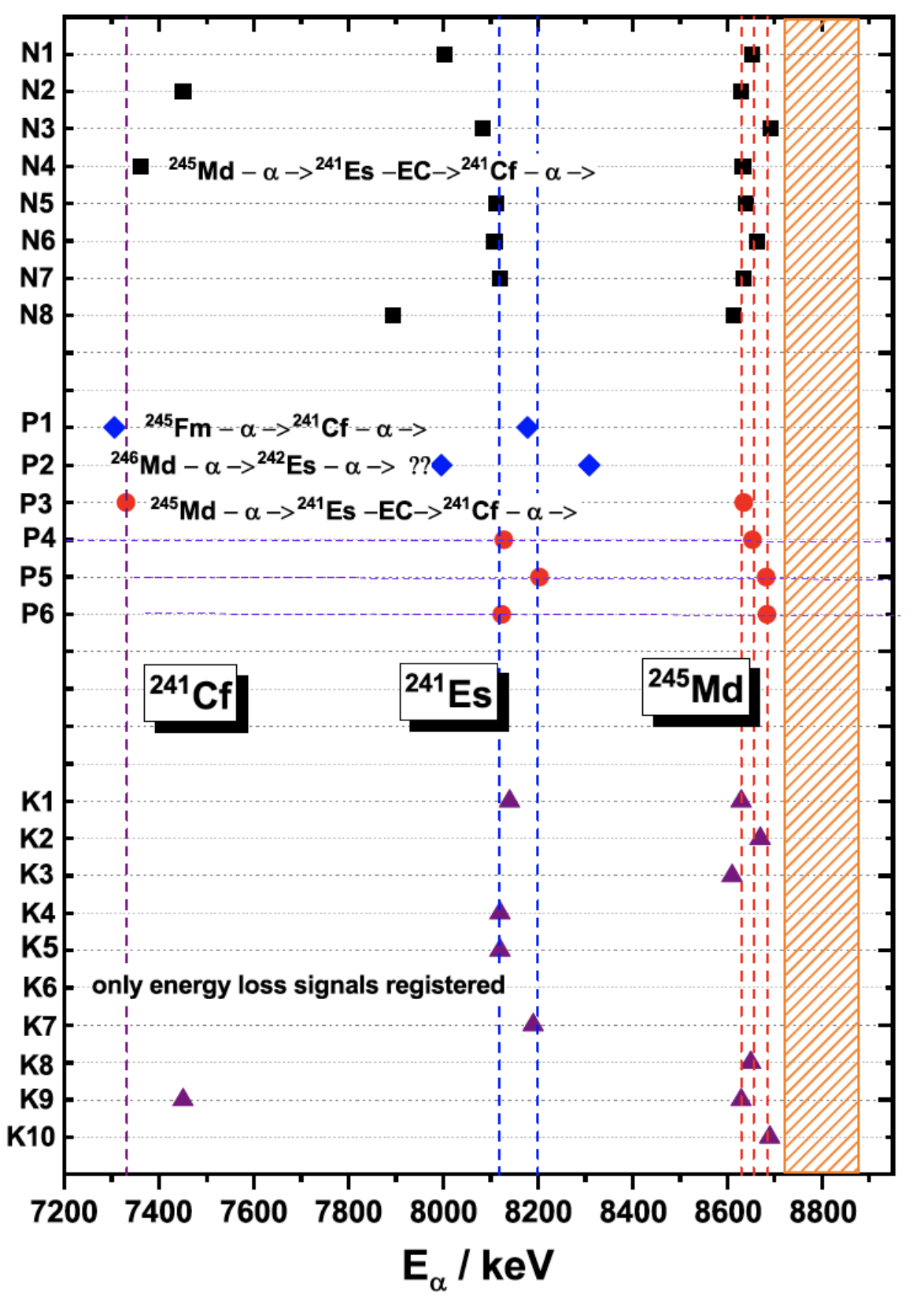}%
\end{center}
\caption{\small Summary of decays attributed to $^{245}$Md in Ref.~\cite{Ninov1996} (squares) as well as in Ref.~\cite{Khuyagbaatar2020} (triangles) together with data reported by Pore et al.~\cite{Pore2020}
$[$circles: events attributed to $^{245}$Md by the present authors, diamonds: events attributed to $^{245}$Fm or (tentatively) to $^{246}$Md$]$. 
The dashed lines are to guide the eyes: the red lines represent the $\alpha$ energies given for $^{245}$Md (8640, 8680 keV) in Ref.~\cite{Ninov1996} and the energy given for $^{244}$Md (8663 keV) in Ref.~\cite{Pore2020}; 
the blue lines represent the $\alpha$ energy for $^{241}$Es (8113 keV) given in Ref.~\cite{Ninov1996} and the highest daughter energy (P5) in Ref.~\cite{Pore2020}; 
the purple line represents the literature value of the $\alpha$ energy of $^{241}$Cf (7335 keV)~\cite{Firestone1996}. 
The orange hatched area marks the range of $\alpha$ energies where the events attributed to $^{244}$Md in Ref.~\cite{Khuyagbaatar2020} were observed. 
(Figure and caption are taken from Ref.~\cite{Hessberger2021})
\label{fig.245Md_Ea}}
\end{figure}
\subsection {Mendelevium - $Z$=101}\label{Md}

Up to now, 17 mendelevium isotopes, from $A$=244 to 260, have been observed and studied experimentally.
Recently, the lightest ones are in the focus of the community due to a debated mass assignment for $^{244,245}$Md. 
A particular feature is the observation of high-$K$ isomers for the odd-even $^{249}$Md and $^{251}$Md, illustrating the role of high-spin orbitals in the vicinity of the Z\,=\,100 and N\,=\,152 shell gaps.\\

{\it $^{\it244,245}$Md}\\ 
Among the recent mendelevium isotopes studied are $^{245}$Md, first observed in 1996 at the velocity filter SHIP of GSI~\cite{Ninov1996}, and $^{244}$Md reported for the first time in 2020~\cite{Pore2020,Khuyagbaatar2020}. 
Two independent studies yielded almost simultaneously results which, due to conflicting assignments, initiated the above-mentioned debate.
One of these experiments, employing the same reaction, $^{40}$Ar+$^{209}$Bi, as the first experiment at GSI, was performed at the Berkeley Gas-filled Separator (BGS)~\cite{Gregorich2013} of the Lawrence Berkeley National Laboratory (LBNL) in Berkley, California, U.S.A., using two complementary detection set-ups~\cite{Pore2020}. 
The decay properties were measured at the BGS focal plane by means of a retractable 32$\times$32 double-sided silicon strip detector (DSSD), while a mass assignment was obtained in a separate measurement using the FIONA set-up, a mass spectrometer consisting of a gas-stopping cell, re-acceleration, a trochoidal spectrometer and particle detection~\cite{Gates2018}. 
Pore et al.~\cite{Pore2020} reported six decay chains which were all assigned to the $\alpha$-decay chain $^{244}$Md$ \rightarrow ^{240}$Es$ \rightarrow ^{236}$Bk. 
The proposed scenario included two $\alpha$ transitions for $^{244}$Md, with $\alpha$-decay energies and half-lives of $E_\alpha$=8663(23)~keV and $T_{1/2}$=0.4$^{+0.4}_{-0.1}$~s, and $E_\alpha$=8306(223)~keV and $T_{1/2}\approx$6~s, respectively. 
In the original paper Pore et al.~suggested the longer-lived state to be the ground state, while in a later paper by the same authors the assignment was changed to the slower transition originating from the isomeric state~\cite{Gates2022}. 
One of the observed decay sequences was assigned to the decay of $^{240}$Es with the $^{244}$Md $\alpha$ decay missing. 
These assignments were motivated by the decay properties of $^{240}$Es from literature~\cite{Konki2017} and the mass determination using FIONA.

The second measurement was performed at the gas-filled separator TASCA of GSI, Darmstadt, Germany~\cite{Khuyagbaatar2020}. 
Khuyagbaatar et al.~employed the reaction $^{50}$Ti+$^{197}$Au at two beam energies (231.5~MeV and 239.8~MeV at center of target), leading to excitation energies $E^*$ of 26.2~MeV and 32.7~MeV, expected to populate the 2- and 3-neutron fusion-evaporation channels $^{245,244}$Md, respectively. 
They observed three decay chains at the lower beam energy and attributed them to the 2$n$ fusion-evaporation channel $^{245}$Md. 
The observed decay properties were in agreement with literature data as reported in Ref.~\cite{Ninov1996}.
At the higher beam energy, a total of ten $\alpha$-decay sequences were extracted. 
Three of them were again in agreement with an assignment to $^{245}$, while the seven others showed similar decay times as the ones assigned to $^{245}$Md, but higher $\alpha$-decay energies of 8.73~MeV to 8.86~MeV.
With a half-life of 0.30$^{+0.1}_{-0.09}$~s, they were attributed to the new isotope $^{244}$Md. 
In particular, the decay properties of the subsequent decays reproduced the results assigned by Konki et al.~to $^{240}$Es~\cite{Konki2017}, while half-lives and $Q_\alpha$ values are almost identical to $^{241}$Es (see Table~\ref{tab:isotope_list}).

These two results are, in their respective interpretation, in contradiction. While the $A$ determination by FIONA suggests that an object with 244 amu was observed at the BGS, the reported decay energy values would suggest the observation of  $^{245}$Md decay events. 
In a comment to these findings, He{\ss}berger et al.~discussed the decay properties of both experiments in the context of the literature values~\cite{Hessberger2021}. 
Fig.~\ref{fig.245Md_Ea} which is taken from~\cite{Hessberger2021}, shows a comparison of the $\alpha$-decay data for the original and the two new measurements and proposed re-assignments of the BGS decay chains to $^{245}$Md and other possible reaction channels (see figure caption). 
As an additional argument, He{\ss}berger et al.~analyzed the fusion-evaporation excitation function in comparison to calculations performed with the statistical model code HIVAP~\cite{Reisdorf1981}, suggesting that the beam energy used by Pore et al.~would favor a 4$n$ fusion-evaporation channel corresponding to $^{245}$Md. The beam energy for the reaction $^{40}$Ar+$^{209}$Bi chosen for the LBNL experiment was given as approximately 220 MeV in Ref.~\cite{Pore2020}, corresponding to $E^*$\,$\approx$\,45~MeV.

To clarify this situation, more statistics, but in particular, instead of the two-step procedure as applied in the case of FIONA, a simultaneous measurement of $A$, $E_\alpha$ and decay times would be extremely useful, using a suited set-up, like, e.g., the Fragment Mass Analyzer FMA at ANL, or in future the separator-spectrometer set-up S$^3$~\cite{Dechery2016}, presently under construction at GANIL/SPIRAL2.

In addition to $\alpha$ decay, fission is a common and expected decay mode in that region of the chart of nuclides. 
Pore et al.~\cite{Pore2020} did not report any fission events, partly due to instrumental insensitivity, while Khuyagbaatar et al.~\cite{Khuyagbaatar2020} observed at total of six and eleven fission events at the two applied beam energies, respectively, part of which could be attributed to the fission activity of $^{245}$Md and to fission of its isomeric state. 
For eight fission events, however, the situation remains somewhat unclear. 

Three possible scenarios are discussed none of which could be ruled out or definitely confirmed. 
The possibility of direct fission decay of $^{244}$Fm populated in a $p2n$ fusion-evaporation reaction (emission of one proton and two neutrons from the excited compound nucleus (CN)) and fission from an excited state in $^{244}$Md are in contradiction to all earlier observations. 
For the first case, the decay times of the observed fission decays (5$^{+3}_{-2}$~ms) and the literature values of $^{244}$Fm (3.21(8)~ms) would match. 
However, $p$-evaporation channels are known to be strongly suppressed in reactions of this kind, as recently shown for example by Lopez-Martens et al., irradiating a $^{209}$Bi target with $^{50}$Ti projectiles for the production $^{256,258}$Rf as $p$ and $p2n$ evaporation channels at the velocity filter SHELS of FLNR JINR, Dubna~\cite{Lopez-Martens2019}.
For the same reaction, one $p$-evaporation event was reported earlier at the velocity filter SHIP of GSI by He{\ss}berger et al.~\cite{Hessberger1985}, who re-analyzed later experiments with the conclusion of 5 events to be attributed to the $p$ evaporation channel for an experiment conducted in 2014~\cite{Hessberger2019} (see also section~\ref{Rf} and \ref{Db}).
For odd nuclei, fission is known to be strongly hindered, and the observation of fission from an excited state in the odd-odd nucleus $^{244}$Md, in particular, in comparison with the fission properties of neighboring nuclei, is rather unlikely. 

As a third scenario, the authors discussed $EC$-delayed fission (ECDF) from a short-lived isomeric state for which often de-excitation by CE is observed. 
As a hint for the existence of such an isomeric state, a low-energy signal was observed in the tail of one evaporation residue (ER) signal trace, registered by the digital electronics system used in this experiment, which was correlated to a $^{244}$Md $\alpha$ decay.
In addition, two fission-like signals of similar decay time were observed in the tail of ER-like events which could originate from such a short-lived state. 
The evidence is, according to Khuyagbaatar et al., still too weak to firmly assign this observation to $^{244}$Md-$^{244}$Fm ECDF. They conclude that all three scenarios could contribute to the observed fission events of unclear origin, which needs certainly further confirmation, given the rather speculative assumptions and counter-indications.\\

{\it $^{\it247}$Md}\\
In 1981 $^{247}$Md was first observed at the velocity filter SHIP of GSI, Darmstadt~\cite{Muenzenberg1981}, Germany. 
In a number of subsequent investigations employing $\alpha$- and $\alpha$-$\gamma$ decay spectroscopy, more details of the low-lying nuclear structure of this odd-$Z$/even-$N$ nucleus, including an isomeric state, were revealed~\cite{Hofmann1994,Hessberger2005,Antalic2010}. 
In a recent decay study at the same instrument~\cite{Hessberger2022}, the excitation energy of the 231(30)~ms isomer could be established with 153~keV. 
For the level structure deduced for the daughter nucleus $^{243}$Es from this decay study see subsection~\ref{Es}.

In addition to population of low-lying levels, He{\ss}berger et al.~~\cite{Hessberger2022} could establish the effect of the quantum structure of the g.s. and isomer, $^{247g.s.}$Md and $^{247m}$Md, on the fission probability of those states. 
The relatively high spin of the Nilsson 7/2$^-$[517] g.s.~configuration results in the low $SF$ branching ratio $b_{SF}$=8.6$\times10^{-3}\pm0.0010$ as compared to the isomeric 1/2$^-$[521] state with $b_{SF}$=0.20(2). 
The corresponding partial spontaneous fission half-lives were established as $\approx$140~s and $\approx$1.2~s for $^{247g.s.}$Md and $^{247m}$Md, respectively.\\

{\it $^{\it248}$Md}\\
New data which are generally in agreement with previous observations on the $\alpha$ and $\beta$ decay of $^{248}$Md as a member of the decay chain of $^{256}$Db were recently accumulated at the BGS of LBNL~\cite{Pore2024}.
Some ambiguities in the assignment of $\alpha$ decays to either $^{248}$Md or its $\beta$ decay daughter $^{248}$Fm from $^{252}$No, the $^{252}$Lr $\beta$ decay daughter, lead to different branching estimates.
The authors of this investigation propose on the basis of two different assumptions for the $^{252}$Lr $EC$ branch with 10\% or 30\%(see subsection~\ref{Lr}, an $\alpha$ branching for $^{248}$Md of 61(16)\% or 68(22)\%, respectively, which in any case overlap within the (large) error bars, and are both slightly higher then the previously adopted value of 58\% (see Table~\ref{tab:isotope_list}).\\

{\it $^{\it249,251}$Md}\\
For $^{249}$Md the possible existence of an isomeric state was first proposed by He{\ss}berger et al., based on the $\alpha$-decay population scheme from two different states in $^{253}$Lr~\cite{Hessberger2001}. 

A recent series of experiments was conducted at the gas-filled separator RITU of the cyclotron accelerator laboratory of the University of Jyv\"{a}skyl\"{a}. 
In this context, Briselet and co-workers reported on in-beam spectroscopy results for $^{249,251}$Md collected with the RITU target area detection system for electron and $\gamma$ spectroscopy SAGE~\cite{Briselet2019,Briselet2020}.
As mentioned in subsection \ref{Es}, these mendelevium isotopes were produced employing targets of $^{203,205}$Tl in the reactions $^{203,205}$Tl($^{48}$Ca,$2n$)$^{249,251}$Md.

In continuation of this study, Goigoux et al.~reported hitherto unobserved $K$ isomers in $^{249,251}$Md~\cite{Goigoux2021}, revealed in the same experiment series, using the combined silicon-germanium detector array GREAT in the RITU focal plane.
They investigated the decay of these two isotopes by means of recoil-$\alpha$-CE-$\gamma$ coincidences. 
For $^{249}$Md, a fast activity of 2.4(3)~ms with a lower excitation energy limit of $E^*\geq$ 910~keV could be established from recoil-$\alpha$-e$^-$ correlations. 
For $^{251}$Md, a state of 1.37(6)~s half-life and a lower excitation energy limit of $E^*\geq$ 844~keV was found applying the same method, however, with the need for the condition of a subsequent $\alpha$ detection in addition to the recoil-e$^-$ correlation, due to the slower decay time overlapping partly with random correlations.
\begin{figure}[ht]
\begin{center}
\includegraphics[width=\columnwidth]{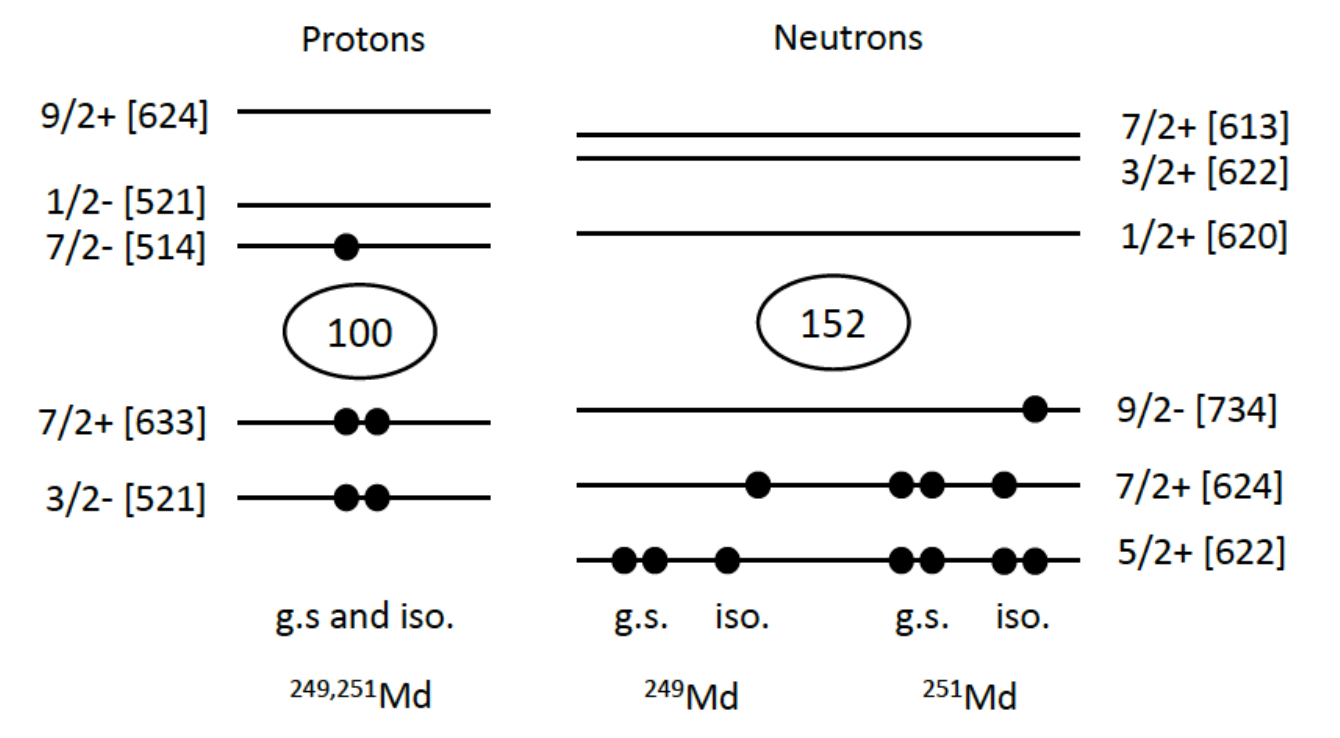}%
\end{center}
\caption{Schematic single-particle configurations for the g.s.~and isomeric state in $^{249,251}$Md.
(Figure and caption are taken from~\cite{Goigoux2021})
\label{fig.249251Mdlevels}}
\end{figure}

For the assignment of the new metastable states to $K$-isomers formed by 3-quasiparticle states, the authors use a microscopic-macroscopic model as described by Muntian et al.~\cite{Muntian2001}, based for the macroscopic component on a Yukawa-plus-exponential folding function~\cite{Krappe1979}, and on a deformed Woods-Saxon single particle potential for the shell correction energies~\cite{Cwiok1987}. 
For both isotopes, the resulting proton configuration is the same, with the unpaired proton in the highest occupied SPL 7/2$^-$[514]. 
With 148 and 150 neutrons, respectively, $^{249}$Md and $^{251}$Md are 2 and 4 neutrons short of the $N$=152 shell gap. Thanks to this, they have access to the three SPLs below the shell gap with relatively high spin values (5/2, 7/2 and 9/2; see Fig.~\ref{fig.249251Mdlevels}) which leads to the high $K$-values generated by the quasiparticle excitations.
A schematic view of the single-particle configurations is given in Fig.~\ref{fig.249251Mdlevels}.

Two approaches within the microscopic-macroscopic model framework, one using a level blocking scheme and a second one applying a quasiparticle method (for details see~\cite{Goigoux2021} and references therein), provide consistently arguments to extract energy and spin/parity arguments for the assignment of the SPLs involved in the 3-quasiparticle configuration.  
For both nuclei, the combination of the unpaired proton in its original g.s. level, breaking the pair in the last occupied neutron level and elevating one of the neutrons into the next higher level turns out to be the most likely configuration for the respective $K$-isomer. 
This leads to $\pi\nu^2$=19/2$^-$ with 7/2$^-$[514]$_\pi$ $\otimes$ 5/2$^+$[622]$-\nu$ $\otimes$ 7/2$^+$[624]$_\nu$ and $\pi\nu^2$=23/2$^+$ with 7/2$^-$[514]$_\pi$ $\otimes$ 7/2$^+$[624]$_\nu$ $\otimes$ 9/2$^-$[734]$_\nu$ for the new 3-quasiparticle $K$ isomers in $^{249}$Md and $^{251}$Md, respectively.\\

{\it $^{\it250}$Md}\\
As $^{246}$Es  and $^{254}$Lr (see subsections~\ref{Es} and ~\ref{Lr}), also the $^{258}$Db decays chain member $^{250}$Md was subject of the investigations reported in Ref.~\cite{Vostinar2019}. Analyzing the decay times of partitions of the relatively broad distribution of $\alpha$-decay energies, the authors could establish, apart from the g.s. decay with a half-life of 25$^{+10}_{-5}$~s, a second activity which was attributed to an isomeric state with $T_{1/2}$=42.2(45)~s at an excitation energy of $E^*$=123~keV. 
The ratio of the relative population of these two states by $\alpha$ decay is found to be different for correlations to g.s. and isomer decays from the earlier members in the decay chain, $^{254}$Lr and $^{258}$Db (see subsections~\ref{Lr} and~\ref{Db}).
This is probably related to the quantum numbers of the involved states, with $\alpha$ decay preferably connecting analog initial and final states. 
The accumulated statistics, however, are still too low to draw firm conclusions on spins and parities.\\

{\it $^{\it257}$Md}\\ 
TKE and mass distributions were reported by A.~Pal et al.~for fission of $^{257}$Md at three excitation energies, 37.5~MeV, 42.9 MeV and 47.6~MeV in a comparative investigation of three nuclei, including also $^{251}$Es and $^{256}$Fm~\cite{Pal2021} (see also subsections \ref{Es} and \ref{Fm}). In contrast to the other two nuclei which show symmetric TKE distributions, $^{257}$Md shows a skewness extending to higher TKE, which is interpreted by the authors as an indication of a new so-called "supershort" fission mode, originating from a compact, non-stretched configuration of the fissioning system at the scission point.

\begin{figure*}[ht]
\begin{center}
\includegraphics[width=0.4 \textwidth]{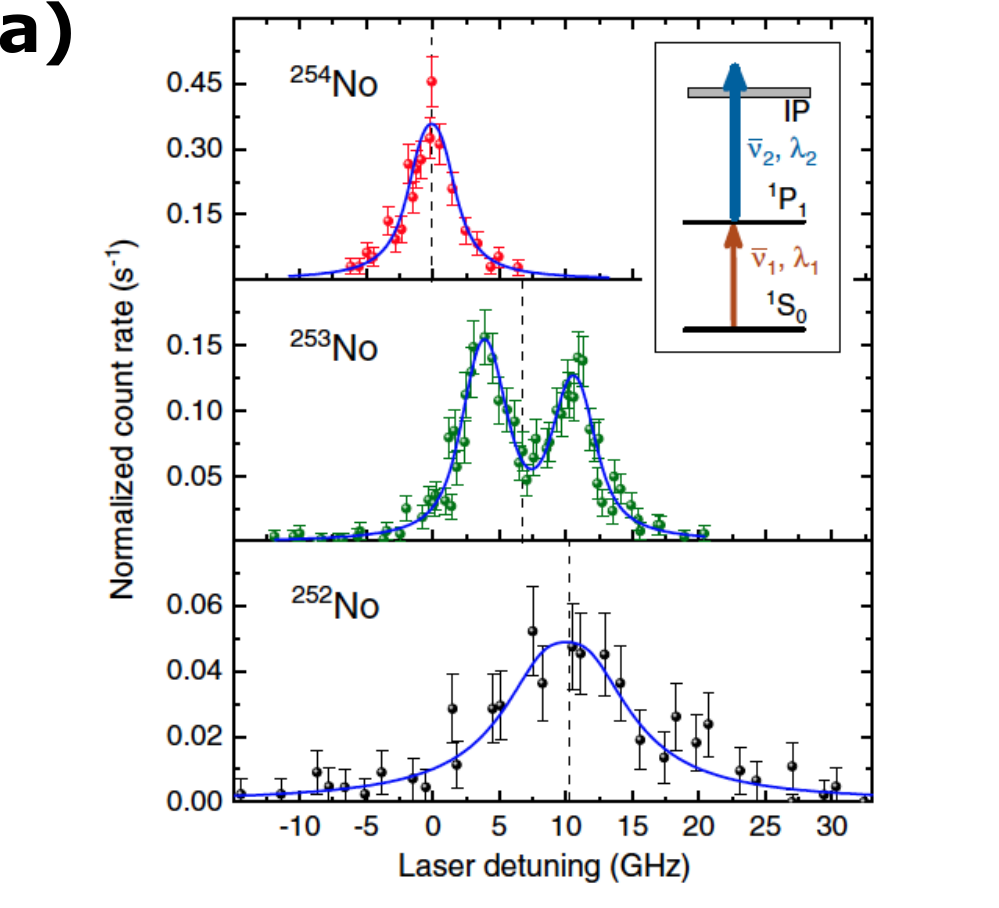}%
\includegraphics[width=0.57 \textwidth]{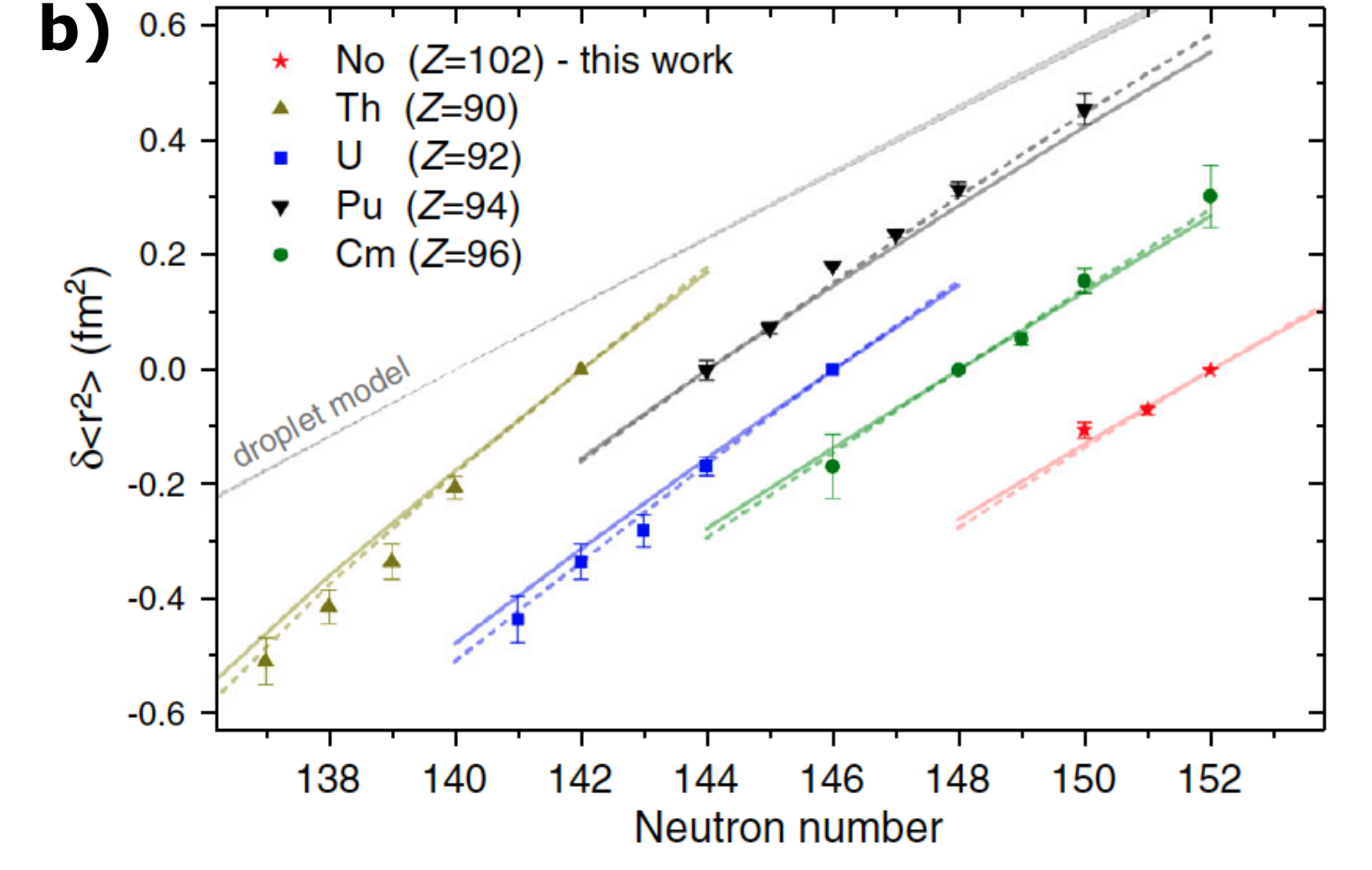}%
\end{center}
\caption{a) Measured (atomic) excitation spectra of the $^1P_1$ level for the isotopes $^{254}$No, $^{253}$No, and $^{252}$No with a best fit to the data (solid line). 
The dashed line represents the center of each resonance, while the solid vertical lines in the $^{253}$No spectrum indicate the position and strength of the individual hyperfine structure components with total angular momentum F=7/2, 9/2, and 11/2 at 3.99~GHz, 4.10~GHz, and 10.74~GHz, respectively. 
The inset shows a schematic ionization scheme. 
b) The change in the nuclear mean square charge radii $\delta \langle r^2 \rangle$, for $^{252-254}$No and even Z actinide nuclei starting from
thorium, is plotted as a function of the neutron number with
arbitrary offset. For each element, the DFT calculations with two
Skyrme energy density functionals, UNEDF1~\cite{Kortelainen2012} (dashed line)
and SV-min~\cite{Kluepfel2009}(solid line), are shown. The slope according to a
schematic droplet model assuming a constant deformation for the
actinide elements referenced to $N$=138 is marked in gray.
(Figures and captions are taken from~\cite{Raeder2018})
\label{fig.No_laser}}
\end{figure*}

\subsection{Nobelium - $Z$=102}\label{No}

Being just a proton pair above fermium, the nobelium isotopes and, in particular, $^{254}$No due to its relatively large production cross-section, are at the center of interest regarding decay spectroscopy as well as in-beam studies for more than two decades.  
In our review from 2017~\cite{Ackermann2017} nobelium isotopes played a central role, in particular, regarding in-beam studies. 
We dedicated a section to what we called {\it "The $^{\it 254}$No breakthrough"} which was a boost for the nuclear structure studies in the fermium/nobelium region of the Segr\`{e} chart and beyond.
In-beam spectroscopy resulted in the observation of rotational bands up to $^{256}$Rf, while DSAS efforts succeeded in revealing features like the $K$ isomers in heavy nuclei up to $^{270}$Ds (see subsections~\ref{K-iso} and~\ref{Ds}).
The relatively high production cross-section of $^{254}$No in the fusion-evaporation reaction $^{208} Pb$($^{48}$Ca, $2n$)$^{254}$No of $\approx$ 2~$\mu$b made detailed in-beam studies of this nucleus possible. 
A natural link between in-beam spectroscopy and DSAS is provided by isomeric states which survive the flight path (typically a couple of $\mu$s) through the separator, connecting the single particle features of a nucleus to its collective behavior. 
Among the 13 known nobelium isotopes from $^{249}$No to $^{262}$No, the $K$-isomers of $^{252}$No and $^{254}$No are textbook cases for this link, as demonstrated in a number of exemplary studies~\cite{Sulignano2007,Sulignano2012,Tandel2006,Clark2010,Hessberger2010}. 

The pertinent role of nobelium isotopes for the nuclear structure of heavy nuclei also triggered the investigation of their fundamental properties.
In a seminal work by Block et al.~\cite{Block2010} the atomic masses of $^{252,253,254}$No had been precisely measured employing the Penning trap installation SHIPTRAP of GSI, where later also the mass of $^{251}$No was measured by Kaleja et al.~\cite{Kaleja2022}. 
As mentioned in the introduction of this review, an overview of mass measurements for the heaviest nuclei can be found in references~\cite{Block2019,Block2019a}. 
Raeder et al.~\cite{Raeder2018} measured the isotopic shift of the laser-excited atomic $^1P_1$ level in these isotopes from which they could establish nuclear charge radii as in the case of fermium isotopes mentioned above (see subsection~\ref{Fm}). 
In Fig.~\ref{fig.No_laser}a, where also the laser excitation spectra are shown for $^{252,253,254}$No, measured charge radii for isotopes of the even-$Z$ elements from curium to nobelium are compared with the predictions of density functional models~\cite{Kluepfel2009,Kortelainen2012}.
 \\

\begin{figure}[ht]
\begin{center}
\includegraphics[width=\columnwidth]{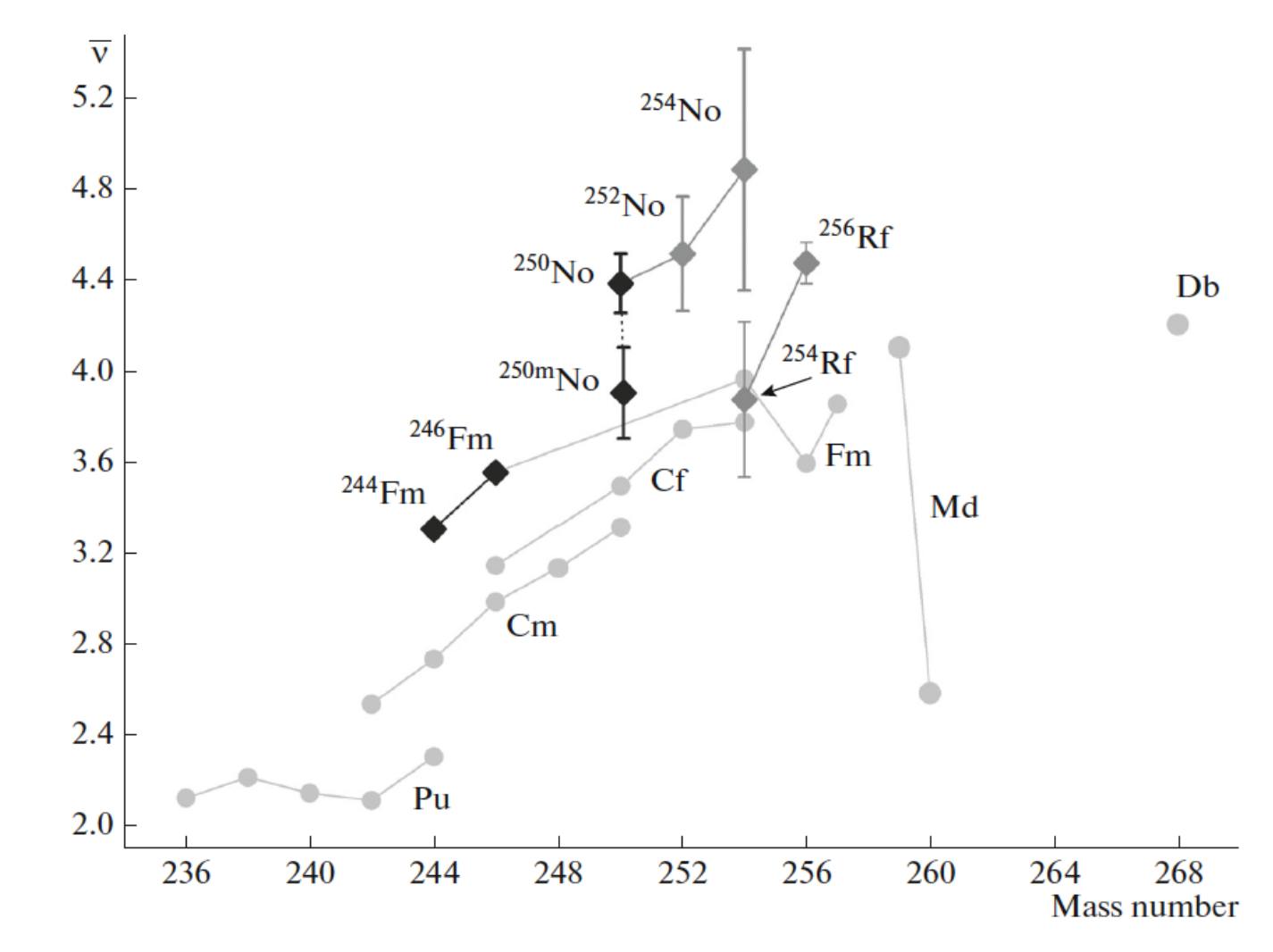}%
\end{center}
\caption{Compilation of the data on the average number of neutrons in spontaneous fissions of different heavy nuclei, including those
for the $^{254}$No nucleus reported in~\cite{Isaev2021}. The measurements carried out with the VASSILISSA and SHELS separators are
denoted by diamonds.
(Figure and caption are taken from~\cite{Isaev2021})
\label{fig.sf_n_mult}}
\end{figure}

{\it $^{\it249}$No}\\ 
In 2021, almost contemporaneously two groups reported the discovery of the lightest of the thirteen known nobelium isotopes, $^{249}$No. 
Khuyagbaatar et al.~\cite{Khuyagbaatar2021} observed three decay chains of three correlated $\alpha$-decays, starting from $^{253}$Rf for which hitherto only a fission branch was known. 
The third $\alpha$-decay energy in this sequence was found to be in good agreement with literature data for $^{245}$Fm (see subsection~\ref{Fm} and references in Table~\ref{tab:isotope_list}) and confirmed this assignment, resulting in the discovery of a new $\alpha$-decay branch for $^{253}$Rf and the new isotope $^{249}$No. 
The latter was directly produced in the reaction $^{48}$Ca+$^{204}$Pb by Svirikhin et al.~\cite{Svirikhin2021} who accumulated a total of about 220 $^{249}$No $\alpha$ decays. 
The decay energies and times for this new isotope reported by the two groups are in good agreement.\\

\begin{figure*}[t]
\begin{center}
\includegraphics[width=0.9\textwidth]{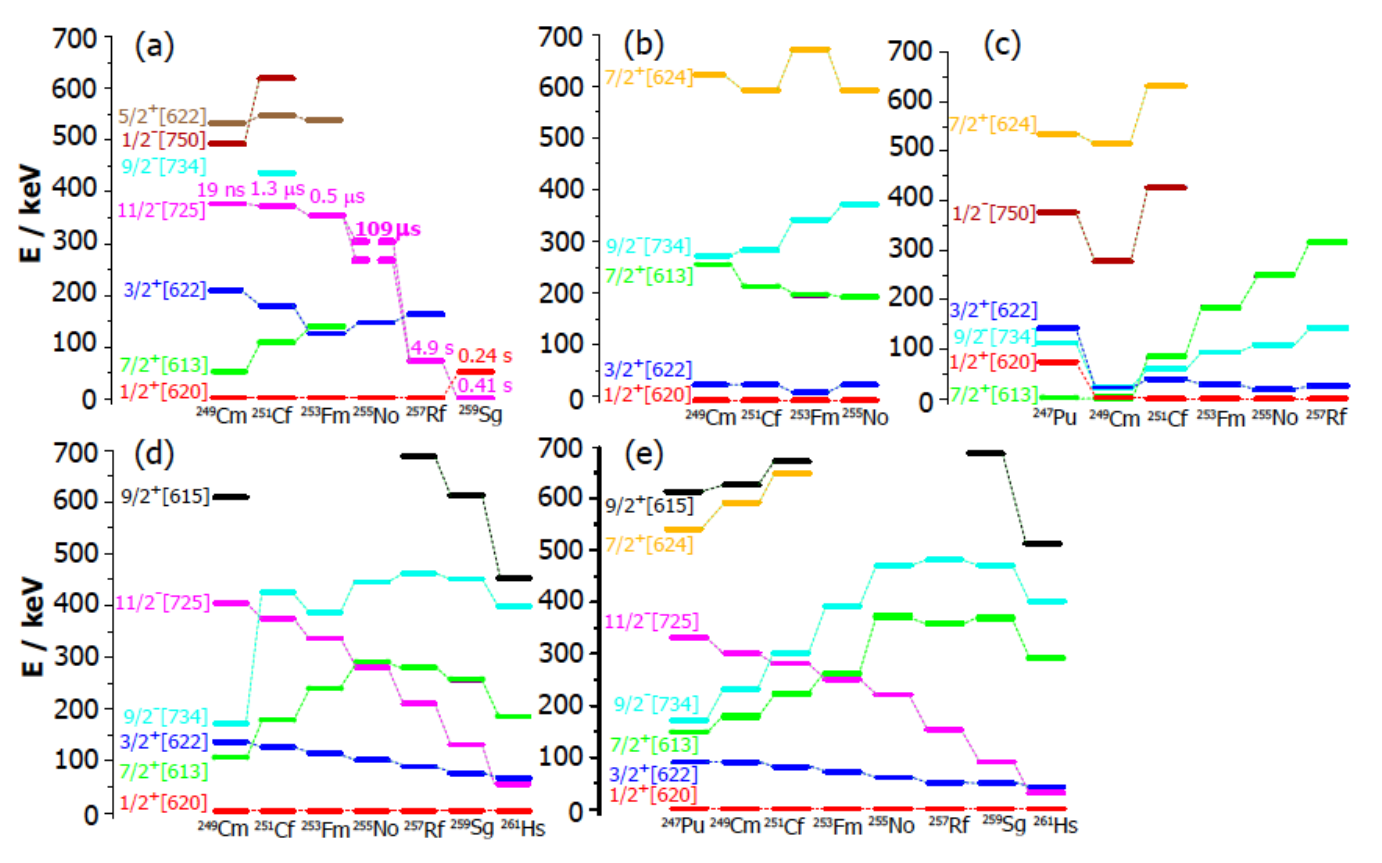}%
\end{center}
\caption{(a) Experimental single particle level systematics of $N$=153 isotones. Theoretical calculations of single particle levels
based on self-consistent models (b)~\cite{Zhang2012} and (c)~\cite{Bender2003b} or microscopic-macroscopic models (d)~\cite{Cwiok1994} and (e)~\cite{Parkhomenko2005} are given for comparison.
Dashed horizontal lines denote the range of the excitation energy for the 11/2$^-$[725] level (260 -- 320 keV) in $^{255}$No assigned in~\cite{Bronis2022}.
(Figure and caption are taken from~\cite{Bronis2022})
\label{fig:153_isotones}}
\end{figure*}

{\it $^{\it250}$No}\\ 
For $^{250}$No, Kallunkathariyil et al.~reported recently on half-life measurements for its ground and $K$-isomeric state performed at the gas-filled separator RITU of the University of Jyv\"{a}skyl\"{a} cyclotron laboratory~\cite{Kallunkathariyil2020}.  
On the basis of $ER$-$CE$-$SF$ coincidence measurements, they found a half-life for the ground state of $T_{1/2}$\,=\,3.8(3)~$\mu$s and for the $K$ isomer of $T_{1/2}$\,=\,34.9$^{+3.9}_{-3.2}$~$\mu$s. 
The rare occurrence of isomeric states being longer lived than the ground state, here for the n-deficient $^{250}$No, is also observed for the superheavy isotopes $^{266}$Hs and $^{270}$Ds (see subsection~\ref{K-iso} and Table~\ref{tab:isotope_list}).

Apart from electron and time spectra, Kallunkathariyil et al.~also show a spectrum of photons in coincidence with $CE$s, which they associate with the de-excitation of the isomer. 
It contains, apart from nobelium $x$-rays, also some strength between 150~keV and 200~keV, just above the nobelium $K_\alpha$ and $K_\beta$ $x$-ray lines. 
In addition, some strength is being observed in the region from 250~keV to 300~keV and around 400 keV.
In an experiment performed recently at the SHELS separator of FLNR JINR in Dubna~\cite{Kuznetsova2020}, Kuznetsova et al.~also reported on $\gamma$ rays observed for the $^{250}$No $K$ isomer, this time in coincidence with the detected $SF$ only. 

With a total of 18,000 SF events, Kuznetsova et al.~observe various $\gamma$ lines in coincidence with $SF$, which could correspond to a consistent scenario of the isomer decay into the g.s.~rotational band.
However, the accumulated statistics with $\approx$\,5 to 10 counts peak height for the observed transitions is too low to confirm possible assignments by more sensitive tests, like e.g.,~$\gamma$-$\gamma$ coincidences. 
A comparison of the results from these two independent studies is difficult as the coincidence criteria are different. 
Ideally, an experiment aiming at higher count rates and combining $CE$-coincidence and $\gamma$ decay spectroscopy could help to reveal the quantum configuration and decay of this neutron-deficient nobelium isotope. 
Such an investigation could possibly help to reveal the mechanism responsible for the enhanced stability of the isomeric state with respect to the ground state, which has been observed in a number of cases.\\

{\it $^{\it251}$No}\\ 
A new isomeric state in $^{251}$No was observed by Lopez-Martens et al.~in the irradiation of an isotopically enriched $^{204}$Pb target by $^{50}$Ti projectiles~\cite{Lopez-Martens2022}. 
The spin and parity assignment of 1/2$^+$ was supported by systematics involving the neighboring  isotopes and, in particular, by the observation of isomeric states in $^{253}$Rf reported in the same paper and in Ref.~\cite{Khuyagbaatar2021} (see subsection~\ref{Rf}).\\ 

{\it $^{\it252,254}$No}\\ 
Despite the efforts and findings regarding $^{254}$No reported in the introduction of this subsection, there are still open questions which are lively debated, like the quasi-particle structure of the 8$^-$ $K$-isomers in $^{254}$No.
A very recent measurement at ANL (reported after the first submission of this review; see below) contributed new important findings in this respect.
Kuznetsova et al.~\cite{Kuznetsova2020} and in a second paper by the same group, Isaev et al.~\cite{Isaev2021} reported on the investigation of the fission properties of $^{252,254}$No at SHELS.
While in the earlier brief report mainly decay times and TKE measurements are mentioned, the more recent work presents more detailed results including for the first time for these isotopes neutron multiplicity distributions.
A summary of experimentally obtained neutron multiplicities for isotopes from plutonium to rutherfordium is shown in Fig.~\ref{fig.sf_n_mult}.

For $^{254}$No Forge et al.~accumulated additional decay data~\cite{Forge2023} in a measurement using the DSAS set-up GABRIELA~\cite{Hauschild2006,Chakma2020} at the focal plane of the velocity filter SHELS~\cite{Popeko2016} of FLNR JINR, Dubna.
They report on the observation of a strong transition from the 8$^-$ isomer, $^{254m1}$No, which was revealed by employing $CE$ correlations. 
Suggesting $E0$ as the most likely transition type, they discuss a possible shape co-existence and presence of a low-energy super-deformed state.

As the most recent addition to the efforts in solving the puzzle of the nature of the second short-lived isomer in $^{254}$No Wahid et al.~reported on new data collected at the gas-filled separator AGFA of the ATLAS facility at ANL~\cite{Wahid2025}.
Based on higher collected statistics with the highest $\gamma$-$\gamma$ coincidence accumulated to date, the authors draw several conclusions. 
From the time difference between the $CE$s feeding the structure below the short-lived 16$^+$ isomer and the 606~keV $\gamma$ transition from its 10$^+$ bandhead to the longer-lived 8$^-$ isomer they establish a half-live of $T_{1/2}$ of 5(2)~ns for this new $K^\pi$\,=\,10$^+$ isomer with the neutron two-quasiparticle configuration $9/2^-$[734]$_\nu$ $\otimes$ $11/2^-$[725]$_\nu$) assigned \cite{Wahid2025,Clark2010}.

While basically confirming the decay scheme presented in 2010 by Clark et al.~\cite{Clark2010}, they claim evidence for a firm assignment of the proton two-quasiparticle configuration $9/2^+$[624]$_\pi$ $\otimes$ $7/2^-$[514]$_\pi$ to the 8$^-$isomer as in proposed bet He{\ss}bere et al.~ also in 2010~\cite{Hessberger2010}.
They base this interpretation on excitation energy arguments discussing detailed model predictions which obtain a too high excitation energy for two-quasineutron state, in contradiction to  Clark et al.~who had proposed such a neutron configuration for this 8$^-$ state, proposing the configurations ($9/2^-$[734]$_\nu$ $\otimes$ $7/2^-$[624]$_\nu$ or $9/2^-$[734]$_\nu$ $\otimes$ $7/2^-$[613]$_\nu$). 
To respond to the argument that this would follow from the two-neutron configuration of the 10$^+$ state feeding this state, Wahid et al.~offer the solution of a small admixture of $9/2^-$[734]$_\nu$ $\otimes$ $7/2^+$[613]$_\nu$ to the configuration of the 8$^-$isomer with the common $9/2^-$[734]$_\nu$ between initial (10$^+$) and final (8$^-$) state, corroborating the decay transition. 

Contrary to an unusually high internal conversion coefficient for the 887~keV transition as reported by Forge et al.~\cite{Forge2023,Forge2023a}, Wahid et al.~list in Table II of Ref.~\cite{Wahid2025} a rather modest contribution by $CE$ emission. 
For the 887.7~keV transition from the 4$^+$ state in the structure below the 8$^-$ isomer at $E^* =$ 1034~keV$\rightarrow$ to the 4$^+$ state of the g.s. rotational band at $E^*$\,=\,146~keV they give a $\gamma$ transition intensity $I_\gamma(887.7~keV)$ of 9.5(5)\% as compared to a total transition intensity $I_{tot}$(887.7~keV) (relative to the 944-keV transition).
\begin{figure}[t]
\begin{center}
\includegraphics[width=\columnwidth]{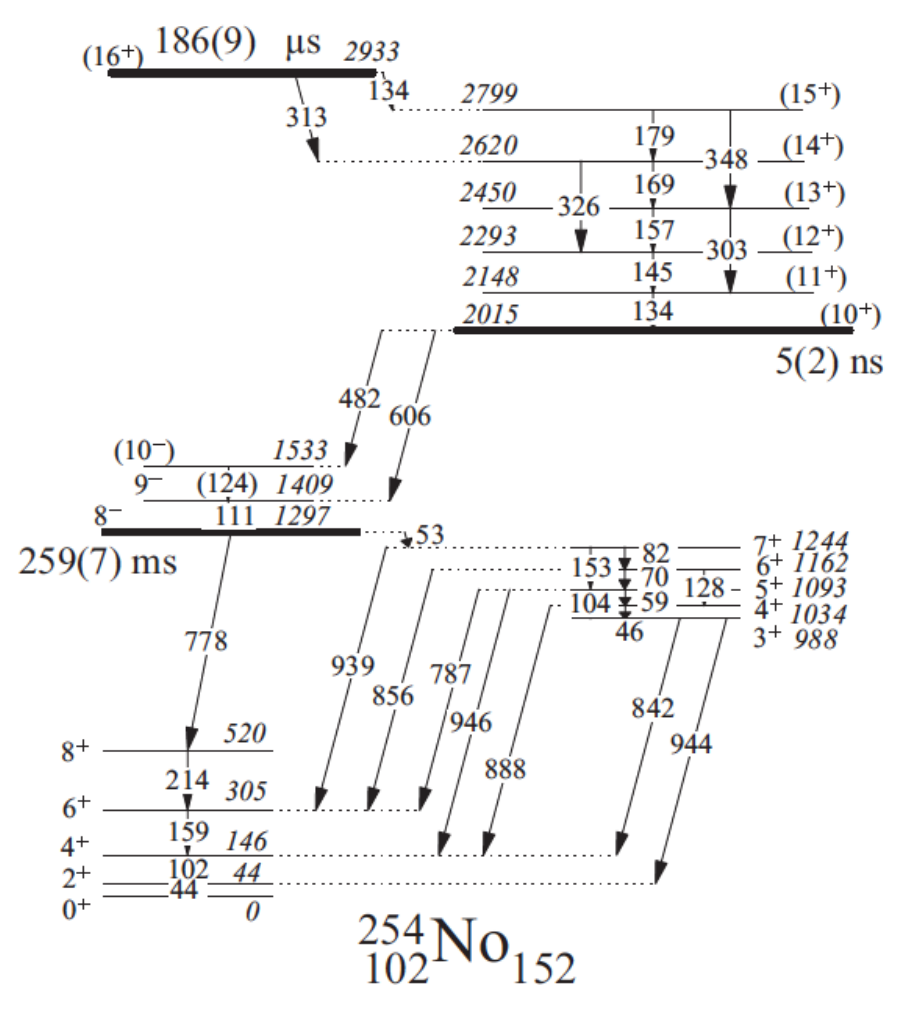}%
\end{center}
\caption{
Partial decay scheme of $^{254}$No established from the present work (\cite{Wahid2025}). The half-life of the $K^\pi$\,=\,10$^+$ at $E^*$\,=\,2015~keV is newly determined while those of the $K^\pi$\,=\,8$^-$ and $K^\pi$\,=\,16$^+$ 
are in agreement with prior work~\cite{Tandel2006,Herzberg2006,Clark2010,Hessberger2010}.
(Figure and caption are taken from~\cite{Wahid2025})}
\label{fig:254No_LS_Wahid}
\end{figure}
The updated decay scheme~\cite{Wahid2025} is shown in Fig.~\ref{fig:254No_LS_Wahid}.\\
 
\begin{figure*}[htb]
\begin{center}
\includegraphics[width=\textwidth]{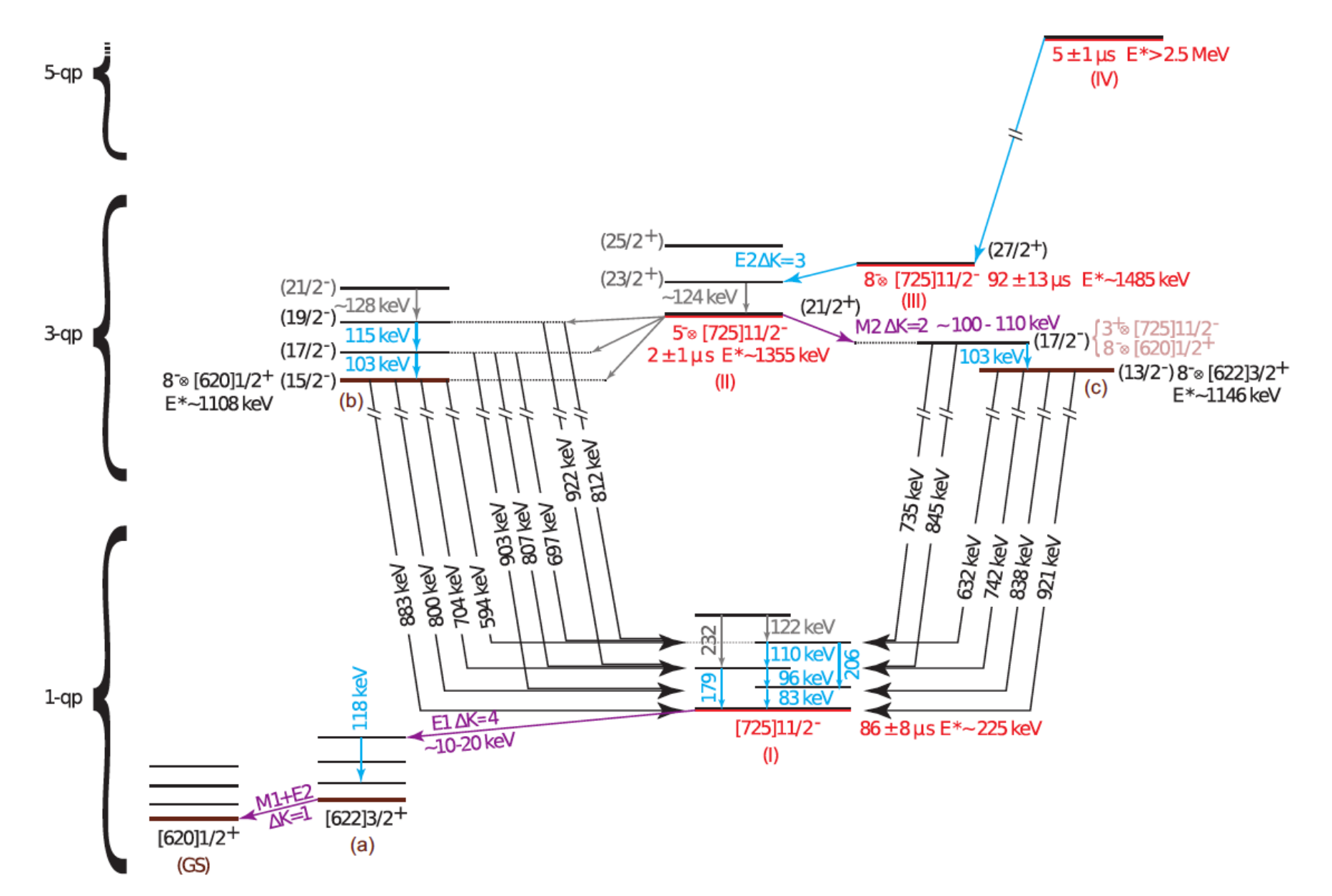}
\end{center}
\caption{Proposed decay scheme of $^{255}$No. 
One can see the four isomeric states observed denoted in red along with the intermediate structures (a)–(c) discussed. 
The observed $\gamma$-ray transitions are denoted in blue while the unobserved ones are denoted in gray and transitions deexciting the isomers are denoted in purple. (Figure and caption taken Ref.~\cite{Kessaci2024})}
\label{Fig:255No_Kessaci}
\end{figure*}
{\it $^{\it253}$No}\\ 
From the isotopic shift, obtained by laser excitation of the atomic $^1p_1$ state in $^{252,253,254}$No and the hyperfine splitting in $^{253}$No Raeder et al.~\cite{Raeder2018} could determine experimentally the magnetic moment $\mu$:
\begin{equation}
	\frac{\mu}{\mu_N} = g_K \frac{I^2}{I+1} + g_R \frac{I}{I+1},
\end{equation}
for $^{253}$No, where $\mu_N = e\hbar/2\rm{m}_pc$ is the nuclear magneton, and $g_K$ and $g_R$ are the intrinsic single-particle and rotational $g$-factor, respectively. 
With the range for $g_R$ of 0.7-1$\times Z/A$ and $g^{\rm exp}_K$= -0.22(5) the g.s. Nilsson configuration 9/2$^-$[734] obtained from in-beam $CE$ and $\gamma$ spectroscopy~\cite{Mistry2018,Herzberg2009} was confirmed while the value $g_K$= -1.2 from an earlier study~\cite{Reiter2005} was ruled out.

The $\alpha$ decay of $^{257m}$Rf with the Nilsson configuration 11/2$^-$[725] is known to populate the ground state and excited states of $^{253}$No with two prominent transitions differing by $\approx$750~keV as observed by He{\ss}berger et al.~\cite{Hessberger2016}. 
Two $\alpha$-$\gamma$ correlations observed in this study could, however, not be assigned to the internal electromagnetic transition which was expected to link the isomer to its homologue state in the decay daughter nucleus $^{253}$No.
Accumulating substantially more statistics, Hauschild et al.~detected an evident $\gamma$ line with $E_\gamma$=750~keV correlated to the $\alpha$ decay transition with the lower decay energy of $E_\alpha$=8986(5)~keV~\cite{Hauschild2022}, which was identified as the searched for 11/2$^-$[725] Nilsson state. This state, originating from the $j$15/2 neutron orbital (see, e.g., ref.~\cite{Chasman1977}), has been observed for all the $N$=153 isotones from curium to seaborgium~\cite{Asai2015}, completed by the recent result  for $^{255}$No discussed below (see Fig.~\ref{fig:153_isotones}). For the $N$=151 isotones which have 9/2$^-$[734] as the g.s.~Nilsson configuration, stemming also from the $j$15/2 neutron orbital, $^{253}$No is the second $N$\,=\,151 isotone after $^{255}$Rf for which 11/2$^-$[725] state is observed.\\

{\it $^{\it255}$No}\\
An isomeric state had been tentatively assigned for $^{255}$No first in 2006~\cite{Hessberger2006a}. 
Re-analyzing these data, employing $CE$s to construct $\alpha$-$CE$-$\gamma$ correlations, Bronis et al.~proposed in 2022 a tentative level scheme including three isomeric states~\cite{Bronis2022}. 
The lowest lying isomeric state, $^{255m1}$No, they tentatively assigned to the $\nu$11/2$^-$[725] Nilsson state, placed in an excitation energy range of 240~keV to 300~keV and decaying with a half-life of 109(9)~$\mu$s.
With this assignment the systematics of $N$=153 isotones from $^{249}$Cm to $^{259}$Sg is completed with the $\nu$11/2$^-$[725] state placed at an excitation energy in between its analog states in $^{253}$Fm and in $^{257}$Rf (see Fig.~\ref{fig:153_isotones}).
For the second isomer $^{255m2}$No with a measured half-life of 77(6)~$\mu$s, a spin of 21/2 or 23/2 and an excitation energy range of 1400~keV to 1600~keV are estimated. 
The third isomer, $^{255m3}$No, is the shortest lived with $T_{1/2}$\,=\,1.2$^{+0.6}_{-0.4}$~$\mu$s and is expected to have an excitation energy of $E^*\geq$1500~keV.

These assignments of isomeric states in $^{255}$No are in fair agreement with results obtained recently at the velocity separator SHELS of the FLNR JINR by Kessaci et al.~\cite{Kessaci2024}.
Fig.~\ref{Fig:255No_Kessaci} shows, in addition to the three isomers reported by Bronis et al., a forth metastable state at higher excitation energy, $E^*$\,>\,2.5~MeV, interpreted as a 5-quasiparticle excitation.
The lowest isomer is also here assigned to the same 11/2$^-$[725] Nilsson level as a 1-quasiparticle excitation.
For isomer II and III in their nomenclature, interpreted as 3-quasiparticle excitations, Kessaci et al.~assigned spin values, based on the 11/2$^-$[725] state coupled to 5$^-$ and 8$-$, i.e. 21/2 and 27/2, respectively, which are close to the tentative suggestions made by Bronis et al.\\

{\it $^{\it256}$No}\\
In an irradiation of $^{238}$U with a $^{22}$Ne beam, Kessaci et al. populated the 4-neutron fusion-evaporation channel $^{256}$No as ER~\cite{Kessaci2021}. 
They observed 15 ER-$e^-$-$\alpha$ correlations with a deduced average half-life of 7.8$^{+8.3}_{-2.6}$~$\mu$s which was interpreted as a high-$K$ isomer decay.
A second scenario assuming a truncation of the time distribution for short times and applying the maximum likelihood estimate as proposed by K.-H. Schmidt et al.~\cite{Schmidt1984} resulted in a slow decay component of 10.9$^{+21.7}_{-4.3}$~$\mu$s and a fast one of $\approx$ 6~$\mu$s. A possible interpretation of the existence of two isomeric states in $^{256}$No has to be confirmed in a future experiment aiming at substantially higher statistics. 
In Table~\ref{tab:isotope_list} the average half-life value was adopted. 
The observation of $\gamma$ and $x$ rays in coincidence with the isomer decays led to a lower limit for its excitation energy of 1089~keV. An assignment of the isomer's quasiparticle excitation configuration with the relative spin and parities needs further experimental efforts.

\subsection{Lawrencium - $Z$=103}\label{Lr}
The heaviest of the fourteen known lawrencium isotopes is the endpoint of the $^{294}$Ts ($Z$=117) decay chain $^{266}$Lr~\cite{Khuyagbaatar2014,Khuyagbaatar2019} and the nuclide with the lowest $Z$ populated in a hot-fusion or $^{48}$Ca-induced reaction, leading to the heaviest nuclides with highest $Z$ and $N$ observed so far. The lawrencium isotopes with their unpaired proton provide a sensitive probe to investigate SPLs and $K$-isomerism, in particular, in the vicinity of the $N$=152 deformed shell gap with $^{255}_{152}$Lr. \\

{\it $^{\it251}$Lr}\\
In 2022, Huang et al.~reported the discovery of the new isotope $^{251}$Lr, produced as the 2$n$ fusion-evaporation channel in the reaction $^{50}$Ti+$^{203}$Tl at the gas-filled separator AGFA of Argonne National Laboratory's linear accelerator facility ATLAS~\cite{Huang2022}. 
In addition to the ground state $\alpha$ decay, they observed an isomeric state decaying as well by $\alpha$ emission. The spin and parity assignments, with 7/2$^+$ for the g.s. and 1/2$^-$ for the excited isomer, and the decay scheme for the new pair $^{251}$Lr-$^{247}$Md are found to be similar to the neighboring isotopes $^{253}$Lr-$^{249}$Md (also investigated in this work) and $^{255}$Lr-$^{251}$Md (see Fig.~\ref{fig.Es-Db_decay_scheme}). 
Very recently, the observation of two $^{251}$Lr $\alpha$ decays as decay daughter of $^{255}$Db was also reported by J.~Pore et al.~\cite{Pore2024a}.\\

{\it $^{\it252}$Lr}\\
In an irradiation of $^{206}$Pb with $^{51}$V projectiles at the BGS of LBNL Pore et al.~accumulated new data for $^{252}$Lr, observed as member of the $^{256}$Db decay chain~\cite{Pore2024}.
The authors report on the detection of a previously unknown $EC$ decay branch for $^{252}$Lr, based on one safe assignment and additional four events which could not be assigned unambiguously. 
Pore et al.~interpret this in terms of a probability range which leads them to propose a range for the $\beta$ decay branching ratio of $b_\beta$\,=\,10\% to 30\%. 
Despite this uncertain assignment, these values are reported in Table~\ref{tab:isotope_list}. 
However, as this assignment is uncertain which is in addition also affected by the low number of events (1 to 5), these numbers should be taken with caution.\\

{\it $^{\it253,254}$Lr}\\
Ito et al. measured the g.s. masses of $^{253,254}$Lr together with other nuclides in the region around $Z$=100 and $N$=152 for the first time using a Muli-Reflection Time-of-Flight Mass Spectrometer (MRToF-MS) installed in the focal plane of the gas-filled separator GARIS of the Nishina Accelerator Centre of RIKEN, Wako, Japan~\cite{Ito2018}.

In 2024, Zhao et al.~reported new $\alpha$-decay data for $^{253,254}$Lr data, collected at the gas-filled separator SHANS of the IMP, Lanzhou, China, Zhao et al.~in the decay chains of the two bohrium isotopes $^{261,262}$Bh, reproducing the literature values~\cite{ZhaoZ2024}. 
\\ 

\begin{figure}[htb]
\begin{center}
\includegraphics[width=\columnwidth]{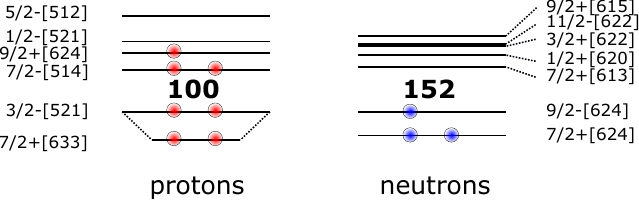}
\end{center}
\caption{Possible ground state configuration for $^{254}$Lr in a single-particle level scheme for protons (left) and neutrons (right) extracted from Ref.~\cite{Chasman1977}.}
\label{Fig:254Lr_excit_scheme}
\end{figure}

{\it $^{\it254}$Lr}\\ 
As $^{246}$Es and $^{250}$Md (see subsections~\ref{Es} and \ref{Md}), $^{254}$Lr is a member of the $^{258}$No $\alpha$-decay chain. 
Vostinar et al. constructed a low-lying level scheme combining states populated by $\alpha$ decay of the ground and isomeric state of $^{258}$Db (see subsection~\ref{Db}) as well as by the observed $\alpha$-decay transitions of $^{254}$Lr itself~\cite{Vostinar2019}. 
The observation of a previously proposed isomeric state could be confirmed by $\alpha$-decay curves for both activities with half-lives of 11.9(9)~s for the g.s. and 20.3(42)~s for the isomer, respectively. 
The literature value for the $^{254}$Lr lies with 18.4(18)~s in between these values. 
For Table~\ref{tab:isotope_list} the values of Ref.~\cite{Vostinar2019} have been adopted.
Spin and parities for $^{254g.s.}$Lr and $^{254m}$Lr were tentatively assigned as $J^\pi$=4$^+$ and 1$^-$, respectively.

The excitation energy of the isomer with $E$*=108~keV has been confirmed by precision mass measurements performed at the Penning trap installation SHIPTRAP at GSI~\cite{Block2020}.

Fig.~\ref{Fig:254Lr_excit_scheme} shows a possible ground-state configuration for $^{254}$Lr with single-particle levels in the vicinity of $Z$=100 and $N$=152 extracted from Ref.~\cite{Chasman1977} for a quadrupole deformation of $\approx$0.25. 
Various scenarios for spin and parity assignments to $^{254g.s.}$Lr and $^{254m}$Lr can be assumed, some of which would put the assignment as proposed by Vostinar et al.~in question. 
For a final conclusion concerning the properties of these states and for a comparison with the various theoretical model approaches, additional data with higher statistics are mandatory.\\

{\it $^{\it257}$Lr}\\ 
Alpha decay of $^{257}$Lr was observed in correlation to $^{257}$Rf and to $CE$s by He{\ss}berger et al.~\cite{Hessberger2016}, which was regarded to be a direct proof for $EC$ decay of both the ground state $^{257g.s.}$Rf and the isomer $^{257m}$Rf which were subject of a recent study by Hauschild et al.~\cite{Hauschild2022} (see subsection~\ref{Rf}).\\

{\it $^{\it258}$Lr}\\ 
Employing the fusion-evaporation reaction $^{248}$Cm($^{23}$Na,5$n$)$^{266}$Bh and the rotating wheel detection system MANON, Haba et al.~collected additional $\alpha$-decay data for $^{258}$Lr as the granddaughter in the $^{266}$Bh decay chain (see subsection~\ref{Bh})~\cite{Haba2020}.\\

{\it $^{\it264}$Lr}\\ 
In a recent experiment with the new SHE-factory facility of FLNR JINR, Oganessian et al.~assigned a new $\alpha$-decay activity to $^{268}$Db in the $^{288}$Mc decay chain~\cite{Oganessian2022c,Oganessian2022d} (see also section~\ref{Db}), which led to the discovery of $^{264}$Lr. 
A 4.9-h $SF$ decay was assigned to the new isotope $^{264}$Lr. 
This activity was previously, before this new measurement at the FLNR JINR SHE-factory, not observed despite the collection of 104 $^{288}$Mc decay chains by the various experiments at the FLNR in 2003 (3 chains)~\cite{Oganessian2004}, in 2010-2012 (31 chains)~\cite{Oganessian2013}, at the gas-filled separator TASCA of GSI (22 chains)~\cite{Rudolph2013}, at the Berkeley gas-filled separator BGS (48 chains)~\cite{Gates2015,Gates2018}. The population of $^{264}$Lr was possible because here the authors investigated for the first time "$\alpha$-like" events in between the $^{272}$Mt and the long-$T_{1/2}$ (28~h) SF events following it and found a new $\alpha$ activity for $^{268}$Db with an $\alpha$ branching of 55$^{+20}_{-15}$\%, an $\alpha$-decay energy of 7.6-8.0~MeV and a new $^{268}$Db half-life of $T_{1/2}$=16$^{+6}_{-4}$h (see subsection~\ref{Db}). 
The half-life for the new fissioning isotope $^{264}$Lr they quote with $T_{1/2}$=4.9$^{+2.1}_{-1.3}$~h.\\

{\it $^{\it266}$Lr}\\
In 2019 Khuyagbaatar et al.\cite{Khuyagbaatar2019} updated the previously published results~\cite{Khuyagbaatar2015} with more details on the decay sequence starting with $^{294}$Ts, which they observed to be terminated by $SF$ of $^{266}$Lr. 
The $\alpha$ decay of $^{270}$Db to $^{266}$Lr, detected here in the two longer decay chains, was not observed in the earlier experiments at the DGFRS of FLNR JINR~\cite{Oganessian2013b} (see also section~\ref{Ts}).

\begin{figure*}[htb]
\begin{center}
\includegraphics[width=0.65\textwidth]{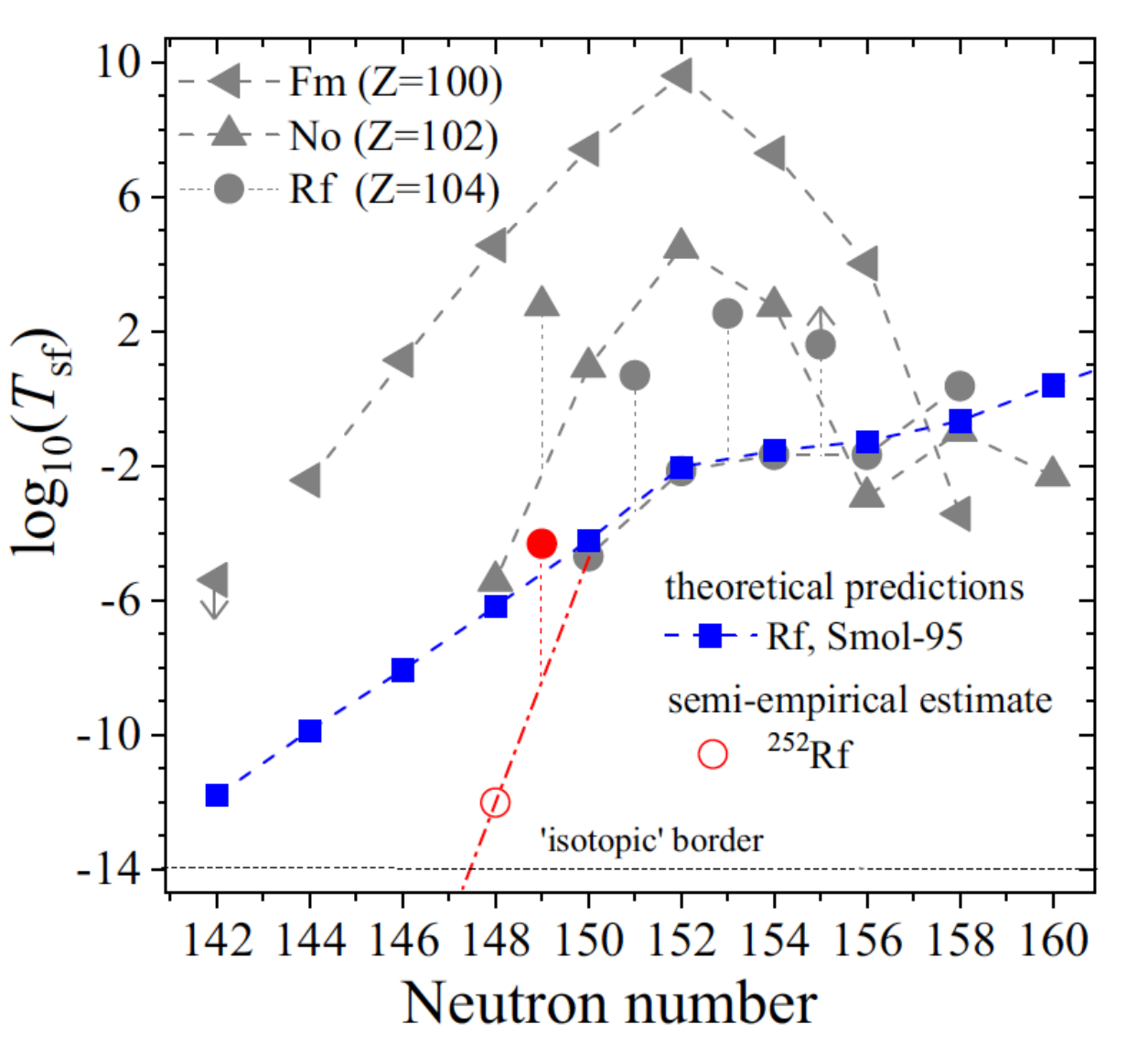}
\end{center}
\caption{Spontaneous fission half-lives of Fm, No, and Rf isotopes~\cite{Hessberger2017}. 
Theoretically predicted half-lives for Rf isotopes~\cite{Smolanczuk1995} are shown by rectangles. 
Dashed lines connect even-even isotopes.
Vertical dotted lines show the fission hindrance factors for odd-A nuclei. 
The dash-dotted line crossing to the isotopic border shows an abruptly falling tendency in half-lives of neutron-deficient Rf isotopes based on an empirically estimated half-life of 1~ps for $^{252}$Rf.
This estimate is based on a 48~$\mu$s half-life of $^{253}$Rf~\cite{Hessberger2017} and an assumed $F_H=10^4$.  
(Figure and caption are taken from Ref.~\cite{Khuyagbaatar2021}). See text and Ref.~\cite{Khuyagbaatar2021} for details.}
\label{fig:Fm-Rf_T_sf}
\end{figure*}

\begin{figure*}[h]
\begin{center}
\includegraphics[width=0.7\textwidth]{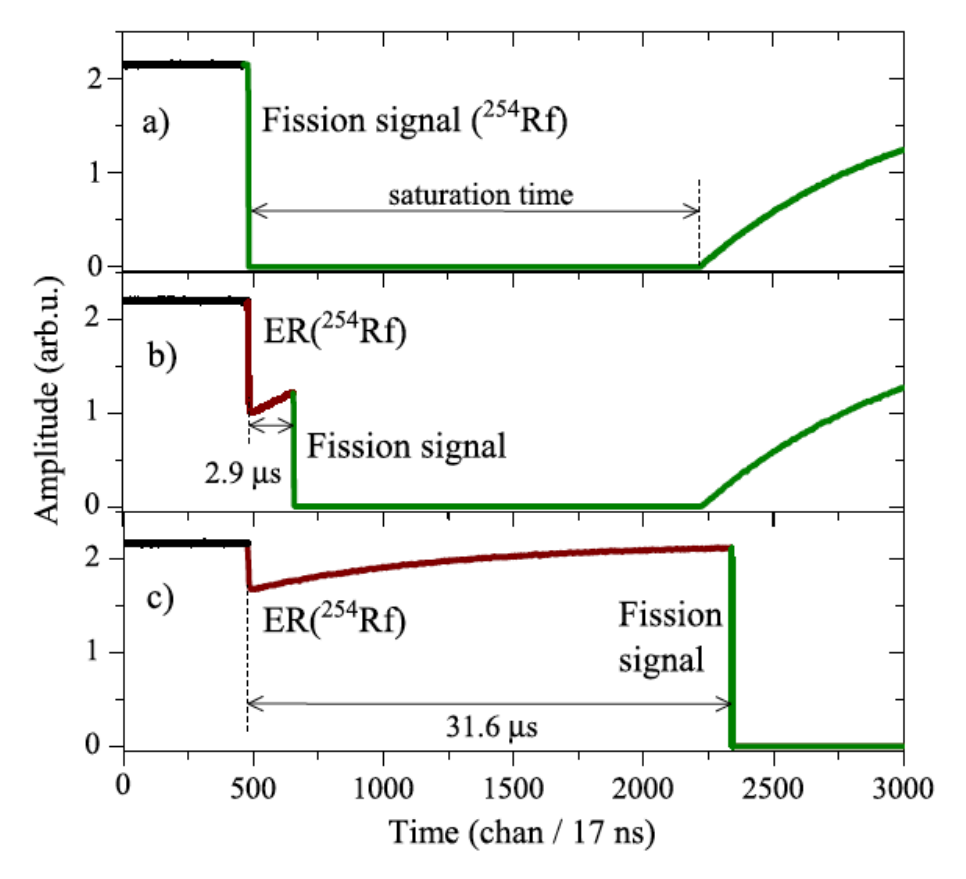}
\end{center}
\vspace{-0,5 cm}
\caption{Examples of different types of traces containing fission events: 
a) Single-signal trace corresponding to $^{254}$Rf. 
b and c) Double-signal traces where the implantation signals (noted by ER) of $^{254}$Rf were followed by fissions after 2.9 and 31.6~$\mu$s, respectively. (Figure and caption are taken from Ref.~\cite{Khuyagbaatar2020b}, where more details are given)}
\label{fig:PSA_fission}
\end{figure*}

\subsection{Rutherfordium - $Z$=104}\label{Rf}

For rutherfordium isotopes, spontaneous fission becomes more important and is influenced by the transition from $N$=152 deformed shell gap towards the next higher one at $N$=162. 
The detailed discussion of $SF$ properties of heavy and superheavy nuclei by F.P.~He{\ss}berger~\cite{Hessberger2017} is addressing the various aspects of fission for nuclear structure and stability.
Rf isotopes play a major role in this comprehensive review of the state-of-the-art of fission studies.
Among the thirteen rutherfordium isotopes $^{256}$Rf, positioned on the $N$=152 deformed shell gap, occupies a special position. 
Apart from being the heaviest nuclide for which a rotational band has been observed (up to a spin of 20~$\hbar$)~\cite{Greenlees2012}, three $K$ isomers with half-lives around 20~$\mu$s were initially reported by Jeppesen et al.~\cite{Jeppesen2009}. 
This has attracted until recently the attention of various collaborations, as we had also reported in our previous review~\cite{Ackermann2017}.

The latest addition to the series of rutherfordium was the discovery and investigation of $^{252}$Rf  by Khuyagbaatar et al., extending the systematics towards the "isotopic border" of shortest fission half-lives~\cite{Khuyagbaatar2025}. It will be discussed below together with the findings for $^{253}$Rf~\cite{Lopez-Martens2022,Khuyagbaatar2021}.\\

{\it $^{\it252,253}$Rf}\\ 
Recently two groups reported on new data collection for $^{253}$Rf~\cite{Khuyagbaatar2021,Lopez-Martens2022}. 
As mentioned in subsection~\ref{No}, Khuyagbaatar et al.~\cite{Khuyagbaatar2021} found a new $\alpha$-decay branch followed also by the $\alpha$ decay of the new isotope $^{249}$No (see subsection~\ref{No}). 
For $^{253}$Rf, they observed two fission activities with fission half-lives of 12.8$^{+7.0}_{-3.4}$~ms and 44$^{+17}_{-10}$~$\mu$s, respectively. 
The longer-lived decay was assigned to the ground state as 7/2$^+$[624] and the short one as the excited 1/2$^+$[631] Nilsson state.

The motivation for the investigation of the fission properties of $^{253}$Rf was based on the fission hindrance, for which a hindrance factor $F_H$ is introduced. 
For a detailed discussion of fission hindrance for nuclei with odd nucleon numbers, see section 5 in F.P.~He{\ss}berger's review on {\em Spontaneous fission properties of superheavy elements}, Ref.~\cite{Hessberger2017}.
This hindrance is assumed to be caused by the possibly complex quantum mechanical configuration due to an unpaired nucleon.
It can be defined on the basis of the $SF$ half-lives $T_{SF}$ of the neighboring isotopes like follows (\cite{Khuyagbaatar2021}):
\begin{equation}\label{Eqn:EH_SF}
	F_H (K^\pi,A) = \frac{T_{SF} (K^\pi,A)}
				      {[T_{SF}(0^+,A-1)T_{SF} (0^+, A+1)]^{1/2}} ~~,
\end{equation}
where $\pi$ and $K$ are the parity and the projection of the total angular momentum of the respective unpaired nucleon (single-particle state) on the symmetry axis of the nucleus.	

This hindrance is, in particular, observed in the rutherfordium isotopic chain in long lifetimes for odd-$A$ isotopes. 
Regarding only even-even isotopes, a specific behavior of $SF$ half-lives is observed around the deformed neutron shell gap at $N$=152. 
For lighter isotopic chains up to nobelium and lawrencium, spontaneous fission can compete with $\alpha$ (and $\beta$) decay more towards the neutron-rich and neutron-deficient tails (see Fig.~\ref{fig:Nchart96_118}). 
This results in a bell-shaped half-life distribution around $N$=152. 
For rutherfordium, fission is a prevailing decay even for isotopes more in the center of the chain with a much smaller retardation at $N$=152. 
For the heaviest rutherfordiums, the effect of the next higher deformed shell gap at $N$=162 becomes evident, resulting in similar or even longer half-lives for $^{260}$Rf and $^{262}$Rf. 
In Fig.~\ref{fig:Fm-Rf_T_sf} (taken from~\cite{Khuyagbaatar2021}), the shown partial spontaneous fission half-lives $T_{SF}$ for fermium to rutherfordium isotopes clearly exhibit this transition. 
With respect to the half-lives of odd-$N$ rutherfordium isotopes, the hindrance is clearly visible and seems to follow a stable trend. 
Employing formula~\ref{Eqn:EH_SF} and with $F_H$=10$^4$ estimated on the basis of spin and parities of lighter $N$=149 isotopes, Khuyagbaatar et al. establish $T_{SF}$ of the at that time yet unknown $^{252}$Rf as 1~ps. 
In addition to the two fission activities, a single short-lived event was interpreted as a fast $CE$, possibly originating from a second high-lying isomer as observed in the next lighter $N$=151 isotone $^{251}$No with an excitation energy $E^*$\,>\,1.7~MeV.

In an independent study, Lopez-Martens et al.~also observed two fission activities for $^{253}$Rf with half-lives 52.8(4.4)~$\mu$s and 9.9(1.2)~ms which are within error bars in agreement with the ones reported in Ref.~\cite{Khuyagbaatar2021}. 
In addition, they observed the electromagnetic decay of a high-lying state ($E^*>$1.02~MeV) with a half-life of 0.66$^{+40}_{-18}$~ms which is proposed to be a $K$-isomer. 
This isomer decaying exclusively to a state followed by the faster fission, led them to the conclusion of an inverted assignment with the short-lived fissioning state being the excited 7/2$^+$[624] Nilsson level, and the slower fission originating from the 1/2$^+$[631] ground state. 
Comparing the partial $SF$ half-life for $^{251}$No (g.s.: 7/2$^+$[624]) with 571~s to the one of the same state in $^{253}$Rf and based on the exclusion of fission for $^{251m}$No, in conclusion from a measurement of the same group~\cite{Svirikhin2021}, they estimate a $T_{SF}$($^{252}$Rf) of 0.69~ps, being close to the value proposed by Khuyagbaatar at al. Therefore, they place $^{252}$Rf at the very limit of nuclear stability.
\begin{figure*}[htb]
\begin{center}
\includegraphics[width=0.75\textwidth]{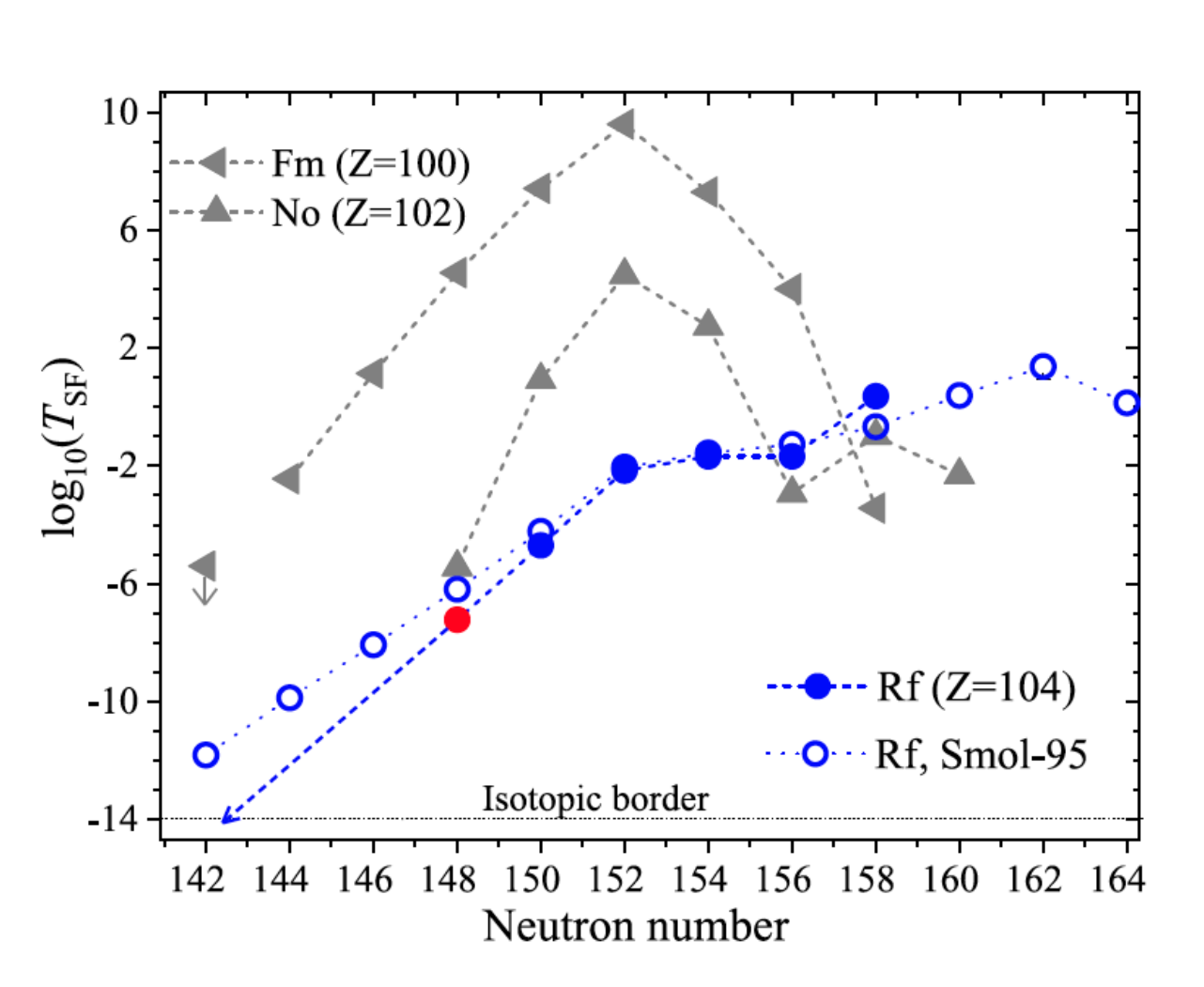}
\end{center}
\caption
{
Update of Fig.~\ref{fig:Fm-Rf_T_sf}: Spontaneous-fission half-lives of Fm, No, and Rf isotopes~\cite{Hessberger2017} are shown by solid symbols. 
Theoretically predicted half-lives for Rf (\cite{Smolanczuk1995}) are shown by open circles. 
The presently (Ref.~\cite{Khuyagbaatar2025}) measured value for $^{252}$Rf is highlighted.  
The horizontal dotted line at 10-14 s indicates an isotopic border for the existence of the chemical elements
See text and Ref.~\cite{Khuyagbaatar2025} for details.
(Figure and caption are taken from Ref.~\cite{Khuyagbaatar2025}).
}
\label{fig:Fm-Rf_T_sf_upd}
\end{figure*}

In a new measurement, reported about a week after the first submission of this review~\cite{Khuyagbaatar2025}, Khuyagbaatar et al.~were successfully synthesizing the new isotope $^{252}$Rf. 
From the measurement of a total of 47 $SF$ events at four different beam energies, and a detailed time and trace analysis, they concluded on the observation of an extremely short-lived ground state with a half-life of 60$^{+90}_{-30}$~ns.
This replaces the hitherto fastest decay of an SHN, which was observed for $^{250}$No with 4.6(2)~$\mu$s.
For technical details of how this information was extracted by the observation of only three $CE$ signals (see Ref.~\cite{Khuyagbaatar2025}).
The observation of such a short decay after in-flight separation, provided by the gas-filled separator TASCA of GSI/FAIR, with a flight time of $\approx$\,0.6~$\mu$s, was only possible due to the existence of a 13$^{+4}_{-3}$~$\mu$s isomer.
This half-life is now shown in the updated half-life systematics in Fig.~\ref{fig:Fm-Rf_T_sf_upd}, indicating a trend much closer to the calculations of Smolanczuk et al.~\cite{Smolanczuk1995} as compared to the expectations reported in Fig.~\ref{fig:Fm-Rf_T_sf}, which was close to the limit of existence for chemical elements of 10$^{-14}$~s (see definition in~\cite{Khuyagbaatar2025}).\\

{\it $^{\it254}$Rf}\\ 
In our earlier review~\cite{Ackermann2017}, we had mentioned $^{254}$Rf as an example for lifetime inversion leading to metastable stable states living longer than the ground state. 
Apart from the long-lived one with $T_{1/2}$=247(73)~$\mu$s David et al.~had observed, thanks to the use of advanced digital electronics, also a short-lived $K$ isomer with $T_{1/2}$=4.7(11)~$\mu$s as compared to the g.s. with $T_{1/2}$\,=\,23(3)~$\mu$s~\cite{David2015}.  
Both isomers were identified by the detection of $CE$s emitted during the de-excitation cascade, and the subsequent g.s. fission. 
In a recent investigation of the products of the fusion-evaporation reactions $^{206}$Pb($^{50}$Ti,$1n$/$2n$)$^{255/254}$Rf, Khuyagbaatar et al.~\cite{Khuyagbaatar2020b} could confirm the short-lived activity in $^{254}$Rf employing pulse-shape analysis, made possible by a digital electronics data acquisition system. 
They were, however, not sensitive to the slow decay. 
An interesting technical detail is the recovery of the $SF$ energy from the large and saturated fission pulse, utilizing the knowledge of the form of the electronic pulse. 
The so-called "time-over-threshold" method allows extending in this way the dynamic range to detect energies from a few hundred keV for the $CE$s to fission fragment signals of $\approx$50-200~MeV with the same preamplifier configuration. 
The method is illustrated in Fig.~\ref{fig:PSA_fission}.

In an in-beam spectroscopy experiment, employing the $\gamma$-detection array GAMMASPHERE, combined with the Argonne Gas-filled Fragment Analyzer (AGFA), despite extremely low statistics, Seweryniak et al.~succeeded in extracting some information on the first members of the ground state rotational band of $^{254}$Rf~\cite{Seweryniak2023}. 
Using the existing knowledge for the rotational bands of neighboring isotopes, in particular, moments of inertia as a function of spin and rotational frequency, and the observation of a cluster of counts which was identified as the 6$^+$ to 4$^+$ transition, they succeeded in locating  the position of $\gamma$ transitions up to 14$^+$ in the $^{254}$Rf $\gamma$ spectrum.
A comparison of the moment of inertia development as a function of the rotational frequency $\omega$ with the $N$~=~150 ($^{250}$Fm, $^{252}$No) and $N$~=~152 ($^{254}$No, $^{256}$Rf) isotones, a behavior similar to that of the lighter $N$~=~150 isotones is found for $^{254}$Rf, confirming the effect of this deformed shell gap on the collective properties of nuclear matter.\\

\begin{figure}[htb]
\begin{center}
\includegraphics[width=\columnwidth]{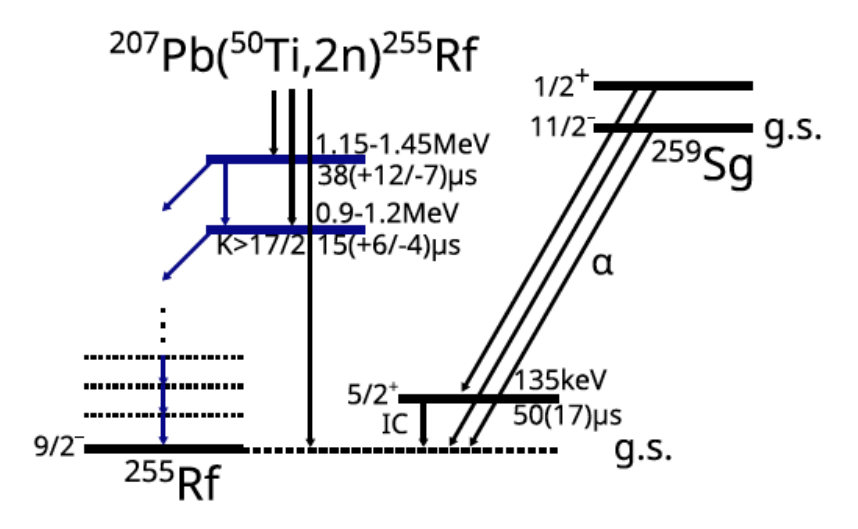}
\end{center}
\vspace{-0,5 cm}
\caption{Proposed decay scheme of the $K$ isomers in $^{255}$Rf, populated in the reaction $^{207}$Pb($^{50}$Ti, $2n$)$^{255}$Rf (left) and in the $\alpha$ decay
of $^{259}$Sg (right). (Figure and caption are taken from Ref.~\cite{Mosat2020})}
\label{fig:255Rf_levels}
\end{figure}
{\it $^{\it255}$Rf}\\
While in the context of the above-mentioned investigation Khuyagbaatar et al., could reproduce within error bars also the literature values for the basic decay properties of $^{255}$Rf, the limited time range of 30~$\mu$s (trace length; see Fig.~\ref{fig:PSA_fission}) of their digital electronics resulted in a truncation of the decay-time spectrum for the detected $CE$s~\cite{Khuyagbaatar2020b}. 
Only the qualitative statement could be made that the observed decay time was longer than 30~$\mu$s which is consistent with the findings of Antalic et al.~\cite{Antalic2015} of $T_{1/2}$($^{255m1}$Rf)\,=\,50(17)~$\mu$s. 

In a study of spontaneous fission for the three rutherfordium isotopes $^{255,256,258}$Rf, Mosat et al.~revisited isomeric structures in $^{255}$Rf employing ER-$CE$-$\alpha$/$SF$ correlations~\cite{Mosat2020}. 
In an energy-decay-time analysis, they revealed two new $K$-isomers with $T_{1/2}$($^{255m2}$Rf)=15$^{+6}_{-4}$~$\mu$s at $E^*$ 0.9-1.2~MeV and $T_{1/2}$($^{255m3}$Rf)=38$^{+12}_{-7}$~$\mu$s at $E^*$ 1.15-1.45~MeV. 
To populate $^{255}$Rf and its low-lying excited states they used the direct production reaction $^{207}$Pb($^{50}$Ti, $2n$) $^{255}$Rf.
After $\alpha$ decay of $^{259}$Sg, these isomeric states were not observed~\cite{Antalic2015}. 
This is illustrated Fig.~\ref{fig:255Rf_levels} taken from~\cite{Mosat2020}.
An analysis of available SPLs was not conclusive, as many different combinations could lead to appropriate high-$K$ configurations to explain the decay hindrance which is causing the observed decay times.
Advances in modern, self-consistent theory, in particular, an improvement of the SPL-energy predictions, are needed to disentangle this puzzle. 
The accumulation of more and more experimental cases of $K$-isomeric structures with their direct link to SPL properties, building up networks in the SPL-energy vs.~the deformation plane of a Nilsson description could be a route to eventually also enable robust predictions for the region of spherical shell-stabilized superheavy nuclei.
\begin{figure}[htb]
\begin{center}
\includegraphics[width=0.75\columnwidth]{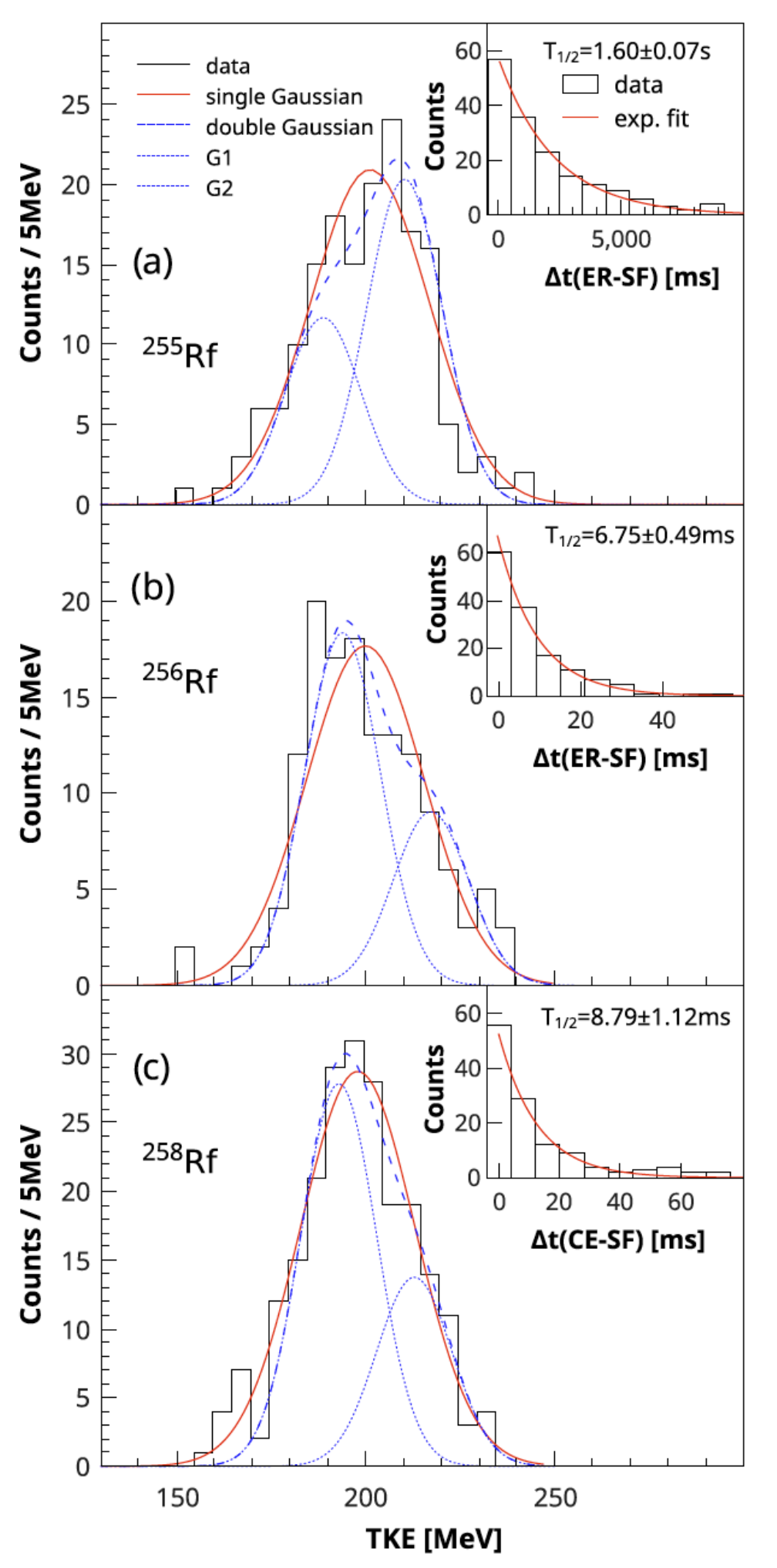}
\end{center}
\vspace{-0,5 cm}
\caption{TKE distributions of SF fragments from STOP-BOX ("STOP" denotes the Si implantation-detector and "BOX" the SI detectors surrounding the "STOP" detector upstream, together forming a box-like configuration open on one side) coincidences (a) for SF of $^{255}$Rf obtained from $ER-SF$ correlations (inset shows $ER-SF$ time differences), (b) for $SF$ of $^{256}$Rf obtained from $ER-SF$ correlations (inset shows $ER-SF$ time differences), and (c) for $SF$ of $^{258}$Rf obtained from STOP-BOX coincidences (inset shows $CE-SF$ time differences in cases when $CE$ was registered). (Figure and caption are taken from Ref.~\cite{Mosat2020})}
\label{fig:255256258Rf_TKL}
\end{figure}
In addition to the investigation of the isomer decay, Mosat et al.~also investigated the Total Kinetic Energy ($TKE$) distribution of the $SF$ decay of $^{255}$Rf employing the method presented by Nishio et al.~in Ref.~\cite{Nishio2007}. 
Despite the observed asymmetry in the $TKE$ distribution, a double-Gaussian fit analysis was not conclusive regarding bi-modal fission which was observed in that region of the Segr\`{e} chart and predicted for $^{256,258}$Rf~\cite{Carjan2015} (see also the respective paragraphs below).
A comparison of the $TKL$ distribution for $^{255}$Rf with $^{256,258}$Rf is shown in Fig.~\ref{fig:255256258Rf_TKL}, taken from~\cite{Mosat2020}.\\

{\it $^{\it256}$Rf}\\    
As mentioned in the introduction of this subsection, the observation of possibly three $K$-isomeric states with half-lives of $\approx$20~$\mu$s has attracted the attention of the community interested in SHN structure. 

Reported first by Jeppesen at al.~in 2009~\cite{Jeppesen2009} and shortly after by Berryman et al., Robinson et al., and Rissanen et al.~\cite{Berryman2010,Robinson2011,Rissanen2013}, it was revisited in 2021 Khuyagbaatar et al.~\cite{Khuyagbaatar2021a}.
In particular, Khuyagbaatar et al.~\cite{Khuyagbaatar2021a}, employing trigger-less digital electronics and pulse-shape analysis (PSA) of registered traces confirm the existence of two isomeric states, despite the much lower statistics as compared to the results from Jeppesen et al.~\cite{Jeppesen2009}.
Apart from being a demonstration of how effective digital electronics and PSA can be applied to short-lived decays, a major aim of this investigation was the evaluation of the $K$-isomer population probability. 
A value of $\geq$28\% for the total population of the metastable states indicates, according to the authors, that the population of high-$K$ isomers in higher-$Z$ nuclei like Sg ($Z$=106) and Hs ($Z$=108) is to be expected.

Also for $^{256}$Rf, Mosat et al. investigated the $SF$ $TKE$ distributions and compared it to $^{255,258}$Rf~\cite{Mosat2020} (see respective paragraphs above and below).

Lopez-Martens et al.~reported recently on the production of $^{256}$Rf as a $p2n$-evaporation channel in the reaction  $^{209}$Bi($^{50}$Ti, $p2n$) together with $^{258}$Rf produced in the same reaction at lower excitation energy  in the $1p$-evaporation channel (see below).\\

{\it $^{\it257}$Rf}\\    

Mosat et al.~confirmed in their study of rutherfordium isotopes ($^{255,256,257}$Rf) the earlier findings for isomeric states, including one in $^{257}$Rf (see also above)~\cite{Mosat2020a}. 
As mentioned in section~\ref{No}, the decay from ground and isomeric states of $^{257}$Rf was employed by Hauschild et al.~to reveal details of the low-lying states in $^{253}$No, including the discovery of the searched-for 11/2$^-$[725] Nilsson state.\\
\begin{figure}[htb]
\begin{center}
\includegraphics[width=\columnwidth]{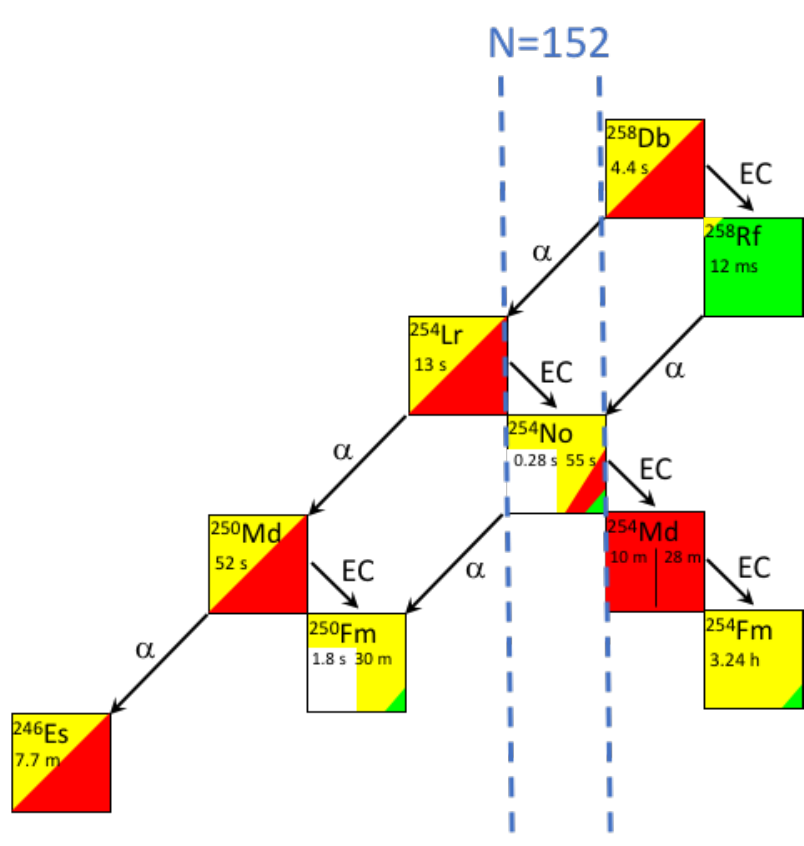}
\end{center}
\caption{Decay scheme of $^{258}$Db (color scheme and data taken from Ref.~\cite{Soti2024}).}
\label{Fig:258Db_decay}
\end{figure}

{\it $^{\it258}$Rf}\\    
As shown in Fig.~\ref{Fig:258Db_decay}, $^{258}$Rf can be produced by $EC$ decay of $^{258}$Db and is mainly decaying by $SF$. 
By delayed $CE$-$\alpha$ coincidences and correlations to $^{254}$No $\alpha$ decay, He{\ss}berger et al.~observed a weak $\alpha$-decay branch with b$_\alpha$=0.049(16)~\cite{Hessberger2016a} which was about an order of magnitude lower than the previously reported value of b$_\alpha$=0.31 by Gates et al.~\cite{Gates2008}. 
The $^{258}$Rf mass can be obtained by the here established $Q_\alpha$ value from the precisely measured mass value of $^{254}$No at the Penning trap set-up SHIPTRAP~\cite{Block2010}. On this basis the two-neutron separation energy trends in the vicinity of the deformed shell gap at $N$=152 and towards $N$=162 are also discussed in comparison with model predictions in the paper by He{\ss}berger et al.~\cite{Hessberger2016a}. 

\begin{figure}[htb]
\begin{center}
\includegraphics[width=\columnwidth]{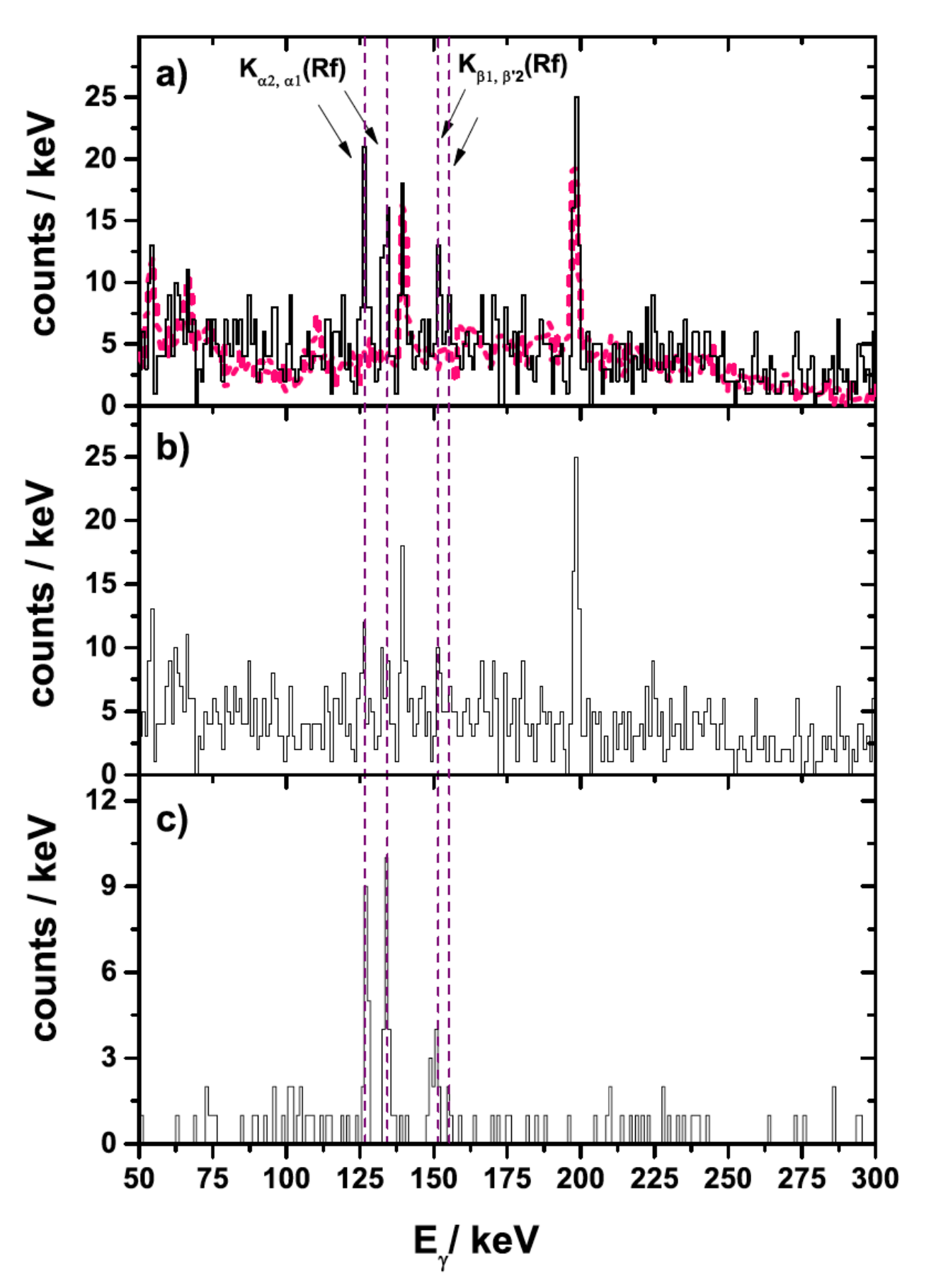}
\end{center}
\vspace{-0,5 cm}
\caption{Photon spectra observed in delayed coincidence with $SF$ events; (a) spectrum of events, recorded between the beam bursts and preceding a SF event within $\Delta t \leq$\,30~ms (black line); the dotted line (pink) shows the spectrum of $\gamma$ background down scaled to similar peak-heights for the background lines at E=139 and 197~keV; (b) same as (a), but for photons registered in anti-coincidence with a $CE$; (c) spectrum of photons registered in coincidence with a $CE$ preceding an $SF$ event within $\Delta t \leq$ 50~ms. 
The dotted lines (purple) mark the energy positions of the $K_{\alpha1,2,\beta1,2} $ $x$-rays of rutherfordium (Z = 104).
No attempts to identify the background lines were made. (Figure and caption are taken from Ref.~\cite{Hessberger2016a})}
\label{fig:258DbECx}
\end{figure}

In addition to the improved $\alpha$ decay data, in this paper $CE$-coincident photons, followed by SF, were studied. Fig.~\ref{fig:258DbECx} shows $EC$-$SF$ correlated photon spectra from the $^{258}$Db decay. Here enhanced rutherfordium $x$-ray lines were observed for photons in coincidence with $CE$, while they were suppressed with at the same time enhanced $\gamma$ transitions appearing in the spectrum in anti-coincidence with $CE$s. 
From $EC$-$\gamma$/$x$-ray-$CE$-$SF$ correlations, the existence and half-lives of two isomeric states in $^{258}$Rf could be established with $T_{1/2}$($^{258m1}$Rf) = 2.4$^{+2.4}_{-0.8}$~ms and  $T_{1/2}$($^{258m2}$Rf) = 15(10)~$\mu$s, respectively.
In particular, the observation of $x$-rays of the decay daughter rutherfordium is interesting with respect to the identification of the atomic charge $Z$ for cases where $SF$ is terminating a decay chain of unknown nuclei (see subsection~\ref{Db}).
 
In an irradiation of $^{209}$Bi with $^{50}$Ti Lopez-Martens et al.~found evidence for proton-evaporation channels ($p$ and $p2n$) leading to $^{258}$Rf (and $^{256}$Rf) as ERs by the observation of short-lived $SF$ events~\cite{Lopez-Martens2019} (see also subsection~\ref{Db}). 
In an earlier measurement in 1985~\cite{Hessberger1985}, He{\ss}berger reported the observation of one SF event with a decay time of 5~ms, which was tentatively assigned to the reaction channel $^{209}$Bi($^{50}$Ti,$p$)$^{258}$Rf at a beam energy of 243~MeV. 
The obtained cross-sections in both experiments, being consistent with each other, are suppressed with respect to neutron-evaporation channels by a factor of 30 to 100 at the given beam energies, with $\approx$\,4~pbarn~\cite{Lopez-Martens2019} to  $\approx$\,19~pbarn~\cite{Hessberger1985,Hessberger2019} for $p$-evaporation channels  and $\approx$\,4~nbarn for the maximum cross-section of the neutron-evaporation channels. 

The detailed structure of the $SF$~$TKL$ distribution of $^{258}$Rf was compared to the one of the two lighter isotopes $^{255,256}$Rf by Mosat et al.~\cite{Mosat2020} (see also above).
The two-Gaussian analysis could possibly indicate bi-modal fission due to the observed skewness for the latter, while the situation is less clear for the more symmetric $TKL$ distribution observed in the case of $^{258}$Rf with only a small improvement of the $\chi^2$ value for the double-Gaussian with respect to the single-Gaussian fit (see Fig.~\ref{fig:255256258Rf_TKL}). 
The two possibility of either single-mode fission or the existence of two modes with similar distributions is supported by theory predictions~\cite{Carjan2015}.\\

{\it $^{\it261}$Rf}\\
In an irradiation on $^{238}$U with $^{40}$Ar projectiles at the new FLNR JINR SHE-factory, Oganessian et al.~observed two $\alpha$-decay chains of type $ER$-$\alpha$-$\alpha$-$SF$, which they assigned to 5$n$ fusion-evaporation channel with the $ER$ $^{273}$Ds, terminated by the spontaneous fission of $^{261m}$Rf, which is consistent with literature data~\cite{Oganessian2024}. 
For a discussion of the whole decay chain in which the population of isomers in all chain members is proposed by the authors of Ref.~\cite{Oganessian2024} see subsection~\ref{Ds}.\\

{\it $^{\it262}$Rf}\\
In a systematic study of the competition between fusion and quasi-fission, $^{262}$Rf was one of the nuclei investigated by Hinde et al.~\cite{Hinde2018}. 
This work was mainly focused on reaction dynamics, being relevant for production and survival of very heavy nuclei. 
As nuclear structure features of the heavy collision participants, in particular nuclear deformation, play a decisive role, this reaction mechanism work with its findings is mentioned in this review on decay spectroscopy. 

Investigating fission mass angular distributions (MAD) for the $^{28}$Si+$^{238}$U and $^{34}$S+$^{232}$Th leading to the excited compound nucleus $^{266}$Sg, and $^{24}$S+$^{238}$U forming $^{262}$Rf, Hinde et al.~have access to the reaction dynamics and can distinguish between fast and slow processes.
From peaked MADs they deduce the existence of a fast quasifission (FQF) mode leading to an asymmetric mass distribution, which can be distinguished from slow quasifission (SQF) that shows a symmetric mass distribution. 
These two quasifission modes compete with the fusion of the colliding system. Systematically analyzing the relative strength of symmetric and asymmetric fission modes and angular distributions as a function of kinetic energy of the colliding system as well as of the atomic charge of the projectile impinging on the deformed actinide targets, the authors discuss the competing processes.\\

{\it $^{\it263}$Rf}\\
At the new FLNR JINR SHE-factory, in a similar investigation as mentioned above for the decay chain from $^{273}$Ds to $^{261}$Rf, Oganessian et al.~observed $^{263}$Rf, terminating three of six decay chains assigned to the 5$n$ fusion-evaporation channel, $^{275}$Ds, of the reaction $^{48}$Ca+$^{232}$Th ~\cite{Oganessian2024}.
This constitutes the discovery of this hitherto not known darmstadtium isotope. 
For this decay chain, the observation of new isomers is also reported which will be discussed in more detail in section~\ref{Ds}.\\

{\it $^{\it265}$Rf}\\
Utyonkov et al.~collected data for three new decay chains originating from $^{285}$Fl and terminating after five subsequent $\alpha$ decays with $SF$ of $^{265}$Rf~\cite{Utyonkov2018} (see also corresponding subsections below). 
These new decay chains mainly confirmed literature data~\cite{Utyonkov2015,Ellison2010} with the  exception of the $\alpha$ decay of $^{269}$Sg in the third of the new chains (see subsection~\ref{Sg}).\\

{\it $^{\it266,267,268,270}$Rf}\\
The decay chains originating from $^{282}$Nh, $^{287,288}$Mc and $^{294}$Ts are terminated by $SF$, which in the most recent publications by Oganessian et al.~are attributed to the $SF$ decay of $^{266,267,268}$Db~\cite{Oganessian2017}. 
However, the decay by $EC$ of these dubnium isotopes cannot be ruled out, as given in earlier papers of the same group, like in the discovery paper for $^{282}$Nh~\cite{Oganessian2007b} or for $^{266,268}$Db decays in a review paper in the same year (2007)~\cite{Oganessian2007}. 
Similarly, for the endpoint of the $^{294}$Ts decay chain in the discovery paper from 2010, $^{270}$Db is given in Ref.~\cite{Oganessian2010} from 2010, while in a review from 2011 the possibility of $EC$ decay to $^{270}$Rf is considered in Fig.~5 of Ref.~\cite{Oganessian2011a}.

As discussed for $^{258}$Db($EC$)$^{258}$Rf the possible population of excited states in the decay daughter could result in $x$-ray emission due to internal conversion ($IC$) which would be of special interest for atomic charge identification for those chains with no connection to known nuclides (see also subsection~\ref{Db}).  
Clear evidence for this process is, however, still missing, and only $^{267}$Rf is confirmed as the endpoint of the $\alpha$-decay sequence starting with $^{291}$Lv~\cite{Oganessian2004a} (see Fig.~\ref{fig:Nchart96_118}). 
The possibility of $EC$ decay for $^{266,268}$Db populating $^{266,268}$Rf is indicated in Fig.~\ref{fig:Nchart96_118} by faint-red triangles (see also discussion in section~\ref{Db} for $^{266,267,268}$Db).\\

{\it $^{\it267}$Rf}\\
The half-life of fissioning $^{267}$Rf has been updated with improved statistics obtained at the new FLNR JINR SHE-factory~\cite{Oganessian2022c} by Oganessian et al.~in the irradiation of $^{242}$Pu and $^{238}$U with $^{48}$Ca projectiles. 
For the plutonium target, six decay chains terminated by $SF$ of $^{267}$Rf, while in the $^{238}$U irradiation one additional decay chain of this type was observed~\cite{Oganessian2022b} (see subsections \ref{Cn} and~\ref{Fl}).

\subsection{Dubnium - $Z$=105}\label{Db}

{\it $^{\it255}$Db}\\
$^{\it255}$Db is the lightest dubnium isotope for which reports of its observation exist. However, the reported properties differ significantly from each other. The earliest paper mentioning this isotope is a conference report from 1976 by G.N. Flerov~\cite{Flerov1976a},  which just lists the branching ratio of 80\% $\alpha$ decay and 20\% SF together with a half-life of 1.5~s in a table without any mention in the text. 
This is probably the same activity which is mentioned by Oganessian et al.~ in the same year in Ref.~\cite{Oganessian1976}, where two activities with half-lives of 1.2$^{+0.6}_{-0.3}$~s and >1~s are given, produced in the same reactions $^{51}$V+$^{207,206}$Pb, respectively. 
In 1986 in a laboratory report by the same group, $^{255}$Db is mentioned with a half-life of 1.6~s~\cite{Oganessian1983}.

A.-P. Lepp\"{a}nnen reported in his Ph.D. thesis in 2005~\cite{Leppanen2005} on two recoil-SF and one recoil-$\alpha$-SF correlation events, which he assigned to the decay of $^{255}$Db with a half-life of 37$^{+51}_{-14}$~ms. 
This half-life differs by orders of magnitude from the earlier reported values. 

Finally, in 2024 Pore et al.~reported on the detection of 55 fission events and three $\alpha$ decays, which they assign to $^{255}$Db produced as the 2$n$ fusion-evaporation channel in the reaction $^{51}$V+$^{206}$Pb~\cite{Pore2024,Pore2024a}.
The average half-life obtained from both activities, with  $\alpha$ and $SF$ branchings of 8(3)\% and 92(3)\%, respectively, $T_{1/2}$~=~2.6$^{+4}_{-3}$~ms is adopted here in Table~\ref{tab:isotope_list}.
\\

{\it $^{\it256}$Db}\\
In the above-mentioned irradiation of a $^{206}$Pb target with $^{51}$V projectiles at the BGS of LBNL (see subsections \ref{Es}, \ref{Fm}, \ref{Md} and \ref{Lr}), a total of 86 decay chains starting at $^{256}$Db were accumulated by Pore et al.~\cite{Pore2024}.
To extract the decay properties of $^{256}$Db and its $\alpha$ decay daughter $^{252}$Lr, the authors considered only 43 events with $\alpha$ decay energies above a certain threshold ($E_\alpha$>8.75~MeV), which is still substantially more than previously collected by He{\ss}berger et al.~\cite{Hessberger2001} and Nelson et al.~\cite{Nelson2008}. 
The decay properties are in agreement with the earlier findings, leading to a better precision of the measured values.

The authors also discuss the possible population and decay of isomeric states in the daughter nucleus $^{252}$Lr. 
To search for such states, they investigate 43 selected events in a restricted $E_\alpha$ window, using $\alpha_1$($^{256}$Db)-$\alpha_2$($^{252}$Lr) time and energy correlation plots which do not show any sign of the presence of an isomeric state. 
Yet, the population and decay of isomeric states at finite excitation energy can proceed via transitions with lower $Q_\alpha$ values. 
The non-considered $\alpha_1$-$\alpha_2$ correlations with $E_{\alpha2}$ below the threshold of 8.75~MeV chosen by the authors could indeed originate from such decay routes.
However, this is not discussed in the paper~\cite{Pore2024}. 
The only statement regarding those events given is that all 86 decay chains were used "To determine the properties of all subsequent daughters in the decay chain, ...".\\

{\it $^{\it256-258}$Db}\\
In an experiment to investigate proton evaporation channels, Lopez-Martens et al.~used the reaction $^{50}$Ti+$^{209}$Bi$\rightarrow^{259}$Db$^*$ to populate $xn$ and $pxn$ evaporation channels (see also section~\ref{Rf}). The data obtained for the three observed $xn$ $ER$s $^{\it256-258}$Db were consistent with the earlier findings of He{\ss}berger et al.~in Ref.~\cite{Hessberger2001}. Some aspects of this measurement are also reported by Kuznetsova et al.~and Yeremin et al.~ in Ref.s~\cite{Kuznetsova2020,Yeremin2019}.

In 2024 Zhao et al.~reported new $\alpha$-decay data for $^{257,258}$Db data, collected at the gas-filled separator SHANS of the IMP, Lanzhou, China, Zhao et al.~in the decay chains of the two bohrium isotopes $^{261,262}$Bh, reproducing the literature values~\cite{ZhaoZ2024}.\\


{\it $^{\it258}$Db}\\
Fig.~\ref{Fig:258Db_decay} shows the decay scheme of $^{258}$Db, illustrating how the various decay paths via $\alpha$ and $\beta$ decay evolve across the $N$\,=\,152 shell gap in the vicinity of the $Z$\,=\,100 isotope $^{252}$Fm. 
Together with its $\alpha$-decay daughter nuclei $^{246}$Es (section~\ref{Es}), $^{250}$Md (section~\ref{Md}) and $^{254}$Lr (section~\ref{Lr}), $^{258}$Db is subject of a detailed decay study reported in two articles focusing separately on its $\beta$($EC$)-~(He{\ss}berger et al.~\cite{Hessberger2016a}) and $\alpha$-decay branch (Vostinar et al.~\cite{Vostinar2019}), respectively. 

Being an odd-odd nucleus in a region with access to high-$J$ orbitals, complex $J^\pi$ (spin and parity) configurations of its low-lying excitation states with consequences for decay-mode competition ($\alpha$-, $\beta$-, $SF$- and intrinsic $\gamma$-decay), $^{258}$Db represents the starting point of a complex decay network (as shown in Fig.~\ref{Fig:258Db_decay}).
The dependence of the various decay modes, as well as decay properties like half-lives and decay energies, provides a rich source of information on the microscopic structure features of the involved nuclei.   

The earlier of the two papers reports on the $^{258}$Db $EC$ decay into excited states and the ground state of $^{258}$Rf~\cite{Hessberger2016a} of the two activities known in $^{258}$Db from earlier $\alpha$-spectroscopy results~\cite{Hessberger2009}, supporting their existence. 
The half-lives deduced for these two states from the measured $\alpha$ decay were $T_{1/2}$ = 4.3(5)~s and 1.9(5)~s. 
The findings for $^{258}$Rf are summarized in the previous subsection~\ref{Rf}. 

The second paper reports on a more detailed analysis of the collected $\alpha$ decays for $^{258}$Db, producing a rather complex spectrum with various components contributing to the two activities. 
From $\alpha$-$\gamma$ coincidences on the basis of decay energy systematics, a tentative level scheme of the low-lying structure in $^{254}$Lr and possible spin and parities in the two states could be inferred. 
In addition, the authors propose the assignment as ground state for $^{258}$Db(2) (the nomenclature of Ref.~\cite{Hessberger2009} was preserved) with $J^\pi$\,=\,(0$^-$) and an improved half-life value of $T_{1/2}$\,=\,2.17(36)~s. $^{258}$Db(1) was then assigned as an isomeric state with $E^*$\,=\,51~keV, $J^\pi$\,=\,(5$^+$) or (10$^-$) and $T_{1/2}$\,=\,4.41(21)~s.\\

{\it $^{\it262}$Db}\\
In 2020, Haba et al. reported on new data for the $\alpha$ decay of $^{262}$Db as the daughter of $^{266}$Bh produced in the reaction $^{248}$Cm($^{23}$Na,5$n$)$^{266}$Bh. The experiment was performed using the rotational wheel detection set-up MANON~\cite{Haba2020} (see also section~\ref{Sg}). $^{262}$Db was first observed in the decay chain of $^{278}$Nh by Morita et al.~\cite{Morita2004b,Morita2007a,Morita2009,Morita2012} and later directly produced in the reaction $^{248}$Cm($^{19}$F,$5n$)$^{262}$Db~\cite{Haba2014}.\\

{\it $^{\it265}$Db}\\
D.~Fernandez et al.~reported in 2023 on fission studies of the compound nucleus $^{265}$Db produced in inverse kinematics, irradiating  $^{238}$U with $^{27}$Al ions\cite{Fernandez2023}. 
Employing the magnetic spectrometer VAMOS++~\cite{Lemasson2023}, the isotopic fission yield distribution was investigated for the highly excited nuclear system. 
Apart from presenting the experimental fission yield distributions, determined by quasi-fission, the authors suggest the observed even-odd staggering for the heavy fragments as a probe to investigate the interplay of quasi-fission and fusion-fission.
No properties of the ground state of $^{265}$Db or its decay are known.\\

\begin{figure}[htb]
\begin{center}
\includegraphics[width=\columnwidth]{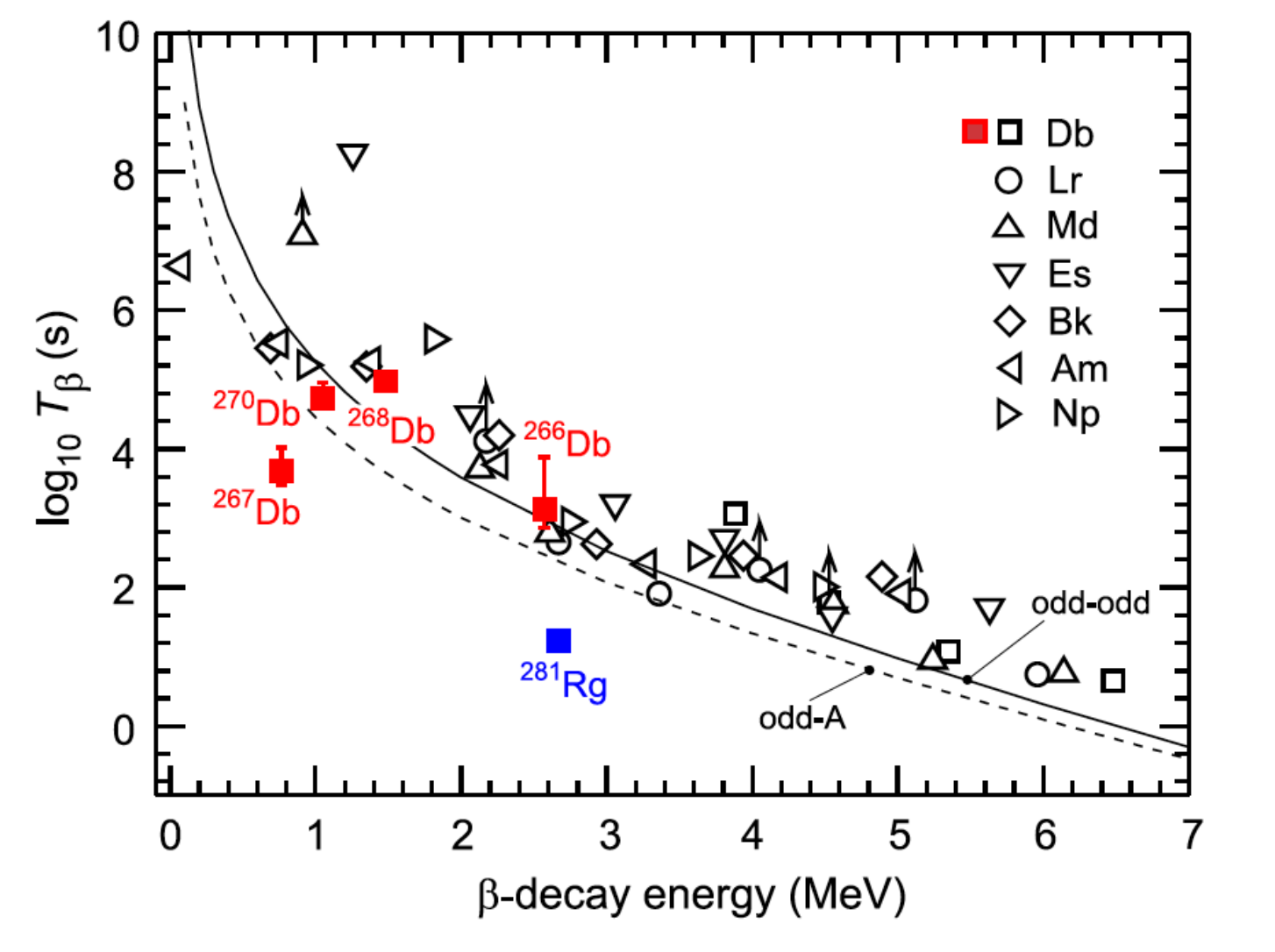}
\end{center}
\vspace{-0,5 cm}
\caption{Partial EC/$\beta^+$-decay half-lives vs. decay energy for known odd–odd isotopes of Np–Db (open symbols). The systematics~\cite{Kolesnikov1980} for odd–odd and odd-A isotopes are shown by lines. Experimental half-lives of terminal isotopes for which spontaneous fission was detected in the decay chains produced in $^{48}$Ca-induced reactions with $^{237}$Np, $^{243}$Am, and $^{249}$Bk targets are given by colored solid squares. EC/$\beta^+$-decay  energies for $^{266-268,270}$Db and $^{281}$Rg were calculated from Ref.~\cite{Muntian2003}. (Figure and caption are taken from Ref.~\cite{Oganessian2015})}
\label{fig:SHN_beta_sys}
\end{figure}
\begin{figure*}[htb]
\begin{center}
\includegraphics[width=0.8\textwidth]{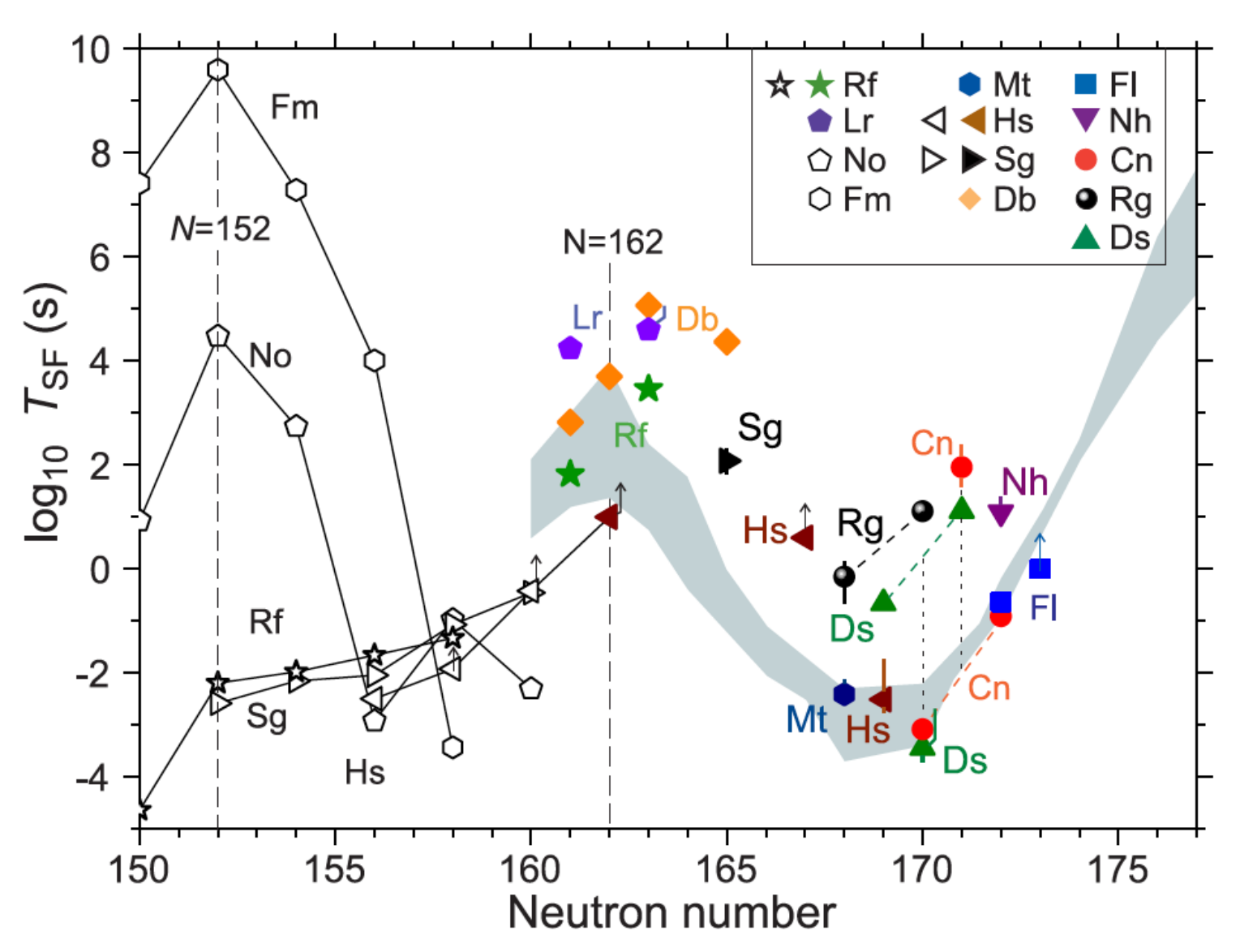}
\end{center}
\vspace{-0,5 cm}
\caption{Common logarithm of partial spontaneous fission half-life
$[{\rm log}_{10} (T_{SF}{\rm [s]})]$ 
vs neutron number for isotopes of elements with $Z$~=~100--114. 
The area of half-lives of even-even isotopes with $Z$~=~104--110, $N$~=~160--164 and $Z$~=~110--114, $N$~=~164--177 predicted in \cite{Lojewski1999} is highlighted in gray.
Partial half-lives of nuclei produced in the $^{48}$Ca-induced reactions with actinide targets are shown by filled symbols. 
(Figure and caption are taken from Ref.~\cite{Oganessian2022d})
}
\label{fig:Fm-Fl_T_sf}
\end{figure*}

{\it $^{\it266-268}$Db}\\
The three isotopes $^{266,267,268}$Db were assigned to SF events, observed as the termination of decay chains of nihonium and moscovium isotopes by Oganessian et al.~\cite{Oganessian2004}, while in Ref.~\cite{Oganessian2007b} the possibility of $\beta$ decay($EC$) is considered for $^{266}$Db, as the detection system used for those measurements was not sensitive to $\beta$ decay (see discussion above). 
In Ref.~\cite{Oganessian2015} the authors discuss the probability of $\beta$ decay of the heaviest isotopes competing with $SF$ observed as chain terminations. 
In Fig.~\ref{fig:SHN_beta_sys} the comparison of the observed half-lives assigned to $SF$ (colored symbols) with measured partial $EC$/$\beta^+$ half-lives shows that for the dubnium isotopes $^{266,268,270}$Db the assigned $T_{SF}$ values are similar to the $T_{\beta}$ value trend of the lighter odd-$Z$ nuclei, indicating a finite probability of $\beta$ decay to occur. 
Part of the observed $SF$ decays would then belong to the rutherfordium daughter isotopes, at least for $^{266}$Db and $^{268}$Db.

For $^{266}$Db one event in addition to the one previously observed in the $^{282}$Nh chain~\cite{Oganessian2007}, was reported in Ref.~\cite{Oganessian2022d} with a half-life shorter by a factor of $\approx$25. 
This is, however, still consistent with a single decay activity according to the Poisson statistics analysis in the prescription of Ref.~\cite{Schmidt2000}, leading to a combined half-life of 11$^{+21}_{-4}$~min (see Table~\ref{tab:isotope_list}).

From the shorter half-lives observed for the two isotopes $^{267}$Db and $^{281}$Rg the authors deduce that $SF$ would be more likely the dominating decay mode.
They also discuss decay time systematics across the $N$=162 shell gap. 
A striking observation are the long half-lives of up to 16$^{+6}_{-4}$ h for $^{268}$Db as given in a recent paper on the first results obtained at the new SHE-factory of FLNR JINR~\cite{Oganessian2022c}.  
Here the authors report on 55 new decay chains obtained for $^{288}$Mc and six new chains for $^{289}$Mc. 

In addition to the known decay features, they observed also $\alpha$ decay of $^{268}$Db leading to the discovery of the new isotope $^{264}$Lr (see subsection~\ref{Lr}). 
This new assignment leads to changes in the decay branching ratios and consequently partial decay half-lives. 
The same $\alpha$ activity was found for $^{268}$Db also in the second irradiation of $^{243}$Am with $^{48}$Ca projectiles during this first series of experiments at the SHE-factory with 55 new $^{288}$Mc decay chains collected (see Ref.~\cite{Oganessian2022d}).

A possible $EC$-decay of the odd-even isotope $^{267}$Db with $T_{1/2}$ = 1.2$^{+58}_{-6}$~h to the daughter nucleus  $^{267}$Rf with a very similar half-life of $T_{1/2}$ = 1.3$^{+23}_{-5}$~h is more unlikely, as supported also by the EC/$\beta^+$-decay half-life systematics in Fig.~\ref{fig:SHN_beta_sys}.
The close-by lawrencium isotopes $^{260,261,262}$Lr exhibit a similar decay mode pattern with $\beta$ decay being observed for the two odd-odd isotopes while the even-odd isotope in between decays exclusively by fission (see Fi.g~\ref{fig:Nchart96_118}).
The half-life trends are also similar with 0.4~h, 1.3~h and 16~h for $^{266,267,268}$Db, and 180~s, 39~min and 4~h for $^{260,261,262}$Lr, respectively.
To disentangle $\beta$ decay ($EC$) from $SF$, correlations by $CE$ emission and the coincident $x$-rays, originating from possibly populated excited states in the daughter nucleus, can be employed, as applied in the case of $^{258}$Db and its $\beta$-decay daughter $^{258}$Rf by He{\ss}berger et al.~in ref.~\cite{Hessberger2016a}.
In addition, the correlated $x$-rays emitted with the $CE$ can provide the identification of the atomic charge $Z$ of the decaying nuclide, which would yield the still missing direct experimental proof for nuclides of elements exclusively produced in hot fusion, for $^{48}$Ca induced reactions with actinide targets.
This concerns all elements from flerovium to oganesson, the five heaviest elements with $Z$=114 to 118.

The $SF$ half-life systematics, including the new data, is shown in Fig.~\ref{fig:Fm-Fl_T_sf} taken from Ref.~\cite{Oganessian2022d}, reporting on this second experiment. 
Extending the investigation to higher beam energies, the $4n$ ($^{287}$Mc) and, for the first time, the 5$n$ fusion-evaporation channel with the new isotope $^{286}$Mc were observed.\\

{\it $^{\it270}$Db}\\
In the irradiation of $^{249}$Bk with $^{48}$Ca, Khuyagbaatar et al.~investigated the decay sequence starting with $^{294}$Ts, which was, as mentioned above (subsection~\ref{Lr}), terminated by $SF$ of $^{266}$Lr~\cite{Khuyagbaatar2019}. The $\alpha$ decay of $^{270}$Db, detected in the two longer decay chains, was not observed in the earlier experiments at the DGFRS of FLNR JINR~\cite{Oganessian2013b} (see also section~\ref{Ts}).
The $Q_\alpha$ value of 8019(30)~keV (see Table~\ref{tab:isotope_list}) deduced from two measured $^{270}$Db $\alpha$ decays is relatively low as compared to the Atomic Mass Evaluation (AME) 2020 prediction of 8310(200)\#~keV~\cite{Wang2021} (see also Fig.~\ref{fig:Qa_94_118}).

\subsection{Seaborgium - $Z$=106}\label{Sg}

For the heaviest nuclei, the nuclear structure information collected is, due to the low production cross-section, often limited to basic decay properties, being most probably rather incomplete. 
Starting with the seaborgium isotopes no information on the $\beta$-decay properties is available for the higher-$Z$ nuclides, as mentioned before, e.g., in the discussion regarding possible $EC$-decay of the heaviest dubnium isotopes in the previous subsection (\ref{Db}), or more generally in section~\ref{weak}.

At the time of the first submission of this review, seaborgium was also the only even-$Z$ element for $Z$ from 100 (fermium) to 110 (darmstadtium), for which no $K$-isomer has been reported yet, before in June 2025 Mosat et al. reported on the first observation of the new isotope $^{257}$Sg~\cite{Mosat2025}. 

In addition, new data, including the first synthesis of $^{268}$Sg, have been recorded for some of the seaborgiums forming the endpoints of decay chains originating from isotopes of heavier elements, mainly due to the first experiments performed at the new SHE-factory of FLNR JINR, as for many of the lighter superheavy elements.\\

{\it $^{\it257}$Sg}\\
In an irradiation of $^{206,208}Pb$, Mosat et al.~observed for the first time the new isotope $^{257}$Sg by detecting its $SF$ as well as a small $\alpha$-decay branch with a branching $SF$-$\alpha$- branching ratio of 90:10. 
The analysis of the $^{208}$Pb-target reaction employing digital electronics and pulse shape analysis (PSA) they revealed strong evidence for the observation of a $K$-isomer in $^{259}$Sg being the first one in a seaborgium isotope.
For more details beyond the data reported in Table~\ref{tab:isotope_list}, see the original publication~\cite{Mosat2025}.\\

{\it $^{\it264}$Sg}\\
For $^{264}$Sg Galkina et al.~investigated fission properties of this nucleus produced in the fusion reaction $^{32}$S+$^{232}$Th at the three different excitation energies $E^*$=45~MeV, 59~MeV, and 75~MeV, around the Coulomb barrier~\cite{Galkina2021}. While a substantial part of the reaction products is found to be due to quasifission, an analysis of the Total Kinetic Energy (TKE), based on the fission mode classification by Brosa et al.~\cite{Brosa1990}, leads the authors to the conclusion that a large contribution to a symmetric mass split of up to 80\% results from compound-nucleus fission.\\

{\it $^{\it265}$Sg}\\
As a $\alpha$-decaying member of the two $^{273}$Ds decay chains, produced at the FLNR JINR SHE-factory in the 5$n$ fusion-evaporation channel of the reaction $^{40}$Ar+$^{238}$U, additional two decay events were collected for $^{265}$Sg with respect to earlier results~\cite{Oganessian2024}. For More details of this experiment see subsection~\ref{Ds}.\\

{\it $^{\it267}$Sg}\\
Also for $^{267}$Sg new data has been reported from an experiment at the new FLNR JINR SHE-factory~\cite{Oganessian2024}.
The deduced $Q_\alpha$ value of 8400(20)~keV is relatively low as compared to the AME 2020 prediction of 8630(210)\#~keV~\cite{Wang2021} (see also Fig.~\ref{fig:Qa_94_118}).
A more detailed discussion on measured properties for $^{275}$Ds and its decay products will be given in section~\ref{Ds}. 
Three of the six $^{275}$Ds decay chains collected in this experiment exhibit $\alpha$ decay of $^{267}$Sg with relatively long decay times, ranging from 760~s to 970~s, while for the three fission events decay times vary from $\approx$1\,~s (2 events) to 360~s.  
The authors propose an isomeric state decaying by fission and the ground state decaying by $\alpha$ emission with a half-life of about 10 min (see Table~\ref{tab:isotope_list}), exceeding the detection time range of 300~s of the earlier chemistry experiment~\cite{Dvorak2006,Dvorak2008} where this nucleus was observed before without an indication for an isomeric state.\\

{\it $^{\it268}$Sg}\\
Irradiating a $^{232}$Th target with $^{48}$Ca projectiles, Oganessian et al.~observed in one of the series of experiments carried out at the new FLNR SHE-factory three decay chains, consisting of two subsequent $\alpha$ decays followed by $SF$ which they assigned to start from the 4$n$ fusion-evaporation residue $^{276}$Ds~\cite{Oganessian2023}. 
All three nuclei in this chain, including as its termination $^{268}$Sg, were hitherto unknown (see subsection~\ref{Ds}).\\

{\it $^{\it269}$Sg}\\
In 2018, Utyonkov et al.~reported on a second investigation of the reaction $^{240}$Pu($^{48}$Ca,$3n$)$^{285}$Fl, with $^{269}$Sg being populated in the decay chain, at a beam energy of $E_{beam}$=250~MeV~\cite{Utyonkov2018} with respect to the earlier measurement at $E_{beam}$\,=\,245 MeV (see subsection~\ref{Fl}).\\

{\it $^{\it271}$Sg}\\
$^{271}$Sg is the last but one member of the $^{287}$Fl decay chain and exhibits $\alpha$-decay as well as $SF$. It was recently again observed at the new SHE-factory of FLNR JINR in an irradiation of $^{242}$Pu~\cite{Oganessian2022b} (see subsection~\ref{Fl}). 
As $^{267}$Rf and $^{275}$Hs, discussed above and below, also $^{271}$Sg is produced following the $\alpha$-decay of $^{279}$Ds with seven and two new decay chains observed in irradiation of $^{242}$Pu and of $^{238}$U, respectively~\cite{Oganessian2022b} (see subsections \ref{Fl} and~\ref{Cn}).
The values in Table~\ref{tab:isotope_list} have been updated accordingly, taking the values from~\cite{Oganessian2022b} which include all previously accumulated data.

\subsection{Bohrium - $Z$=107}\label{Bh}

Apart from a re-evaluation of an earlier measurement for $^{274}$Bh, for the bohrium isotopes new data were collected in experiments at the gas-filled separator SHANS of the Institute for Modern Physics (IMP), Lanzhou, China, for $^{261,262}$Bh~\cite{ZhaoZ2024}, and at the new SHE-factory of FLNR JINR with the respective bohrium isotopes being members of the decay chains of heavier nuclei.\\

{\it $^{\it261,262}$Bh}\\
At the gas-filled separator SHANS of the IMP, Lanzhou, China, Zhao et al.~ synthesized the two bohrium isotopes $^{261,262}$Bh in the reaction channels $^{209}$Bi($^{54}$Cr,$xn$)$^{263-x}$Bh, with $x$\,=\,1 and 2, collected at two beam energies a total of 243 and 26 new decay chains for $^{262}$Bh and $^{261}$Bh, respectively. The new data are in agreement with literature values and, in particular, the two decay activities for $^{262}$Bh were reproduced. \\

{\it $^{\it270}$Bh}\\
In the first campaign of experiments at the Dubna SHE-factory, Oganessian et al.~reported in 2022 on the discovery of $^{286}$Mc, observing a decay chain of which $^{270}$Bh is the last $\alpha$-decaying member~\cite{Oganessian2022d} (see subsection~\ref{Mc}), while it was first observed in the $\alpha$-decay sequence of $^{282}$Nh in 2007~\cite{Oganessian2007b}. 
In addition to the earlier observed event, a second $^{270}$Bh $\alpha$ decay was reported in Ref.~\cite{Oganessian2022d}.\\

{\it $^{\it271}$Bh}\\
Similar as for $^{270}$Bh, Oganessian et al.~collected also for $^{271}$Bh new data in the decay products following $^{287}$Mc (see subsection~\ref{Mc}), which they reported in 2022~\cite{Oganessian2022d}, again doubling the number of observed events from two to four.\\

{\it $^{\it272}$Bh}\\
Until the last $^{288}$Mc synthesis experiment performed at the FLNR SHE-factory~\cite{Oganessian2022d}, $^{272}$Bh was the last $\alpha$-decaying member of the decay chain, before the $^{268}$Db $\alpha$ decay was extracted from the new data (see subsection~\ref{Db}). In an earlier re-investigation, performed at the Berkeley Gas-filled Separator (BGS), Gates et al.~had used the troichoidal mass spectrometer to establish the atomic mass number A of $^{288}$Mc~\cite{Gates2018}, where they also detected $^{272}$Bh as the last $\alpha$ decay of the two observed chains (see subsection~\ref{Mc}).\\

{\it $^{\it274}$Bh}\\
In 2019, Khuyagbaater et al.~published more details with respect to the first publication in 2015~\cite{Khuyagbaatar2015} on the reaction $^{48}$Ca+$^{249}$Bk to synthesize $^{294}$Ts~\cite{Khuyagbaatar2019}. 
$^{274}$Bh is a member of the decay chain (see subsection~\ref{Ts}).

\subsection{Hassium - $Z$=108}\label{Hs}

For the new hassium data, the same is valid as stated in the introduction of the previous subsection~\ref{Bh} for the bohrium isotopes.\\

{\it $^{\it269}$Hs}\\
As the $\alpha$-decay daughter of $^{273}$Ds, $^{269}$Hs and its $\alpha$ decay were discussed in a recent paper of an experiment at the new FLNR JINR SHE-factory~\cite{Oganessian2024}. 
Despite the fact that only for one chain an indication of this decay by an escape-$\alpha$ signal was observed, the authors discuss the presence of an isomeric state with spin 1/2 and a half-life of 2.8$^{+3.6}_{-1.3}$~s above the ground state with spin 9/2 and a half-life of 13$^{+10}_{-4}$~s (see Table~\ref{tab:isotope_list}) on the basis of existing earlier data for the $^{277}$Cn decay chain~\cite{Hofmann2002,Sumita2013} and its direct production~\cite{Dvorak2006,Dvorak2008}. For more details on the properties the whole decay chain, see subsection~\ref{Ds}.\\

{\it $^{\it271}$Hs}\\
A similar scenario as for $^{269}$Hs has been proposed for $^{271}$Hs, again observed at the FLNR JINR SHE-factory in six decay chains following the $\alpha$ decay of $^{275}$Ds which was newly synthesized as the 5$n$ fusion-evaporation channel of the reaction $^{48}$Ca+$^{232}$Th~\cite{Oganessian2024}.
The authors propose again an isomeric state with low spin (3/2) and $T_{1/2}$~=~7.1$^{+8.4}_{-2.5}$~s above a ground state with high spin (11/2) and $T_{1/2}$~=~46$^{+56}_{-16}$~s (see Table~\ref{tab:isotope_list}).
More details on the interpretation of the decay chain and the population of isomeric states by $\alpha$ decay is discussed in section~\ref{Ds}.\\

\begin{figure*}[htb]
\begin{center}
\includegraphics[width=0.8\textwidth]{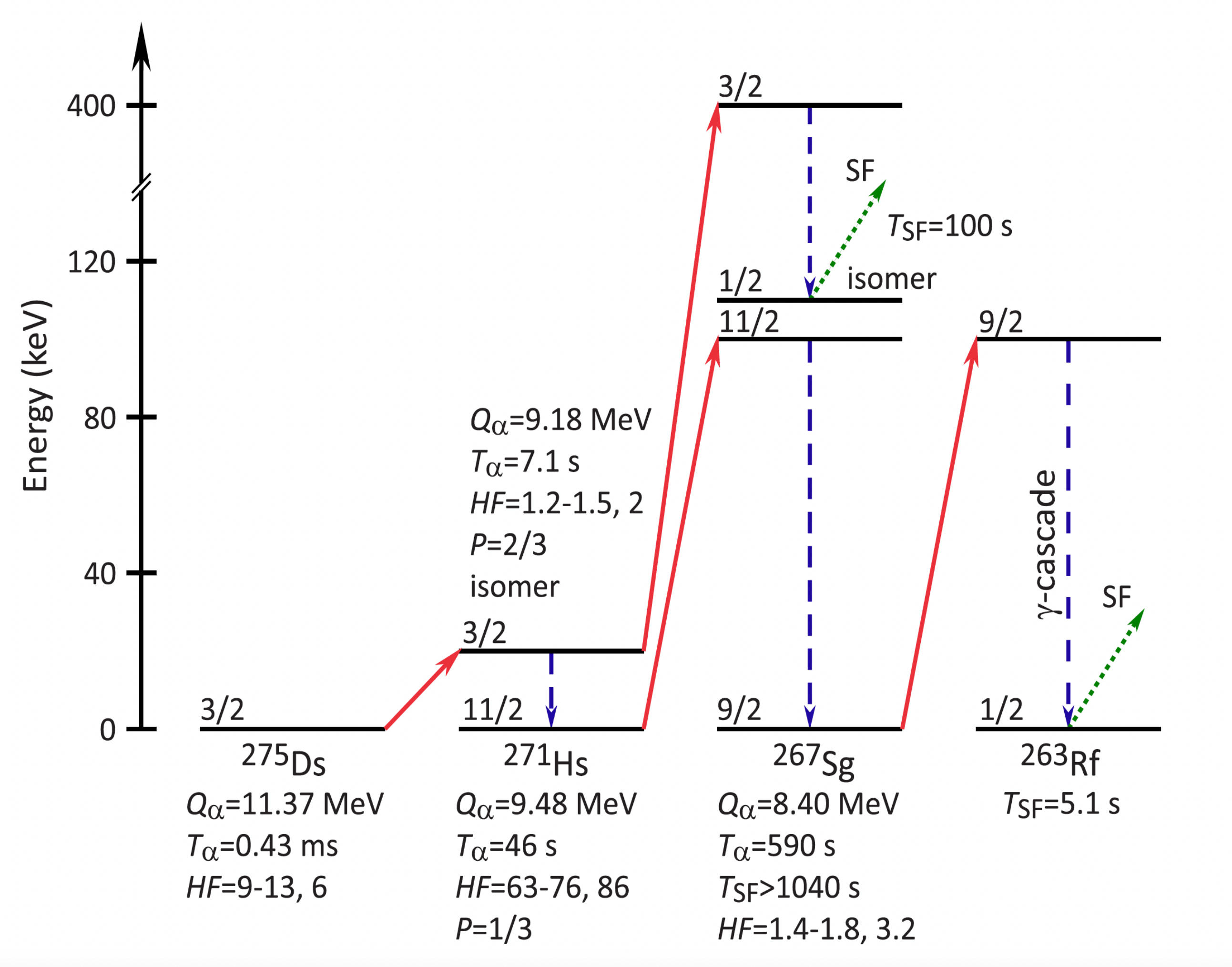}
\end{center}
\vspace{-0,5 cm}
\caption{A proposed experimental and calculated decay scheme for $^{275}$Ds based on present and previously published data~\cite{Dvorak2006,Dvorak2008}. 
Alpha or $\gamma$ transitions to approximate one-quasiparticle states and $SF$ decays are shown by red (solid), blue (dashed), and green (dotted) lines, respectively. 
Alpha-decay energies ($Q_\alpha$) and half-lives are provided for states decaying by $\alpha$ decay ($T_\alpha$) or spontaneous fission ($T_{SF}$).
Hindrance factors $H_F$~=~$T^{exp}_{1/2}$/$T^{calc}_{1/2}$ were derived from experimental and calculated half-lives according to Ref.s~\cite{Oganessian2004a,Qi2009,Xu2022,Ismail2022,Luo2022} (see discussion in~\cite{Oganessian2024}).
Population probabilities P of energy levels for $^{271}$Hs are based on experimental data~\cite{Dvorak2006,Dvorak2008} and this paper~\cite{Oganessian2024}.
(Figure and caption are taken from Ref.~\cite{Oganessian2024})}
\label{fig:275Ds_decay}
\end{figure*}

{\it $^{\it272}$Hs}\\
Irradiating a $^{232}$Th target with $^{48}$Ca projectiles, Oganessian et al.~observed three decay chains consisting of two subsequent $\alpha$ decays, followed by a $SF$, which they assigned to start from the 4$n$ fusion-evaporation residue $^{276}$Ds populated in this reaction~\cite{Oganessian2023}. 
All three nuclei in this chain, including as its termination $^{268}$Sg, were hitherto unknown (see subsections~\ref{Sg} and~\ref{Ds}).\\

{\it $^{\it273}$Hs}\\
As mentioned earlier for the lighter isotopes, Utyonkov et al.~collected data for three new decay chains originating from $^{285}$Fl and terminating after five subsequent $\alpha$ decays, of which $^{273}$Hs is the 4$^{\rm th}$ member, with $SF$ of $^{265}$Rf~\cite{Utyonkov2018} (see also corresponding subsections below and above).\\

{\it $^{\it275}$Hs}\\
Also for $^{275}$Hs, as a member of the decay chains of $^{283}$Cn and $^{287}$Fl, new $\alpha$-decay events were observed at the new SHE-factory of FLNR JINR in the irradiation of $^{242}$Pu and of $^{238}$U targets with $^{48}$Ca projectiles ~\cite{Oganessian2022b} (see subsections~\ref{Cn} and~\ref{Fl}).\\

{\it $^{\it277,278}$Hs}\\
In 2019, Khuyagbaatar et al.~observed four decay chains produced in the reaction $^{48}$Ca+$^{249}$Bk of which they assigned two each as starting from $^{294}$Ts and $^{293}$Ts, respectively. For the latter, they assumed the possibility of $^{277}$Mt $\beta$ decay leading to $^{277}$Hs, similar to what was discussed for $^{266,267,268}$Db in section~\ref{Db}.


\subsection{Meitnerium - $Z$=109}\label{Mt}

Also for meitnerium new or updated data have been reported for five meitnerium isotopes as members of decay chains originating from heavier nuclei, for measurements performed by Oganessian et al.~at the new Dubna SHE-factory, as well as for earlier experiments re-evaluated by Khuyabaatar et al.\\

{\it $^{\it274,275,276}$Mt}\\
For the three meitnerium isotopes $^{274,275,276}$Mt as members of the decay chains of the respective moscovium isotopes, additional data were accumulated at the new SHE-factory at FLNR JINR in 2022~\cite{Oganessian2022d} (see subsection~\ref{Mc}). In this experiment for $^{274}$Mt no $\alpha$ decay energy, in addition to the previously observed one event was measured, and only indirect information on its decay time was deduced.\\

{\it $^{\it277,278}$Mt}\\
In 2019, Khuyagbaatar et al.~\cite{Khuyagbaatar2019} reported, in addition to the earlier published $^{294}$Ts decay data~\cite{Khuyagbaatar2014}, hitherto unpublished data for $^{293}$Ts. $^{278}$Mt is an $\alpha$-decaying member of the $^{294}$Ts decay chain and $^{277}$Mt terminates the $^{293}$Ts decay chain, while the authors suggest, in line with what was discussed in subsection~\ref{sf-b}, alternatively $EC$ decay to $^{277}$Hs which then decays by $SF$~\cite{Khuyagbaatar2019} (see subsection~\ref{Ts}).

\subsection{Darmstadtium - $Z$=110}\label{Ds}
The possibly most interesting new data is the synthesis of new darmstadtium isotopes $^{275}$Ds, adding another stepping stone in between the island populated mainly by reactions of $^{48}$Ca projectiles on actinide target nuclei. 
A complete connection to the "mainland" with the heaviest nuclei produced in cold fusion reactions, resulting in ER at low excitation energy, would strengthen the up to now missing direct $Z$-identification.\\

{\it $^{\it273}$Ds}\\
Similar to the interpretation of the six decay chains observed for $^{275}$Ds (see below) and obviously encouraged by the apparent consistency, Oganessian et al. constructed a similar decay scenario for the decay sequence starting from $^{273}$Ds. 
To this end, they used two decay chains of type $ER$-$\alpha$-$\alpha$-$SF$, which they observed in the irradiation of $^{238}$U with $^{40}$Ar projectiles at the FLNR JINR SHE-factory~\cite{Oganessian2024}, combined with available literature data. 
In Fig.~5 and Table 3 of Ref.~\cite{Oganessian2024} they propose new isomers in all four members, $^{273}$Ds, $^{269}$Hs, $^{265}$Sg and $^{261}$Rf, of the decay sequence (see also sections~\ref{Rf}, \ref{Sg} and \ref{Hs}).
This scenario also suggests that $SF$ occurs for $^{261}$Rf only for the low-spin 3/2 isomeric state and is suppressed for the high-spin (11/2) ground state (see also the discussion below for the $^{275}$Ds decay chain and $^{267}$Sg).\\

{\it $^{\it275}$Ds}\\
The hitherto unknown isotope $^{275}$Ds was synthesized for the first time in a recent experiment at the FLNR JINR SHE-factory in the reaction $^{48}$Ca+$^{232}$Th~\cite{Oganessian2024}.
A total of six decay chains were accumulated of which three were of type $ER$-$\alpha$-$\alpha$-$SF$, ending in fission of $^{267}$Sg. 
The same number of decay chains was recorded showing the decay sequence $ER$-$\alpha$-$\alpha$-$\alpha$-$SF$, terminated by $SF$ of  $^{263}$Rf. (see also sections~\ref{Rf}, \ref{Sg} and \ref{Hs}). 

For the second decay chain member, $^{271}$Hs, two different decay times correlated to two different $\alpha$ decay energies were observed.
For $^{267}$Sg the three observed $SF$ events exhibited decay times similar to what had been observed in chemistry experiments before~\cite{Dvorak2006,Dvorak2008}, while the decay times of the three $\alpha$ decays of this isotope were detected with an average decay time of about 10 minutes, a time interval much longer then the maximum time range of 300~s accessible for the chemistry studies reported by Dvorak et al.~in 2006 and 2008~\cite{Dvorak2006,Dvorak2008}.

These findings were interpreted on the basis of model predictions for $Q_\alpha$ values, decay times, the competition between $\alpha$ decay and $SF$ and spin values of ground and excited states, referring to model calculations within the Di-Nuclear System (DNS) and the Two-Center Shell Model (TCSM) approach~\cite{Rogov2019,Rogov2021,Adamian2016}.
In a recent paper, the same authors discussed $SF$ and $\alpha$ decay from $K$-isomeric states, calculating the properties for the heaviest nuclear for which such states have been observed experimentally, $^{266}$Hs and $^{270}$Ds (see Ref.~\cite{Ackermann2024}).

As a result, the possible decay sequence from $^{275}$Ds to $^{263}$Rf, shown in Fig.~\ref{fig:275Ds_decay} (taken from~\cite{Oganessian2024}), has been constructed.  
Here the authors follow the assumption that $\alpha$ decay between analog spin states
is favored and $SF$ from high-spin states is hindered. This leads to new isomeric states in $^{271}$Hs and $^{267}$Sg, respectively, while in this scenario the isomer in $^{267}$Sg disintegrates by $SF$ only, and the ground state decays exclusively by $\alpha$ emission.  
This situation is similar to other cases, apart from $^{261}$Rf in the previous paragraph, like, e.g., $^{259}$Sg~\cite{Antalic2015} or $^{247}$Md~\cite{Hessberger2022}, where $SF$ is also observed to decay mainly from the excited low-spin state (see also the discussion in subsection~\ref{a-sf}.

To conclusively address the scenarios proposed for $^{273,275}$Ds here, apart from a larger body of data, a more self-consistent model approach than DNS or TCSM, explaining these intriguing observations based on firmer grounds, would be highly desirable.\\

{\it $^{\it276}$Ds}\\
Before the discovery and investigation of $^{275}$Ds, $^{276}$Ds had already been synthesized at the FLNR JINR SHE-factory as the 4$n$ fusion-evaporation channel of the same reaction, $^{48}$Ca+$^{232}$Th, at lower beam energies~\cite{Oganessian2023}.
The three beam energies used in total allowed the construction of an excitation function for an excitation energy range from $E^*$\,=\,35~MeV to 50~MeV, with the maximum located at 40~MeV.

In total seven events consisting of four $ER$-$SF$ and three $ER$-$\alpha$-$\alpha$-$SF$ decay sequences were collected.
The isotopes $^{272}$Hs and $^{268}$Sg were also observed for the first time.
The authors support the isotope assignment by comparing properties like $\alpha$ decay energies and partial $\alpha$ and $SF$ half-lives to neighboring isotopes. 
In particular, the systematic behavior of partial $SF$ half-life as shown in Fig.~\ref{fig:Fm-Fl_T_sf} (Fig.~3 in Ref.~\cite{Oganessian2022d}) and Table II in Ref.~\cite{Oganessian2022d}~show the onset of a rising $\log_{10} T_{SF}$ value towards lower neutron numbers with $\log_{10} T_{SF}$~=~-0.357~s for $^{276}$Ds (166 neutrons) as compared to the value for $^{279}$Ds (169 neutrons) of -0.658 s approaching the value for $^{279}$Rg (168 neutrons) of -0.148~s, lying, however, a bit below the general trend.
An update of Fig.~\ref{fig:Fm-Fl_T_sf} was not provided by Oganessian et al.~in Ref.~\cite{Oganessian2024}.\\

{\it $^{\it277}$Ds}\\
Utyonkov et al.~collected additional data for $^{277}$Ds, being an $\alpha$-decaying member of the $^{285}$Fl decay chain~\cite{Utyonkov2018} (see subsection~\ref{Fl}).\\

{\it $^{\it279}$Ds}\\
Also for $^{279}$Ds, new $\alpha$-decay events were observed at the new SHE-factory of FLNR JINR in the irradiation of $^{242}$Pu and of $^{238}$U, as member of the decay chains of $^{283}$Cn and $^{287}$Fl~\cite{Oganessian2022b} (see subsections~\ref{Cn} and~\ref{Fl}). 
The authors claim that the $SF$ branch of $^{279}$Ds was observed in terms of ER-$SF$ correlations, where one or two $\alpha$ decays for the $^{238}$U and $^{242}$Pu irradiation, respectively, were missing due to the limited $\alpha$-particle detection efficiency.    
\\

{\it $^{\it280}$Ds}\\
By observing the hitherto unknown $\alpha$-decay branch of $^{284}$ in the $^{288}$Fl decay chain (see also subsection \ref{Fl} and \ref{Cn}), S{\aa}mark-Roth in 2021 at the GSI gas-filled separator TASCA discovered the $SF$-decaying, new isotope $^{280}$Ds~\cite{Samark-Roth2021}.

\subsection{Roentgenium - $Z$=111}\label{Rg}
For three roentgenium isotopes, new or updated data were collected, being members of decay sequences of heavier nuclides.\\

{\it $^{\it279}$Rg}\\
Being the third member of the $^{287}$Mc decay chain, $^{279}$Rg was also observed in the recent experiment at the FLNR JINR SHE-factory~\cite{Oganessian2022d}. 
Apart from two additional $\alpha$-decay events, one $SF$ was observed for the first time for this roentgenium isotope, leading to an $\alpha$-decay branch of 87$^{+5}_{-19}$\% (see Table~\ref{tab:isotope_list}).
Following the authors' argumentation, the $SF$ assignment is supported by a missing $\alpha$ decay being very unlikely, and by decay time systematics.
The partial $SF$ half-life resulting from the observation of the new $SF$ branch in $^{279}$Rg is included in the half-life systematics in Fig.~\ref{fig:Fm-Fl_T_sf}.\\

{\it $^{\it281,282}$Rg}\\
In 2019, Khuyagbaatar et al.~reported the $^{294}$Ts decay hitherto unpublished data for $^{293}$Ts. $^{281,282}$Rg are the first $\alpha$-decaying members of the decay chains of these nuclei~\cite{Khuyagbaatar2019} (see subsection~\ref{Ts}).

\subsection{Corpernicium - $Z$=112}\label{Cn}
For the last but one element for which an isotope was produced by cold fusion, new data were accumulated for all five isotopes synthesized in hot fusion reactions.\\  

{\it $^{\it281}$Cn}\\
In 2018, Utyonkov et al.~reported additional data for $^{281}$Cn, being an $\alpha$-decaying member of the $^{285}$Fl decay chain~\cite{Utyonkov2018} (see subsection~\ref{Fl}).\\

{\it $^{\it282,283,284}$Cn}\\
As mentioned repeatedly above, Oganessian et al.~irradiated at the new SHE-factory of FLNR JINR targets of $^{242}$Pu and of $^{238}$U with $^{48}$Ca, each at two beam energies~\cite{Oganessian2022b} (see also section~\ref{Fl}).

$^{282}$Cn with its 100\% $SF$ branch, was only observed as $\alpha$-decay daughter of the 4$n$ ER $^{286}$Fl, for the plutonium target.
Fourteen new $SF$ decays were accumulated.
In the same experiment, $^{283}$Cn was produced directly as the $3n$ ER in the reaction with the uranium target, and as the first member of the $\alpha$-decay chain of  $^{287}$Fl.
For the direct production, 16 new decay chains were observed, improving the uncertainties of experimental decay properties, including decay energies, half-lives and branching ratios as shown in Table~\ref{tab:isotope_list}.
The deduced $Q_\alpha$ value of 9667(15)~keV for $^{283}$Cn is relatively low as compared to the AME 2020 predictions~\cite{Wang2021} (see also Fig.~\ref{fig:Qa_94_118}).

In an investigation of flerovium decay chains, new data have been collected in two experimental runs at the GSI gas-filled separator TASCA for $^{282,284}$Cn~\cite{Samark-Roth2021,Samark-Roth2023} and $^{285}$Cn~\cite{Cox2023} (see also section~\ref{Fl}).
Also, the $Q_\alpha$ value for those two isotopes is lower than the values provided by Ref.~\cite{Wang2021}. 
This leads, together with the value from Ref.\cite{Oganessian2022b} for $^{283}$Cn to a relatively pronounced gap between the flerovium and copernicium $Q_\alpha$ values in this region (see Fig.~\ref{fig:Qa_94_118}).

Terminating the $^{290}$Lv of the decay chain by $SF$ $^{282}$Cn was observed in a very recent experiment at the BGS of LBNL~\cite{Gates2024} (see subsection~\ref{Lv}).
\\

\subsection{Nihonium - $Z$=113}\label{Nh}

{\it $^{\it282}$Nh}\\
The lightest nihonium isotope, produced directly in a hot fusion reaction, i.e., $^{237}$Np($^{48}$Ca,$3n$)$^{282}$Nh,  was recently also observed as the $\alpha$-decay daughter in the decay chain of $^{286}$Mc~\cite{Oganessian2022c} (see subsection~\ref{Mc}.\\

{\it $^{\it285,286}$Nh}\\
$^{285,286}$Nh are the second $\alpha$-decaying members of the decay chains of $^{293}$Ts and $^{294}$Ts for which Khuyagbaatar et al.~published updated and new data in 2019~\cite{Khuyagbaatar2019} (see subsection~\ref{Ts}).

\subsection{Flerovium - $Z$=114}\label{Fl}
Flerovium is the first element that was produced in hot fusion only. 
Flerovium isotopes were also among the first superheavy nuclei being investigated at the new Dubna SHE-factory.
But also at GSI and at LBNL, directly and as a decay product of the first livermorium synthesized in Berkeley, flerovium isotopes were investigated recently.\\

\begin{figure*}[htb]
\begin{center}
\includegraphics[width=\textwidth]{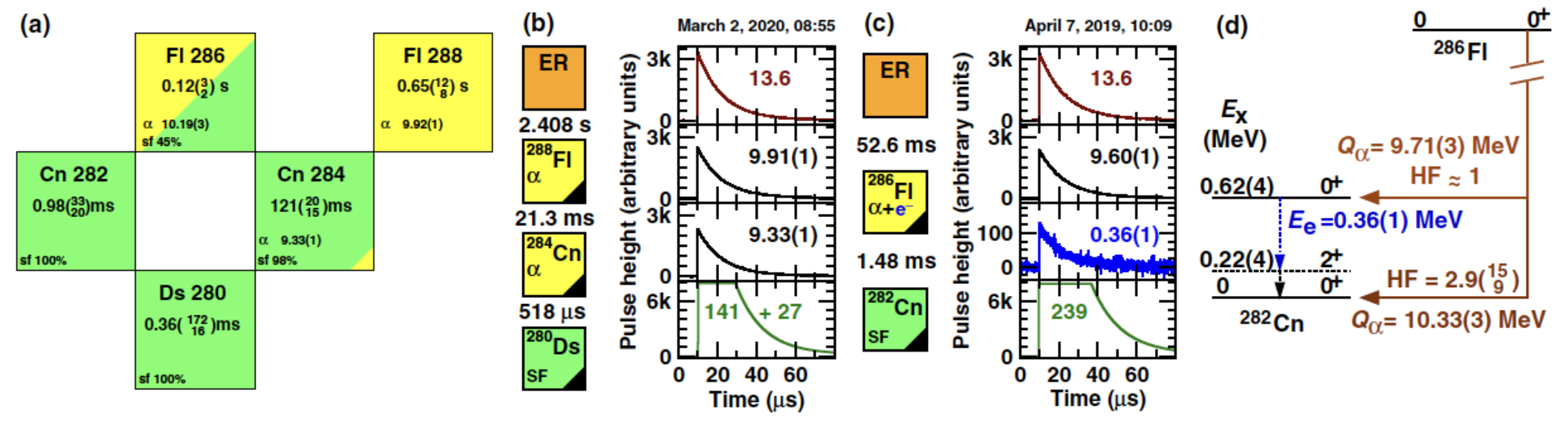}
\end{center}
\caption{(a) Summary of the decay chains of $^{286;288}$Fl (cf. Supplemental Material (of Ref.~\cite{Samark-Roth2021}) and Table II). 
(b),(c) Digitized preamplifier pulses of events associated with decay chains 08 and 02. 
Numbers in the panels are calibrated energies in mega-electron volts. Correlation times are given between recoil implantation (orange squares), $\alpha$ decays (yellow squares), and fission (green squares). 
The 0.36(1)-MeV event in an upstream DSSD (blue) is in prompt coincidence with the 9.60(1)-MeV event in the implantation DSSD. 
Black triangles in the lower right corner of a square indicate detection during beam-off periods. 
(d) Suggested decay sequence of chain 02 through excited states in $^{282}$Cn. Tentative levels and transitions are dashed.
(Figure and caption are taken from~\cite{Samark-Roth2021})}
\label{fig:286Fl_CE}
\end{figure*}

{\it $^{\it284,285}$Fl}\\
Results on the decay properties of the lightest flerovium isotopes $^{284,285}$Fl produced as the 4$n$ and 3$n$ fusion-evaporation channels, respectively, in the reaction $^{48}$Ca+$^{240}$Pu, were first reported by Utyonkov et al.~in 2015~\cite{Utyonkov2015}. While the lighter of the two decays by $SF$, the $ER$ $^{285}$Fl is followed by five subsequent $\alpha$ decays and the $SF$ of $^{264}RF$, terminating the decay chain.

In a new experiment, in order to favor the population of the 4$n$ fusion-evaporation channel, $^{285}$Fl, Utyonkov et al.~chose a beam energy of  $E_{beam}$\,=\,250~MeV~\cite{Utyonkov2018}, corresponding to the higher of the two beam energies (245~MeV ad 250~MeV) used in the first experiment~\cite{Utyonkov2015}.
The 3$n$ channel was indeed not observed at this energy in the earlier measurement, but with the higher beam dose applied for the new run, three $^{285}$Fl decay chains were observed, which show consistent properties with respect to the three events observed earlier at the lower beam energy.
In addition, three $ER$-$SF$ correlations were detected and assigned to $^{284}$Fl.\\

{\it $^{\it286,287,288,289}$Fl}\\
The irradiation of $^{242}$Pu and $^{238}$U targets with $^{48}$Ca beam, at the new SHE-factory of FLNR JINR at two beam energies provided new decay data for $^{287}$Fl and $^{286}$Fl, produced as $3n$ and $4n$ evaporation channels, respectively~\cite{Oganessian2022b}.
The various members of both decay chains with up to six subsequent $\alpha$ decays for $^{287}$Fl, terminated by fission of  $^{267}$Rf for the longest chains, and short decay sequences of $\alpha$-$SF$ and $SF$ of the ER for $^{286}$Fl, have been presented in the respective subsections (see subsections \ref{Rf}, \ref{Sg}, \ref{Hs}, \ref{Ds} and \ref{Cn}).

Concluding on specific decay properties observed in the $^{287,286}$Fl and $^{283}$Cn decay chains, in Ref.~\cite{Oganessian2022b} the possible occurrence of isomeric states is discussed. 
For $^{279}$Ds, e.g., based on the observation of a difference in the newly measured experimental $Q_\alpha$ value of 9.531(15)~MeV, as compared to the theoretical value of 10.11(12)~MeV~\cite{Wang2021}, Oganessian et al.~suggest that $SF$ occurs from an excited, low-spin isomeric state and is hindered for the supposedly high-spin ground state. 
A theoretical scenario supporting this interpretation had been suggested by Kuzmina et al. in Ref.~\cite{Kuzmina2012}.

The competition between $SF$ and $\alpha$ decay, with fission being hindered by the complex quantum structure of particular states, has been discussed in subsection~\ref{a-sf}, and for various examples through this review like, e.g., the case of $^{247}$Md in subsection~\ref{Md}~\cite{Hessberger2022}.
Similarly, in Ref.~\cite{Oganessian2022b} the possible existence of isomeric states in $^{286}$Fl and $^{282}$Cn is discussed on the basis of three alpha decays which exhibit a by $\approx$200~keV\ lower $\alpha$-decay energy, as compared to the main activity.
These findings show the potential offered by facilities like the SHE-factory and others providing high-intensity beams (see section~\ref{outlook}), like here with up to 6~particle\,$\mu$A of $^{48}$Ca beam intensity.
With these capabilities, it will be possible extending the study of the heaviest SHN towards more detailed spectroscopy, providing access to more complex nuclear structure features.

The spontaneous fission decay of $^{286}$Fl has also been seen in the decay chain of $^{294}$Og detected by Brewer et al.~in their first attempt to synthesize oganesson isotopes employing a target composed of a mixture of californium isotopes~\cite{Brewer2018} (see subsection~\ref{Og}).

$^{286}$Fl has very recently also been detected at the BGS of LBNL as $\alpha$-decaying member of the $^{290}$Lv decay chain~\cite{Gates2024} (see subsection~\ref{Lv}).

In 2021 and 2023, S${\mathring{\rm a}}$mark-Roth et al.~reported on experiments employing the fusion-evaporation reactions $^{48}$Ca + $^{242,244}$Pu to study the decay sequences of flerovium isotopes~\cite{Samark-Roth2021,Samark-Roth2023}.
In two experimental runs at the gas-filled separator TASCA of GSI, they collected a total of 29 decay chains in the 3$n$ and 4$n$ fusion-evaporation channels for the isotopes $^{286}$Fl (2 chains), $^{288}$Fl (12 chains) and $^{289}$Fl (15 chains). 
For details see Table 1 in Ref.~\cite{Samark-Roth2023}.
The smoothly increasing trend of $Q_\alpha$ values for the isotope series $^{284}$Cn-$^{288}$Fl-$^{292}$Lv is, according to the authors, indicative of the absence of a possible shell gap at $Z$\,=\,114 (at $N$\,=\,174) as predicted by some theoretical models as the next proton shell gap beyond $^{208}$Pb~\cite{Samark-Roth2021}.

A specific feature observed in one of the decay chains for $^{286}$Fl is the detection of a 360-keV $CE$, marking the population and de-excitation of an excited state in $^{282}$Cn (see Fig.~\ref{fig:286Fl_CE}). 
On the basis of triaxial beyond-mean-field calculations (TBMF)~\cite{Egido2020,Egido2021} and employing GEANT4 simulations to simulate the decay pattern involving this 360~keV $CE$ four scenarios are discussed. 
All of them suggest an excited state populated by the $^{286}$Fl $\alpha$ decay having spin and parity of 0$^+$ (scenario i-iii) or 2$^+$ (scenario iv)~\cite{Samark-Roth2023}.
For decay chain 8 (see supplemental material of Ref.~\cite{Samark-Roth2021}) which is assigned to $^{288}$Fl as $ER$, a second $\alpha$ is observed instead of $SF$ as seen for $^{284}$Cn in all other chains. 
This sequence is then terminated by $SF$ of  $^{280}$Ds.
The TBMF calculations which are applied in two versions, one based on a Gogny, the other on a UNIDEF-Skyrme interaction, underline the importance of triaxiality for the ground and exited states, explicitly 0$+$ states, of livermorium and flerovium isotopes and their decay products.

The details of the fifteen decay chains observed for $^{289}$Fl are discussed in detail in Ref.~\cite{Cox2023}.\\

\subsection{Moscovium - $Z$=115}\label{Mc}
Moscovium isotopes, in particular, $^{288}$Mc produced in $^{48}$Ca-induced reactions, have proven to be good candidates for the investigation of the heaviest SHN, due to their large production cross-sections of up to the order of 15~pbarn.
From the nuclear structure point of view, they are promising candidates as well, considering the long $\alpha$-decay chains reaching down to lawrencium isotopes ($Z$\,=\,103).\\

{\it $^{\it286}$Mc}\\
The lightest moscovium isotope was first produced recently at the SHE-factory of JINR/FLNR employing the new gas-filled separator DGFRS-2~\cite{Oganessian2022a} as 5$n$ fusion-evaporation channel of the reaction $^{48}$Ca + $^{243}$Am, by the observation of one decay chain ending with SF of $^{266}$Db~\cite{Oganessian2022d}.\\

{\it $^{\it287}$Mc}\\
In Ref.~\cite{Oganessian2022c}, the authors report also on new data for the decay of $^{287}$Mc, for which four new chains were observed, including the observation of $SF$ for $^{279}$Rg in one of them.\\

{\it $^{\it288,289}$Mc}\\
In the two experiments at the new SHE-factory of FLNR JINR irradiating $^{243}$Am targets with $^{48}$Ca projectiles, Ref.s~\cite{Oganessian2022c,Oganessian2022d}, 55 new  $^{288}$Mc decay chains were observed for each of them, produced as the 3$n$ fusion-evaporation channel. 
This adds up to a total production of about 210 decay sequences observed for this nucleus, including measurements at GSI and LBNL~\cite{Rudolph2015,Gates2015,Forsberg2016,Gates2018}.
The new data confirm the earlier observation.
The sequence of nuclei populated by subsequent $\alpha$ decay down to $^{268}$Db which terminates the sequence, according to present assignments, by $SF$ (see subsections~\ref{Db} and, in particular,~\ref{sf-b} for the discussion of a possible $SF$-$EC$ decay competition).

In addition to those longer decay chains, a total of 24 shorter ones of type $ER$-$\alpha$-$SF$ or $ER$-$\alpha$-$\alpha$-$SF$ have been observed in the various experiments.
The origin and interpretation of these chains, including scenarios which propose $\beta$ decay ($\beta^+$/$EC$) of $^{288}$Mc's decay daughter and granddaughter nuclei, $^{284}$Nh and $^{280}$Rg~\cite{Forsberg2016}, or the assignment of $^{289}$Mc as the decaying ER~\cite{Oganessian2022d}, are still under debate, even after the new findings at the SHE-factory. 
Despite this situation and valid counterarguments opposing the assignment by Oganessian et al., this latter interpretation has been adopted in Table~\ref{tab:isotope_list}. 

As a possible solution, the authors of Ref.~\cite{Oganessian2022d} propose the employment of instrumentation providing mass identification, like the trochoidal mass spectrometer FIONA of LBNL~\cite{Gates2018} or the MASHA separator at FLNR JINR~\cite{Rodin2014}.
The separator-spectrometer setup S$^3$~\cite{Dechery2015}, approaching the completion of its installation, together with highest beam intensities from the LINAC of GANIL SPIRAL2 and its upgrade, the new injector NEWGAIN~\cite{Ackermann2021}, being presently constructed, would be an ideal tool to tackle this problem (see section~\ref{outlook}).\\

{\it $^{\it289,290}$Mc}\\
In 2019, Khuyagbaatar et al.~reported details for the two $^{294}$Ts decay chains published earlier~\cite{Khuyagbaatar2014} and hitherto unpublished data for $^{293}$Ts. $^{289,290}$Mc are the first $\alpha$-decaying members of the decay chains of these nuclei~\cite{Khuyagbaatar2019} (see subsection~\ref{Ts}).

\subsection{Livermorium - $Z$=116}\label{Lv}
Also for the second heaviest even-$Z$ element, new data have been accumulated for its hitherto lightest isotope $^{290}$Lv, as a member of the $^{294}$Og decay chain, as well as in direct production, and two new even light isotopes, $^{288,289}$Sg have been discovered at the new FLNR JINR SHE factory.\\

{\it $^{\it288,289}$Lv}\\
In July 2025, Oganessian et al.~reported on the discovery the two lighter isotopes $^{288,289}$Sg~\cite{Oganessian2025}.
This is the most recent addition to the list of isotopes discussed in this review.
For more details on these isotopes, beyond the data given in Table~\ref{tab:isotope_list} see the original publication~\cite{Oganessian2025}.\\

{\it $^{\it290}$Lv}\\
In an interesting approach, $^{290}$Lv was detected as a member of the $^{294}$Og decay chain by Brewer et al.~in their first attempt to synthesize oganesson isotopes, employing a target composed of a mixture of californium isotopes~\cite{Brewer2018} (see subsection~\ref{Og}).\\

Very recently; $^{290}$Lv has also been synthesized at the BGS of LBNL.
Irradiating a $^{244}$Pu target with a $^{50}$Ti beam, produced with the 88-inch cyclotron of LBNL, Gates et al.~observed two decay chains of this nucleus with a production cross-section of 0.44~pbarn~\cite{Gates2024}. 
For both events; the same decay sequence was observed, consisting of $\alpha$ decay of $^{290}$Lv and $^{286}$Fl, terminated by $SF$ of $^{282}$Cn.

\subsection{Tennessine - $Z$=117}\label{Ts}
For the last discovered element with the second-highest atomic number; only one activity has been reported recently, studying both known tennessine isotopes.\\

{\it $^{\it293,294}$Ts}\\
In 2014, Khuyagbaatar et al.~had reported two long decay chains, observed at the GSI gas-filled separator TASCA in an irradiation of $^{249}$Bk with $^{48}$Ca beam, consisting of seven consecutive $\alpha$ decays following the ER $^{294}$Ts and being terminated by $SF$ of $^{266}$Lr~\cite{Khuyagbaatar2014}. 
These two chains are by one $\alpha$ decay longer than the four decay sequences observed at the DGFRS of FLNR JINR, which end with $SF$ of $^{270}$Db~\cite{Oganessian2013b}.

In a paper in 2019, more details of this experiment are discussed and two shorter decay chains from the same irradiation assigned to $^{293}$Ts are presented~\cite{Khuyagbaatar2019}. 
These chains running down to $^{277}$Mt are in agreement with two out of the sixteen decay sequences observed also at the DGFRS of FLRN JINR~\cite{Oganessian2013b}, where the additional 14 decay chains are by one member shorter and end with $SF$ of $^{273}$Rg.

In line with what was discussed earlier in this review, e.g., for the dubnium isotopes $^{266-268}$Db, Khuyagbaatar et al.~consider the possibility that $^{277}$Mt undergoes $EC$ decay and the observed $SF$ belongs to $^{277}$Hs, which would be in agreement with the known properties of this nucleus.\\

\subsection{Oganesson - $Z$=118}\label{Og}
Finally, for the only one known isotope of the heaviest element oganesson, $^{294}$Og~\cite{Oganessian2015}, only the attempt to use a composite target, holding a cocktail of different californium isotopes was published recently~\cite{Brewer2018}. \\

 {\it $^{\it294}$Og}\\
The nucleus with the highest atomic charge $Z$\,=\,118 and together with $^{294}$Ts the highest mass number $A$\,=\,294, $^{294}$Og was observed four times~\cite{Oganessian2015}, before Brewer et al.~followed various theoretical predictions for synthesis cross-sections of up to $\approx$80~pbarn of oganesson isotopes with mass numbers from 292 to 298, employing the californium isotopes $^{249}$Cf to $^{252}$Cf as target material~\cite{Brewer2018}. 

In a first attempt, they used a mixed target consisting of a combination of $^{249}$Cf and $^{251}$Cf. 
For the beam energy, two values were chosen, with 252~MeV favoring the 3$n$ fusion-evaporation channel and 258~MeV populating most likely the 4$n$ channel.
One decay chain was observed, with three subsequent decays assigned to the sequence of a $^{294}$Og $\alpha$ decay, followed by $\alpha$ decay of $^{290}$Lv and terminated by $SF$ of $^{286}$Fl, increasing the total number of observed $^{294}$Og nuclei to five.



\onecolumn
\begin{center}\scriptsize
\begin{longtable}{c|ccccccl}
\caption {
List of basic decay properties, including year of discovery and relevant literature references, for all hitherto experimentally known isotopes ($^AZ$) and isomeric states ($^{Am}Z$) from $Z$=99 to 118. 
Most of the data were retrieved from the various services offered by the National Nuclear Data Center (NNDC) of Brookhaven National Laboratory (BNL) via its website: \href{https://www.nndc.bnl.gov/}{https://www.nndc.bnl.gov/} as well as from the NUBASE2020 data (\href{https://www-nds.iaea.org/relnsd/nubase/nubase_min.html}{https://www-nds.iaea.org/relnsd/nubase/nubase\_min.html}) 
as reported in the Atomic Mass Evaluation 2020 documents (decay data from Ref.~\cite{Kondev2021} and $Q_\alpha$ values from~\cite{Wang2021}). 
Exceptions are indicated in the footnotes to this table. 
In cases of asymmetric error bars for values based on low statistics, NUBASE symmetrizes and gives the central values of the uncertainty range as nominal ones. 
Here, I keep the original value and asymmetric error bars as given in the original publication.
Half-lives $T_{1/2}$, and g.s.~spins and parities,  $J_{gs}^\pi$ (where available), are also given for isomeric states. 
For excitation energies of isomeric states, please refer to the quoted literature or to the respective (sub)section in this review. 
Symmetric error bars are given in brackets for the last digits of the corresponding value, while for asymmetric ones, they are quoted in the value units. 
$Q_\alpha$ values marked by $\#$ are not purely experimental but are estimated on the basis of the values of neighboring nuclei (see explanation in~\cite{Wang2021}). 
For isomers, the excitation energy $E^*$ of the metastable state is shown instead of $Q_\alpha$, if known.
The year of discovery has been verified with the data reported in~\cite{Thoennessen2016}, including updates from the Discovery of Nuclides Project at \href{https://people.nscl.msu.edu/~thoennes/isotopes/}{https://people.nscl.msu.edu/$\sim$thoennes/isotopes/}.
For references that were originally published in Russian language their corresponding publications are given, which provide the translation in English.
Due to space limitations, for isotopes with more than six references, the reference list in the last column is continued in the following lines.
\label{tab:isotope_list}}\\
\end{longtable}
\begin{tablenotes}\scriptsize
\item{$^a$} values are taken from~\cite{Khuyagbaatar2020} (see subsection~\ref{Md} and discussion in Ref.~\cite{Hessberger2021})
\item{$^b$} nomenclature taken from Ref.~\cite{Antalic2010}. The authors did/could not deduce which level is the ground state and which is the isomer. The nominal g.s. $Q_\alpha$ values is given while isomer level energies are not known.
\item{$^c$} value is taken from Ref.~\cite{Hessberger2022}
\item{$^d$} values are taken from~\cite{Svirikhin2021}  while a conservative error for $Q_\alpha$ which was not given, was deduced from Fig.~1 in~\cite{Svirikhin2021}
\item{$^e$} first discovery report~\cite{TerAkopian1975} not confirmed and disproved by new results~\cite{Oganessian2001,Belozerov2003,Peterson2006}
\item{$^f$} earlier assignment to $^{249}$No~\cite{Belozerov2003,Popeko2003,Yeremin2003} (see subsection~\ref{No} and discussion in Ref.~\cite{Peterson2006})
\item{$^g$} values are taken from Ref.~\cite{Kessaci2021}
\item{$^h$} value is taken from Ref.~\cite{Pore2024a}
\item{$^i$} values are taken from Ref.~\cite{Vostinar2019}
\item{$^j$} values are taken from Ref.~\cite{Oganessian2022d}
\item{$^k$} values are taken from Ref.~\cite{Khuyagbaatar2025}
\item{$^l$} values are taken from Ref.~\cite{Khuyagbaatar2021}, confirming the original tentative assignment of an 11$^{+6}_{-3}$~ms activity in Ref.~\cite{Hessberger1997} (see subsection~\ref{Rf})
\item{$^m$} values are taken from Ref.~\cite{Mosat2020}
\item{$^n$} values are taken from Ref.~\cite{Oganessian2024} 
\item{$^o$} values are taken from Ref.~\cite{Oganessian2022b}
\item{$^p$} values are taken from Ref.~\cite{Pore2024}
\item{$^q$} values are taken from Ref.~\cite{Oganessian2023} 
\item{$^r$} nomenclature taken from Ref.~\cite{Haba2012}
\item{$^s$} values (and nomenclature) are taken from Ref.~\cite{ZhaoZ2024}; due to a large scattering of the experimental $\alpha$ decay energies the $Q_\alpha$ values from~\cite{Wang2021} are kept in Fig.~\ref{fig:Qa_94_118}.
\item{$^t$} value as reported in Ref.~\cite{Utyonkov2018}
\item{$^u$} value as reported in Ref.~\cite{Hofmann2000}  
\item{$^v$} value deduced from Ref.~\cite{Kaji2017}
\item{$^w$} value taken from Ref.~\cite{Oganessian2025}
\vspace{-0.3 cm} 
\end{tablenotes}   
\end{center}
\twocolumn

\newpage
\section{Outlook}\label{outlook}

The long history of the search for the heaviest nuclear species, almost coinciding with the birth of nuclear physics through the discovery of radioactivity by Henri Becquerel in 1896, has provided us with many fascinating discoveries and exciting insights into fundamental properties of nuclear matter. 
Throughout these last almost 130 years, not only 32 new heavy elements ($Z$\,>\,82) have been found, seven discovered in the gap between bismuth and uranium and 26 newly synthesized from neptunium to oganesson, but more importantly, many fundamental properties of nuclear matter have been revealed in the study of this  exotic species, advancing science as a whole. 
Here the decay processes, $\alpha$-, $\beta$-, $gamma$-, $IC$-decay, and $SF$, first exciting new features by themselves, have developed into powerful tools to understand basic properties like nuclear excitation, deformation and shapes, isomerism and the stability of nuclear matter itself. 
The first nuclear isomer was found in a heavy nucleus, $^{234}$Pa, by chemists, which explains the notion borrowed from chemistry~\cite{Ackermann2017}. 
The findings in (heavy) nuclei led to many applications in medicine and applied science.

SHE and SHN research, throughout the years and decades, always reached a point when progress seemed to come to a halt. 
A step in technological development and advancement of the experimental toolset, introducing accelerators, detection techniques and efficient separation methods, often led then 
to a substantial jump in experimental progress, enlarging our understanding of the heaviest nuclei.

At present, with the advance of acceleration technology, promising and partly delivering already the capability of an order of magnitude higher beam intensity for the low cross-section processes governing SHE/SHN research. 
New separators, developed further for efficient separation, mass identification and handling of those high beam intensities, including target technology, are starting operation now. 
The advancement of detection set-ups with digital signal processing and the introduction of new methods, borrowed from neighboring fields, like the classic connection to nuclear chemistry, but also the recent introduction of methods coming from atomic physics, like precise mass measurements in Penning traps and MR-ToF mass spectrometers, or laser spectroscopy, have the potential to push fundamental SHN/SHE research to new horizons. 
Atomic and nuclear properties like binding energies, ionization potentials and atomic excitation levels, as well as the single-particle and collective nuclear excitations, will foster the progress towards the eventual localization for the next proton and neutron shell closures. 

New installations are partly already available at accelerator facilities worldwide. 
The quest for the discovery of new elements following traditional pathways is in the focus of some SHE research teams, like the group working at the gas-filled separators DGFRS~\cite{Tsyganov1999} of FLNR JINR in Dubna, Russia, or GARIS at RIKEN in Wako, Japan \cite{Morita1992}.  
The introduction of novel techniques like ion traps for high-precision mass measurements and laser spectroscopy like the velocity filter SHIP of GSI/FAIR, Darmstadt, Germany, is pursued by others. 

The basic nuclear structure features are in the focus of research programs on nuclear spectroscopy, in-beam and, in particular, decay spectroscopy after separation (DSAS) at installations like the mass spectrometer FMA \cite{Davids1989} and the gas-filled separator AGFA~\cite{Seweryniak2013} at ANL in Lemont, IL, U.S.A., RITU~\cite{Leino1995} at the cyclotron laboratory of the University of Jyv\"{a}skyl\"{a} in Finland, the gas-filled separator BGS equipped with the FIONA mass separator at LBNL, Berkeley, CA, U.S.A.~\cite{Gates2022}, the gas-filled separator TASCA at GSI/FAIR~\cite{Semchenkov2008} and the velocity filter SHELS at FLNR JINR, Dubna, Russia~\cite{Yeremin2015}. 

For the major working horses, the accelerators, the next-generation high-intensity stable beam facilities are already or will be coming online soon, like the SHE-factory of FLNR JINR with the new gas-filled separator DGFRS2-2~\cite{Oganessian2022a} with its early achievements mentioned in this review, the linear accelerator RILAC with its upgraded performance in terms of beam intensities at RIKEN, Tokyo, Japan, or the HELIAC project of GSI/FAIR, which is still in an early stage with the first components being tested. 
The next facility in line to actively attack the challenge of SHN research is the SPIRAL2 LINAC at GANIL, which has started operation of its first phase being limited to the highest intensities to projectile masses $A$\,$\lessapprox$\,40.
The complete capabilities of this installation will be available with the second phase of the project, the new injector NEWGAIN~\cite{Ackermann2021} will then provide the highest intensities for all ions, and together with the separator-spectrometer set-up S$^3$~\cite{Dechery2015} will be one of the worldwide most competitive facilities for the investigation of SHN. 
Combined with the comprehensive separation and detection installation S$^3$, equipped with the detection array for Spectroscopy and Identification of Rare Isotopes Using S$^3$ (SIRIUS) as well as the S$^3$ Low Energy Branch (S$^3$ LEB)~\cite{Ajayakumar2023} offering tools for laser spectroscopy, mass measurement and the setup for laser/trap-assisted DSAS SEASON, it will be ready to advance SHE/SHN research to the next level.

In conclusion, building on the experience of more than a century, with the installations starting operation in these days worldwide, SHE/SHN research faces possibly the next
major jump in experimental progress, with the potential to substantially contribute to the fundamental understanding of nuclear matter and the nature of the strong force.

	\newpage
	\section*{Acknowledgments}
	Over the decades, I profited enormously from the vast knowledge and experience which my teachers, supervisors and collaborators shared generously with me. 
    The recent years saw some of them go, leaving us with partly unrecoverable vacancies which we will have a hard time to fill. 
    After Sigurd Hofmann had left us in 2022, this year, 2024, saw Gottfried Münzenberg and Peter Armbruster go, all of them supporting me and my scientific development, starting from my early days as a diploma student in Darmstadt. 
    
    To Christophe Theisen I owe a lot and I mentioned it already in the dedication at the beginning of this review. 
    
    There are numerous scientists, students, engineers, technicians and administrators without whom our work and our achievements, and in particular mine, would not be possible. 
    For this review, I would like to express my deepest gratitude to three colleagues who shaped my understanding of our common research in many discussions and collaborative efforts, often leading to acknowledged publications. 
    Fritz He{\ss}berger, my mentor from my first insecure steps on, and throughout my whole career, Stanislav Antalic, my collaborator in DSAS projects and together with Fritz co-author of our last review on SHN isomers, and  Christelle Stodel, my local colleague and daily advisor at GANIL, with her patience and friendly understanding, and her help with some of the graphics in this review. 
    I thank Julien Piot for carefully reading the document.
    
    Finally, I would like to thank my wife, Elena Litvinova for her patient support, understanding and advice when it comes to the real, i.e., the theory questions. 


	\newpage
	\bibliography{DA}
	


	
	

\end{document}